\documentclass[12pt]{article}
\pdfoutput=1
\usepackage{graphicx}
\usepackage{epstopdf}
\usepackage{amsmath}
\usepackage{amsfonts}
\usepackage{amssymb}
\usepackage{color}
\usepackage{mathrsfs}
\usepackage[utf8]{inputenc}
\usepackage{subcaption}
\usepackage{cite} 

\usepackage[font={footnotesize}]{caption}

\pdfoutput=1

\usepackage[english]{babel}

\usepackage[colorlinks,bookmarks]{hyperref}
\definecolor{linkblue}{rgb}{0,0,0.8}
\definecolor{linkgreen}{rgb}{0,0.5,0}

\hypersetup{pdfpagemode=UseNone, pdfstartview=FitH, linkcolor=linkblue, %
            citecolor=linkgreen, urlcolor=linkblue}

\newcommand{\website}{\url{http://web.stanford.edu/~senatore/}}

\numberwithin{equation}{section}

\newcommand{\hinvMpc}{\,h\, {\rm Mpc}^{-1}\,}

\def\calH{{\cal{ H}}}

\def\vk{{\vec k}}
\def\vq{{\vec q}}
\newcommand{\tT}{\Theta}
\newcommand{\td}{\delta}

\def\d{{\partial}}
\def\nn{{\nonumber}}
\def\xfl{{\vec x_{\rm fl}}}
\def\vdm{{\vec v_{\rm dm}}}

\def\r{{ _{\rm rel}}}
\def\R{{{\rm R}}}

\def\C{{{\rm C}}}

\newcommand{\bea}{\begin{eqnarray}}
\newcommand{\eea}{\end{eqnarray}}
\newcommand{\be}{\begin{equation}}
\newcommand{\ee}{\end{equation}}

\newcommand{\vnl}{v_{\rm NL}}
\newcommand{\kfs}{k_{\rm fs}}

\newcommand{\knl}{k_{\rm NL}}

\newcommand{\qvec}{\vec{q}}

\newcommand{\Comment}[1]{{}}

\def\thefootnote{\fnsymbol{footnote}}

\begin{document}

\setcounter{page}{1} \baselineskip=15.5pt \thispagestyle{empty}

\begin{flushright}
\end{flushright}

\begin{center}

{\Large \bf  The Effective Field Theory of Large-Scale Structure\\ extended to Massive Neutrinos   \\[0.7cm]}
{\large  Leonardo Senatore$^{1}$ and Matias Zaldarriaga$^{2}$}
\\[0.7cm]

{\normalsize { \sl $^{2}$  SITP and KIPAC, Department of Physics and SLAC, \\
Stanford University, Stanford, CA 94305}}\\
\vspace{.3cm}

{\normalsize { \sl $^{2}$ School of Natural Sciences, \\ Institute for Advanced Study, Olden Lane, Princeton, NJ 08540}}\\
\vspace{.3cm}

\end{center}

\vspace{.8cm}

%
%
%

\hrule \vspace{0.3cm}
{\small  \noindent \textbf{Abstract} \\[0.3cm]
\noindent  We develop a formalism to analytically describe the clustering of matter   in the mildly non-linear regime in the presence of massive neutrinos.  
 Neutrinos, whose free streaming wavenumber ($\kfs$) is typically longer than the non-linear scale ($\knl$) are described by a Boltzmann equation coupled to the effective fluid-like equations that describe dark matter. We solve the equations expanding in the neutrino density fraction $(f_\nu)$ and in $k/ \knl$, and add suitable counterterms to renormalize the theory. This allows us to describe the contribution of short distances to long-distance observables.  Equivalently, we construct an effective Boltzmann equation where we add additional terms whose coefficients renormalize the contribution from short-distance physics. We argue that neutrinos with $\kfs\gtrsim \knl$ require an additional counterterm similar to the speed of sound ($c_s$) for dark matter. We compute the one-loop total-matter power spectrum, and find that it is roughly equal to $16f_\nu$ times the dark matter one for $k$'s larger that the typical $\kfs$. It is about half of that for smaller $k$'s. The leading contribution results from the back-reaction of the neutrinos on the dynamics of the dark matter. The counterterms contribute in a hierarchical way: the leading ones can either be computed in terms of $c_s$, or can be accounted for by shifting $c_s$ by an amount proportional to $f_\nu$.

 \vspace{0.3cm}
\hrule

\def\thefootnote{\arabic{footnote}}
\setcounter{footnote}{0}

%
%
%
%
\newpage
\tableofcontents
\section{Introduction}

By now, there is a well established effort to apply effective field theory techniques to make analytical predictions for Large-Scale Structure (LSS) observables. Overall this  research program aims to describe the evolution of dark matter, baryons and other more exotic components  such as dark energy. It goes under the name of the Effective Field Theory of Large Scale Structure (EFTofLSS)~(see for example~\cite{Baumann:2010tm,Carrasco:2012cv,Porto:2013qua,Senatore:2014via,Carrasco:2013sva,Carrasco:2013mua,Pajer:2013jj,Carroll:2013oxa,Mercolli:2013bsa,Angulo:2014tfa,Baldauf:2014qfa,Senatore:2014eva,Senatore:2014vja,Lewandowski:2014rca,Mirbabayi:2014zca,Foreman:2015uva,Angulo:2015eqa,McQuinn:2015tva,Assassi:2015jqa,Baldauf:2015tla,Baldauf:2015xfa,Foreman:2015lca,Baldauf:2015aha,Baldauf:2015zga,Bertolini:2015fya,Bertolini:2016bmt,Assassi:2015fma,Lewandowski:2015ziq,Cataneo:2016suz,Bertolini:2016hxg,Fujita:2016dne,Perko:2016puo,Lewandowski:2016yce,Lewandowski:2017kes,delaBella:2017qjy}). The idea that unifies all of these efforts is  that analytic predictions for LSS observables on large scales can be obtained with {\it arbitrary accuracy} by perturbatively expanding in the ratio of the long wavelength associated to the non-linear scale over the long wavelength of interest. This can be achieved by systematically accounting for the effect of uncontrolled non-linear short-distance physics through the addition of suitable counterterms to the equations that describe the dynamics of the system.

There is one known ingredient of the universe that has not yet been described in the context of the EFTofLSS, neutrinos. Due to their non-vanishing mass, neutrinos become non-relativist at late times and cluster, affecting the clustering of the other components on  large distances. Though the overall effect is suppressed by the smallness of their contribution to the overall energy density (about 1\%), it is  expected that upcoming LSS observations will become sensitive to the effect of massive neutrinos (see for example~\cite{Audren:2012vy,Cerbolini:2013uya}). In that case, if we are able to predict the effect of neutrinos with sufficient accuracy, we could determine their masses from cosmological observations. This would be an amazing result.

There have been so-far two main approaches to describe the effect of neutrinos on LSS. One has been to develop numerical simulations that include a population of neutrinos (see for example~\cite{Villaescusa-Navarro:2013pva,Baldi:2013iza,Zennaro:2016nqo,Banerjee:2016zaa}), and the other has been to model them with analytical methods of various levels of accuracy (see for example~\cite{Hu:1998kj, Saito:2008bp, Wong:2008ws, Lesgourgues:2009am, Upadhye:2013ndm, Shoji:2010hm, Lesgourgues:2011rh, Shoji:2009gg, Blas:2014hya}). There are also mixed approaches that solve numerically for dark matter and model neutrinos analytically~\cite{AliHaimoud:2012vj}.

In this paper we take a different approach that  involves only controlled approximations ({\it i.e.} approximations that are proportional to a small parameter and for which one can, at least in principle, compute the subleading corrections). We describe the neutrinos using a Boltzmann equation, and solve for it by expanding  both in the smallness of their relative energy density  and also in the ratio of the wavenumber of interest over the wavenumber associated to the non-linear scale. In solving the Boltzmann equation at non-linear level, we encounter products of gravitational fields evaluated at the same location which are therefore sensitive to unknown short distance non-linearities. We renormalize these local products of fields by adding  suitable counterterms that allow us to correctly parametrize the effect of short distance non-linearities at long distances. Additionally, it is important that the gravitational field that enters in the Boltzmann equation is described using the EFTofLSS counterterms, so that it has the correct behavior on long distances. For `fast' neutrinos, that is neutrinos whose free streaming wavenumber $\kfs$ is longer than the wavenumber associated to the non-linear scale, $\knl$, each interaction corresponds to a small correction, and we can therefore safely truncate the expansion. Instead, for neutrinos that have a velocity so slow that $\kfs\gtrsim\knl$, each interactions involving fields with wavenumber higher than $\knl$ corresponds to an order one perturbation for which we cannot trust the resulting perturbative expansion. This leads us to account for this different behavior between `slow' and `fast' neutrinos by adding a sort of speed-of-sound counterterm for slow neutrinos.

Dark matter can be described at long distances as a fluid-like system, endowed with a stress tensor that accounts for the effect of short distance non-linearities. Through the gravitational force the clustering of neutrinos  affects the dark matter dynamics.  We account for this effect by perturbatively solving the coupled (neutrino Boltzmann)+(dark-matter effective-fluid) equations. The presence of the neutrino density in the Poisson equation introduces new non-linearities in the momentum equation for dark matter. As we did for the solution of the neutrino Boltzmann equation, we also renormalize the resulting local products of fields by adding suitable counterterms. Since these local products are the same that appeared in the Boltzmann equation, no new parameter is actually added.

Concerning the Boltzmann equation, instead of renormalizing the product of fields that appear in its solution, one could wonder if we can develop a set of modified equations whose  solution automatically encoded the effect of uncontrolled short-distance physics. This is what we have for dark matter, whose dynamical equations take the form of an effective fluid-like system. The difficulty in doing this for neutrinos in general stems from  the fact that fast neutrinos travel a distance much longer than the non-linear scale in a Hubble time: naively, the resulting equations would  not just be non-local in time as the dark matter ones but also non-local in space.  The spatial non-locality scale is of order of the free streaming length making the equations  hard to deal with. For fast neutrinos, we are able to overcome this issue by expanding not only in the smallness of the gravitational interactions, but also in the small ratio of the non-linear scale over the distance traveled in a Hubble time. This allows us to develop an `Effective Boltzmann equation' that we derive in App.~\ref{app:EffectiveBoltzmann}. In the main text we focus on the approach that we described in the two paragraphs above, that renormalizes directly the products of fields appearing in the solution of the ordinary Boltzmann equation.

We then proceed to compute the one-loop power spectrum for the total matter and present the results in Sec.~\ref{sec:results}. Very roughly speaking, we find that at wavenumbers greater than the neutrino's free streaming length, the contribution of neutrinos scales approximately as $16 f_\nu$, while at lover wavenumbers is about half of that. The enhancement with respect to the naive $f_\nu$ is due to the boost of order the logarithm of the redshift of matter-radiation equality, that originates from the modification, at linear level, of the Poisson equation due to the presence of neutrinos. However, this is not quite all: the contribution of the other diagrams counts for about 30\% of the result.  

We find that the leading counterterm is associated to the usual dark matter  $c_s$. Even though the coefficient is the same as the  dark-matter $c_s$ counterterm the functional form of the resulting power spectra is different. The contribution associated  the the renormalization of the new local products of fields that appear only because of the presence of neutrinos is highly subleading apart for a contribution that can be accounted for by shifting the $c_s$ coefficient of  the ordinary dark-matter  counterterm by an amount of order $f_\nu$. This is in accord to the fact that fast neutrinos are rather insensitive to scales much shorter than their free streaming length, a fact that the effective Boltzmann equation makes evident. We leave to future work a comparison with simulations and with the phenomenological models available in the literature. If general, however, we anticipate that since neutrinos represent a small contribution to the overall energy density, one can tolerate larger mistakes in the theoretical predictions than for dark matter before the inferences from measurements begin to be systematically affected.

\section{Set-up Equations}

We wish to provide a set of equations that allows us to compute the correlation function of observables in LSS accounting for the presence of neutrinos. We do not intend to solve the system exactly, rather we aim at providing a system of equations that can be solved perturbatively and that on large scales can reach arbitrary precision by going to sufficiently high order in the expansion parameter.  

If we assume that we have arbitrarily accurate knowledge of the physics down to the smallest astrophysical scales, then the evolution of neutrinos would be described by a collisionless Boltzmann equation. For applications to LSS, we will be interested on length scales much smaller than the Horizon, and in the regime where the neutrinos are already non-relativistic. We define the comoving momentum as $\vec q=a m\, d\vec x/d\tau $ where $\tau$ is the conformal time, and the comoving velocity as $\vec v=\vec q/m=a \, d\vec x/d\tau$ (notice that this definition of velocity is not the standard $d\vec x/d\tau$). We denote the proper number density of neutrinos with coordinate position $\vec x$ and comoving velocity $\vec v$ as $f(\vec x,\vec v,\tau)$. In this limit the Boltzmann equation becomes (see for example~\cite{Bertschinger} or~\cite{AliHaimoud:2012vj} for a pedagogical re-derivation): 
\be\label{eq:Boltzone}
\frac{\d f(\vec x,\vec v,\tau)}{d\tau}+ \frac{v^i}{a} \frac{\d f(\vec x,\vec v,\tau)}{\d x^i}-   a \frac{\d\Phi}{\d x^i}\, \frac{\d f(\vec x,\vec v,\tau)}{\d v^i}=0\ ,
\ee
where $\Phi$ is the gravitational potential. There are relativistic corrections to~(\ref{eq:Boltzone}) but they are negligible~\footnote{If we look for example at~\cite{LoVerde:2013lta} we notice a correction in $\dot \Phi$. The ratio of what we keep over the new term is
\be
\frac{v \dot \Phi}{\d_i \Psi}\sim \frac{v H }{k }\sim \left(\frac{H}{k}\right)\left(\frac{v }{H}\right)\sim \left(\frac{H}{k}\right)\left(\frac{H}{\kfs}\right)\ ,
\ee
where $\kfs$ is the wavenumber associated to the free streaming length. 
For the $k$'s of interest are $ \left(\frac{H}{k}\right)\lesssim 10^{-2}$, and for the masses of interest, $\left(\frac{H}{\kfs}\right)\lesssim 10^{-2}$. So $\frac{v \dot \Phi}{\d_i \Psi}\lesssim  10^{-4}$, which makes it safely negligible. Of course, we need to start to solve our equations at late enough times so that the relativistic corrections in $v$ are negligible.}.  In Sec.~\ref{sec:perturbativebegin} we will describe how to correctly parametrize the effects of the uncontrolled short distance non-linearities on the perturbative solution to this Boltzmann equation. 

For dark matter we know that at distances longer than the non-linear scale we can describe it as an effective fluid-like system. Therefore the equations of motions of dark matter in the presence of a neutrinos are given by~\cite{Baumann:2010tm,Carrasco:2012cv,Carrasco:2013mua}: 
\bea\label{eq:newdmequations2} \label{nonlincont}\label{nonlineuler} \label{eq:newdmequations}
&&\frac{\d^2}{a^2}\Phi= \frac{3}{2}H_0^2 \frac{a_0^3}{a^3} \left(\Omega_{\rm dm,0} \delta_{\rm dm}+\Omega_{\nu, 0} \delta_\nu\right)\simeq \frac{3}{2}H_0^2 \frac{a_0^3}{a^3} \Omega_{\rm NR,0} \left(\delta_{\rm dm}+f_\nu \left(\delta_\nu-\delta_{\rm dm}\right)\right)\\ \nn
&&\frac{1}{a}\frac{\d \delta_{\rm dm}}{\d \tau} + \frac{1}{a\rho^{(0)}_{\rm dm}}\d_i\pi^i_{\rm dm}=0\ , \\ \nn
&&\frac{1}{a}\frac{\d\pi^i_{\rm dm}}{\d\tau} + 4 H\pi^i_{\rm dm}+\frac{1}{a}\d_j\left(\frac{\pi^i_{\rm dm}\pi^j_{\rm dm}}{\rho^{(0)}_{\rm dm}(1+\delta_{\rm dm})}\right)+\frac{1}{a}\rho^{(0)}_{\rm dm} (1+\delta_{\rm dm}) \d^i\Phi= -\frac{1}{\, a}\d_j\tau^{ij}\ .
\eea
where $\pi^i_{\rm dm}$ is the dark matter momentum, and $\Omega_{\rm NR,0}$ the present energy fraction in non-relativistic matter: $\Omega_{\rm NR,0}=(\rho^{(0)}_{\rm dm}(t_0)+\rho^{(0)}_{\rm \nu}(t_0))/\rho_{\rm tot}(t_0)$. $\tau_{ij}$ is the effective stress tensor for dark matter~\cite{Baumann:2010tm,Carrasco:2012cv}. It can be expressed at leading order as (see for example~\cite{Carrasco:2013mua})
\be
\tau_{ij}(\vec x)=\int^\tau d\tau'  \left[K_{1}(\tau,\tau')\, \delta_{\rm dm}(\xfl(\vec x,\tau,\tau'),\tau') +\ldots \right]\ .
\ee
where $\ldots$ stands for higher order and higher derivative terms and
\be\label{eq:xfl}
\xfl(\vec x, t,t') = \vec x - \int_{\tau(t')}^{\tau(t)} d\tau''\; {\vdm}(\tau'',\vec x_{\rm fl}(\vec x, \tau,\tau''))\ ,
\ee
with $\vdm$ being the dark matter velocity, defined as $\vdm =d\vec x/d\tau$ (this is the standard definition for velocity, different than the one we are using for neutrinos). $\xfl(\vec x, \tau,\tau')$ represents the location at time $\tau'$ of the fluid element that at time $\tau$ is at location $\vec x$. It is defined iteratively in the dark matter velocity. The effective stress tensor is sufficient to correctly parametrize the effect of uncontrolled short distance non-linearities on the perturbative solution for dark matter when the universe contains only dark matter. The generalization to a Universe with baryons was given in~\cite{Lewandowski:2016yce} where it was  shown for example that stellar and galactic dynamics can be correctly parametrized in the EFTofLSS. In Sec.~\ref{sec:Pdmdm_counter} we will describe how the presence of neutrinos forces the addition of new counterterms beyond the stress tensor.

For definiteness, in this paper we will focus on the one-loop power spectrum, though our formalism applies to general correlation functions.  We write
\be
\rho(\vec x,t)=\rho_{\rm dm}(\vec x,t)+ \rho_\nu(\vec x,t)\quad\Rightarrow\quad
\delta=(1-f_\nu)\delta_{\rm dm}+ f_\nu\, \delta_\nu=\delta_{\rm dm}+ f_\nu\left(\delta_\nu-\delta_{\rm dm}\right) \ , 
\ee
where $\rho_\nu(\vec x,t)=\rho_\nu^{(0)}(t)\,\int d^3 v\; f(\vec v,\vec x,t)$ (we chose to normalize to one the neutrino distribution $f(\vec v,\vec x,t)$),   $\delta_{\rm dm}={\rho_{\rm dm}}/{\rho^{(0)}_{\rm dm}(t)}-1$,  $\delta_{\rm \nu}={\rho_{\rm \nu}}/{\rho^{(0)}_{\rm \nu}(t)}-1$ and  $ \delta_{}={\rho}/{\rho^{(0)}}-1$. We define 
\be
f_\nu=\rho_{\rm \nu}^{(0)}/(\rho_{\rm dm}^{(0)}+\rho_{\rm \nu}^{(0)})\ .
\ee 
We can therefore write
\bea
&&\langle\delta(\vec k,t)\,\delta(\vec k', t)\rangle=\langle\delta_{\rm dm}(\vec k,t)\,\delta_{\rm dm}(\vec k', t)\rangle+2 f_\nu  \langle\delta_{\rm diff}(\vec k,t)\,\delta_{\rm dm}(\vec k', t) \rangle\\ \nn
&&\ \ \qquad\qquad\qquad=(2\pi)^3 \delta_D^{(3)}(\vec k+\vec k')\left( P_{\rm dm,\,dm}(k)+2 f_\nu\,P_{\rm \rm diff,\,dm}(k)\right)\ ,
\eea
where we defined
\be
\delta_{\rm diff}=\delta_{\nu}-\delta_{\rm dm}\ .
\ee
$\delta_{\rm diff}$ has the nice property that at linear level it is zero at distances longer than the free streaming length  of the neutrinos. As we will see, computing correlation functions involving $\delta_{\rm diff}$ is as easy as (or rather as difficult as) computing correlation functions involving $\delta_\nu$. Furthermore we notice that the solution for $\delta_{\rm dm}$ is the same as the one for~$\delta_\nu$, once we substitute in the solution for~$\delta_\nu$ a distribution with vanishing temperature. So, in practice the solution for~$\delta_{\rm diff}$ is the same as for~$\delta_{\nu}$ with the replacement of the Fermi distribution for neutrinos with the difference of two Fermi distributions, one with  zero temperature. For ease of notation whenever the context makes it clear we will call $\delta_{\rm diff}$  the neutrino-$\delta$, ($\delta_\nu$), though we always compute correlation functions of $\delta_{\rm diff}$. 

In order to find our perturbative solution to the Boltzmann equation, we will need expressions for the Fourier transform of the unperturbed distribution of the neutrinos ($\tilde f^{[0]}_\nu$) and dark matter~($\tilde f^{[0]}_{\rm dm}$): 
\be\label{eq:tildefnu}
\tilde f^{[0]}_\nu(q)=\int d^3 v \; e^{-i\,\vec q\cdot\vec v} \; f^{[0]}(v)\ .
\ee
$\tilde f^{[0]}_{\rm dm}$ is simply
\be
\tilde f^{[0]}_{\rm dm}(q)=1\ ,
\ee 
because dark matter is taken to be exactly cold.

 The Fourier transform of the relative distribution for the difference of the neutrinos and dark matter ($\tilde f^{[0]}$) is 
\be\label{eq:newftilde}
\tilde f^{[0]}(q)=\tilde f^{[0]}_\nu(q)-\tilde f^{[0]}_{\rm dm}(q)=\tilde f^{[0]}_\nu(q)-1\ ,
\ee
Clearly, the evaluation of the diagrams for $\delta_{\rm diff}$ is computationally as hard as the one for the diagrams involving $\delta_\nu$

For the Fermi distribution one can find a series expansion for $\tilde f^{[0]}_\nu$~\cite{Bertschinger}. Here we will use the fitting function provided in~\cite{AliHaimoud:2012vj} which is accurate to better than $3\%$ in the whole range of arguments and speeds up the numerical evaluation: 
\be
\tilde f^{[0]}_{\nu_i}(q)=\frac{0.0407 \left(q \frac{k_B T_{\nu,0}}{m_{\nu_i} c}\right)^4+0.0168 \left(q \frac{k_B T_{\nu,0}}{m_{\nu_i} c}\right)^2+1}{1.6787 \left(q \frac{k_B T_{\nu,0}}{m_{\nu_i} c}\right)^{4.1811}+0.1467 \left(q \frac{k_B T_{\nu,0}}{m_{\nu_i} c}\right)^8+2.1734 \left(q \frac{k_B T_{\nu,0}}{m_{\nu_i} c}\right)^2+1}\ ,
\ee
where $T_{\nu,0}$ is the temperature of the unperturbed neutrino distribution, $m_{\nu_i}$ is the mass of the nutrinos of type $i$. $c$ is the speed of light and $k_B$ is the Boltzmann constant, that we did not set to one to help the reader perform the conversion. Since there are three neutrinos, we have:
\be
\tilde f^{[0]}_{\nu}(q)=\frac{1}{\sum_j m_{\nu_j}}\sum_i  m_{\nu_i}\;\tilde f^{[0]}_{\nu_i}(q)\ .
\ee

\section{Perturbative Solutions and their UV sensitivity\label{sec:perturbativebegin}}

We now begin to describe the perturbative solution for the neutrino Boltzmann equation~(\ref{eq:Boltzone}). It is very useful to immediately make an important expansion. The gravitational potential $\Phi$ is sourced by neutrinos only in an amount proportional to $f_\nu\sim 10^{-2}$. If we work at leading order in~$f_\nu$, we can regard the gravitational field in (\ref{eq:Boltzone}) as an external source, {\it i.e.} a term whose dynamics is not affected by the presence of neutrinos. This is the limit we will adopt in this paper, and we leave going to higher order in $f_\nu$ to future work~\footnote{Going to higher order in~$f_\nu$ appears to be unfortunately almost a purely academic question, at least for the time being. At leading order in~$f_\nu$, our results for the total matter power spectrum will be inaccurate by order~$f_\nu^2\delta^2\lesssim {\cal O}(10^{-4},10^{-5})$, where $\delta$ is the dark matter overdensity at the length scale of interest. In the future, we will be very happy to make calculations accurate to this order if experiments have sensitivity to such small corrections.

The extension to higher order should be conceptually straightforward. It should be enough to write all fields as an expansion in $f_\nu$, as for example for $\Phi$ one would write $\Phi=\Phi_0+ f_\nu \Phi_1+ f_\nu^2 \Phi_2+\ldots$, and solve all the equations at the relevant order in $f_\nu$. Of course, the number of diagrams would probably quickly become very large.}.

It is worthwhile to first develop some intuition on what controls the perturbative expansion. Solving the Boltzmann equation amounts to solving for the trajectory of the particles that started in the initial conditions at a given location with a given velocity. Of course, we can also consider the backward problem: given, at a certain time, a particle with a certain position and velocity, we can ask what was its initial position and velocity. Let us focus on the backward problem and let us consider the deviation from a straight trajectory of such a particle when it crosses a region of size $L\sim 1/k$. Let us consider only one Hubble time and for this reason, let us neglect in this schematic representation all factors of the scale factors. The particle trajectory is schematically given by the following iterative solution which accounts from the deviation from the gravitational force
\bea
&&x(t)-x(t_{\rm in})\sim \int^t dt_1\; \left(v-\int^{t_1}dt_2\;\d\Phi\left(x-\int^{t_2}dt_3\left(v- \int^{t_3} dt_4\; \d\Phi\left(x-\ldots\right)\right)\right)\right)\\ \nn
&&\quad\qquad\qquad\sim v \Delta t + \d\Phi\,  (\Delta t)^2+ \d^2\Phi\, v \,(\Delta t)^3+\ldots\ .
\eea
The ratio of the first two terms, that we denote by $\delta x^{(1)}$ and $\delta x^{(2)}$ respectively, tells us how much the gravitational force affects the trajectory. We obtain
\be
\frac{\delta x^{(2)}}{\delta x^{(1)}}\sim \frac{\d\Phi\, \Delta t}{v}\ .
\ee
The crossing time of a region of size $1/k$ is given by the minimum between an Hubble time (which the the whole time we have at our disposal), and $1/(k v)$: $\Delta t\sim{\rm Min}[1/H, 1/(k v)]$. There are therefore two cases to consider. For slow velocities, $v< H/k$, then the whole region is not crossed even in an Hubble time, but it is crossed just a subregion of size~$v/H\sim1/\kfs$. This is the case when the free-streaming wavenumber is larger than $k$, $\kfs\gtrsim k$. In this case, we can go to a local inertial frame and Taylor expand the gravitational potential just around its minimum and evaluate it at a distance of order $1/\kfs$: $\d\Phi\sim\d^2\Phi \cdot \Delta x\sim \d^2\Phi /\kfs$. Furthermore, we have $\Delta t\sim 1/H$. We therefore have 
\be
\left.\frac{\delta x^{(2)}}{\delta x^{(1)}}\right|_{\rm \kfs\gtrsim k}\sim  \frac{\d^2\Phi}{\kfs} \cdot\frac{1}{v H}\sim \frac{\d^2 \Phi}{H^2}\sim \delta(k)\ ,\quad{\rm for}\quad \kfs\gtrsim k\ , 
\ee
where we have used that $\d^2\Phi\sim H^2\delta$. If instead $\kfs\lesssim k$, the $1/k$ region is crossed in  $\Delta t\sim 1/(k\,v)$, and we obtain
\be
\left.\frac{\delta x^{(2)}}{\delta x^{(1)}}\right|_{\rm \kfs\lesssim  k}\sim  \frac{\d\Phi}{v} \cdot\frac{1}{v k}\sim \frac{ \Phi}{v^2}\sim \delta(k) \left(\frac{\kfs}{k}\right)^2\ ,\quad{\rm for}\quad \kfs\lesssim k\ .
\ee
In our universe, where $\Phi$ is approximately scale invariant, this parameter is less than one at all wavenumber of interests. In fact, if we suggestively define
\be
 \vnl\equiv \frac{H}{\knl}\ ,
\ee
we have that $\Phi\sim \vnl^2$, and the expansion parameter for $\kfs\lesssim k$ reads $\vnl^2/v^2$.

We can see that we can naturally split the neutrinos into two populations, characterized by different expansion parameters. We can define `fast' neutrinos as those whose $\kfs\lesssim \knl$ (or $v\gtrsim \vnl$). For these, the perturbative solution corresponds to expanding in a parameter which is always smaller than one, be it either $\delta(k)$, for $k\lesssim \kfs$,  or $ \frac{ \Phi}{v^2}$, for $k\gtrsim \kfs$, even when considering wavenumbers inside the non-linear scale.  Instead, there are slow neutrinos, defined so that $\kfs\gtrsim \knl$ (or $v\lesssim \vnl$), for which there are wavenumbers, $k\gtrsim \knl$, for which the expansion parameter is larger or order one.

Let us now proceed with the perturbative solution of the Boltzmann equation. Indeed, after Taylor expanding in $f_\nu$,~(\ref{eq:Boltzone}) is very prone to a perturbative solution. We write $f=f^{[0]}+f^{[1]}+f^{[2]}+\ldots$, where the superscript ${}^{[i]}$ refers to how many factors of the gravitational force $\vec\d\Phi$ have been inserted in the solution. $f^{[0]}$ is the homogenous solution, which does not depend on $\vec x$ nor on the direction of~$\vec v$. The zeroth order Boltzmann equation tells us that $f^{[0]}$ cannot depend on time $\tau$: $f^{[0]}(\vec x,\vec v, \tau)=f^{[0]}(v)$ (this is the reason for the unusual choice of the definition of velocity in $f$). In general if we go to Fourier space we have:
\be\label{eq:Boltzone_f}
\frac{\d f(\vec k,\vec v,\tau)}{\d\tau}+i \frac{\vec v\cdot \vec k}{a} f(\vec k,\vec v,\tau)-   a \left[\frac{\d\Phi}{\d x^i}\, \frac{\d f(\vec x,\vec v,\tau)}{\d v^i}\right]_k=0\ ,
\ee
where  $\left[{\cal O}(\vec x)\right]_{\vec k}$ means that we take the $\vec k$-component of the Fourier transform of ${\cal O}(\vec x)$.
The perturbative solution reads
\be\label{eq:perturbativeseries}
f^{[n]}(\vec k, \vec v, \tau)=\int_0^\tau d\tau'  \; G_R(\tau, \tau'; \vec v, k)  \left[ a(\tau') \left(\frac{\d}{\d x^i}\Phi(\vec x,\tau')\right) \frac{\d}{\d v^i} f^{[n-1]}(\vec x,\vec v, \tau')\right]_{\vec k}\ ,
\ee
where $G_R(\tau, \tau'; \vec v, \vec k)$ is the retarded Green's function from $\tau'$ to $\tau$. The Green's function is known analytically~\cite{Bertschinger,AliHaimoud:2012vj}:
\be
G_R(s, s'; \vec v, k)= e^{- i\, \vec k\cdot\vec v\, (s-s')}\ ,\qquad s(\tau)=\int^\tau \frac{d\tau'}{a(\tau')}\ .
\ee $s$ is the so-called `superconformal' time.
We therefore can more explicitly write
\be\label{eq:pert2}
f^{[n]}(\vec k, \vec v, s)=\int_0^\tau ds'\, a(s')^2  \;  e^{- i\, \vec k\cdot\vec v\, (s-s')} \left[  \left(\frac{\d}{\d x^i}\Phi(\vec x,s')\right) \frac{\d}{\d v^i} f^{[n-1]}(\vec x,\vec v, s')\right]_{\vec k}\ .
\ee

The compactness of Eq.~(\ref{eq:pert2}) hides two of its important limitations. First, the right-hand side of (\ref{eq:pert2}) will in general involve convolutions of fields evaluated at arbitrary high wavenumbers. In general, it is not clear that we have accurate knowledge of these fields at high wavenumber. In particular, the gravitational potential that is computed analytically by the EFTofLSS, which is the approach we will take in this paper, is usually accurate only for wavenumbers $k\ll \knl$, with $\knl$ being the wavenumber associated to the nonlinear scale. Therefore, we should formulate equations that can correct for these mistakes. A second and more important  limitation is the fact that any perturbative expansion has a radius of convergence outside of which  they do not converge anymore towards the correct answer~\footnote{In reality asymptotic series have vanishingly small radius of convergence. However, for a given value of the expansion parameter, they converge towards the exact result up to some distance beyond which non-perturbative effects become important and the answer begins to diverge from the correct one. In these cases, we can simply treat asymptotic series as normal perturbative series until the order in perturbation theory when non-perturbative effects become important.}. The convolutions in (\ref{eq:pert2}) will make high momentum contribution which can be  outside the radius of convergence affect the large scales.  We wish to correct~(\ref{eq:pert2}) in order to account for these effects as well. The reality of these limitations and how to adjust for them will become more evident as we proceed.

In fact, let us ignore for the moment this issue and try to solve perturbatively (\ref{eq:Boltzone}) using (\ref{eq:pert2}). At first order, we have
\bea\label{eq:first_order_sol}
f^{[1]}(\vec k, \vec v, s)= e^{- i\, \vec k\cdot\vec v\, (s-s_i)}\; \delta f (\vec k,\vec v,s_i)+\int_{s_i}^s ds'\, a(s')^2  \;  e^{- i\, \vec k\cdot\vec v\, (s-s')}   i \frac{ \vec v}{v}\cdot \vec k\, \Phi(\vec k,s') \frac{\d f^{[0]}(v)}{\d v}\ .
\eea
where we used that for a function of the modulus of $\vec v$, we have $\frac{\d f^{[0]}(v)}{\d v^i}= \frac{v^i}{v} \frac{\d f^{[0]}(v)}{\d v}  $. The first term is the contribution from the initial conditions, with $s_i$ being the initial time of our evaluation, and $ \delta f (\vec k,\vec v,s_i)$ the perturbation to the neutrino distribution at that time. The second term is the perturbation forced at later times by the dark matter gravitational potential.  It is very easy to estimate that if we start our evaluation at an early enough times, the contribution from the initial conditions is negligible. This is because neutrinos tend to cluster less than dark matter especially at early times when their velocity is higher. Even assuming that $\frac{\delta\rho_\nu}{\rho_\nu}(\vec k, s_i)\sim \frac{\delta\rho_{\rm dm}}{\rho_{\rm dm}}(\vec k, s_i)$ at the initial time, the induced overdensity at the wavenumbers of interest from the initial conditions at the present time is $\lesssim 10^{-3}\frac{\delta\rho_{\rm dm}}{\rho_{\rm dm}}(a_0)$ already if the initial redshift $z_i$  is taken to be $z_i\sim 10$ (and smaller if we start at higher redshifts)~\footnote{In practice, we will use $z_i=100$ for our numerical evaluations.}. We will therefore neglect the initial conditions for the rest of the paper, and write, for convenience, $s_i=0$ in the limit of integration of the forced solution.  This amounts to a mistake that is smaller than the $f_\nu^2$ terms that we neglect.

At second order we therefore have
\bea\label{eq:second_order_sol}\nn
&& f^{[2]}(\vec k, \vec v, s)=\int_0^s ds_1\, a(s_1)^2  \;  e^{- i\, \vec k\cdot\vec v\, (s-s_1)} \left[  \left(\frac{\d}{\d x^i}\Phi(\vec x,s_1)\right) \frac{\d}{\d v^i} f^{[1]}(\vec x,\vec v, s_1)\right]_{\vec k}\\  
&&\quad =\int \frac{d^3 q_1}{(2\pi)^3}\int \frac{d^3 q_2}{(2\pi)^3}\; (2\pi)^3\delta^{(3)}_D(\vec k-\vec q_1-\vec q_2)\\ \nn
&&\qquad\quad \int_0^s ds_1\, a(s_1)^2  \;  e^{- i\, (\vec q_1+\vec q_2)\cdot\vec v\, (s-s_1)}   i\, q_1^{i_1}\Phi(\vec q_1,s_1) \\ \nn
&&\qquad\qquad \frac{\d}{\d v^{i_1}} \int_0^{s_1} ds_2\, a(s_2)^2  \;  e^{- i\, \vec q_2\cdot\vec v\, (s_1-s_2)}   i\,  q_2^{i_2}\, \Phi( \vec q_2,s_2) \frac{\d f^{[0]}(v)}{\d v^{i_2}}\ .
\eea
For several reasons, it is useful to go up to third order, so we write
\bea\label{eq:third_order_sol}\nn
&& f^{[3]}(\vec k, \vec v, s)=\int_0^s ds_1\, a(s_1)^2  \;  e^{- i\, \vec k\cdot\vec v\, (s-s_1)} \left[  \left(\frac{\d}{\d x^i}\Phi(\vec x,s_1)\right) \frac{\d}{\d v^i} f^{[2]}(\vec x,\vec v, s_1)\right]_{\vec k}\\  \nn
&&\quad =\int \frac{d^3 q_1}{(2\pi)^3}\int \frac{d^3 q_2}{(2\pi)^3}\int \frac{d^3 q_3}{(2\pi)^3}\; (2\pi)^3\delta^{(3)}_D(\vec k-\vec q_1-\vec q_2-\vec q_3)\\ \nn
&&\qquad\quad \int_0^s ds_1\, a(s_1)^2  \;  e^{- i\, (\vec q_1+\vec q_2+\vec q_3)\cdot\vec v\, (s-s_1)}   i\, q_1^{i_1}\Phi(\vec q_1,s_1) \\ \nn
&&\qquad\qquad \frac{\d}{\d v^{i_1}} \int_0^{s_1} ds_2\, a(s_2)^2  \;  e^{- i\, (\vec q_2+\vec q_3)\cdot\vec v\, (s_1-s_2)}   i\,  q_2^{i_2}\, \Phi( \vec q_2,s_2) \frac{\d f^{[1]}(\vec q_3,\vec v, s_2)}{\d v^{i_2}}\\
&&\quad =\int \frac{d^3 q_1}{(2\pi)^3}\int \frac{d^3 q_2}{(2\pi)^3}\int \frac{d^3 q_3}{(2\pi)^3}\; (2\pi)^3\delta^{(3)}_D(\vec k-\vec q_1-\vec q_2-\vec q_3)\\ \nn
&&\qquad\quad \int_0^s ds_1\, a(s_1)^2  \;  e^{- i\, (\vec q_1+\vec q_2+\vec q_3)\cdot\vec v\, (s-s_1)}   i\, q_1^{i_1}\Phi(\vec q_1,s_1) \\ \nn
&&\qquad\qquad \frac{\d}{\d v^{i_1}} \int_0^{s_1} ds_2\, a(s_2)^2  \;  e^{- i\, (\vec q_2+\vec q_3)\cdot\vec v\, (s_1-s_2)}   i\,  q_2^{i_2}\, \Phi( \vec q_2,s_2) \\ \nn
&&\qquad\qquad \quad \frac{\d}{\d v^{i_2}} \int_0^{s_2} ds_3\, a(s_3)^2  \;  e^{- i\, \vec q_3\cdot\vec v\, (s_2-s_3)}   i\,  q_3^{i_3} \Phi(\vec q_3,s_3) \frac{\d f^{[0]}(v)}{\d v^{i_3}}\ .
\eea
We can clearly see the simple iterative structure. We will write this iterative structure in terms of the dark matter density instead of the gravitational potential using that at leading order in $f_\nu$
\be
\frac{\d^2}{a^2}\Phi\simeq \frac{3}{2}H_0^2 \frac{a_0^3}{a^3} \Omega_{\rm dm,0} \delta_{\rm dm}\ .
\ee 
 
In Fig.~\ref{fig:pertrubative_nu_1}, we diagrammatically represent the perturbative solutions for $f$ when we use the linear solutions for the dark matter field. Other contributions to the non-linear solutions are obtained when we use the non-linear solutions for dark matter into the perturbative solutions for $f$, as represented in Fig.~\ref{fig:pertrubative_nu_2}. Our notation for the $\delta_{\rm diff}$ solutions is as follows: $f^{[n,i_1\ldots i_n]}$ represents the solution obtained by using the $f^{[n]}$ solution for the distribution function and replacing each $\Phi$ in it with  $\Phi^{(i_1)}$ to $\Phi^{(i_n)}$. We denote with $\Phi^{(i)}$ the gravitational potential at order $i$ in perturbation theory. The time ordering of the different insertions is also encoded in the notation,  $\Phi^{(i_1)}$ is associated to the latest inserted interaction while $\Phi^{(i_n)}$ is associated to the first interaction. The order of the resulting solution in $i_1+\ldots+i_n$.
\begin{figure}[htb!]
\centering
\includegraphics[trim={0cm 5cm 3cm 5cm},width=6cm]{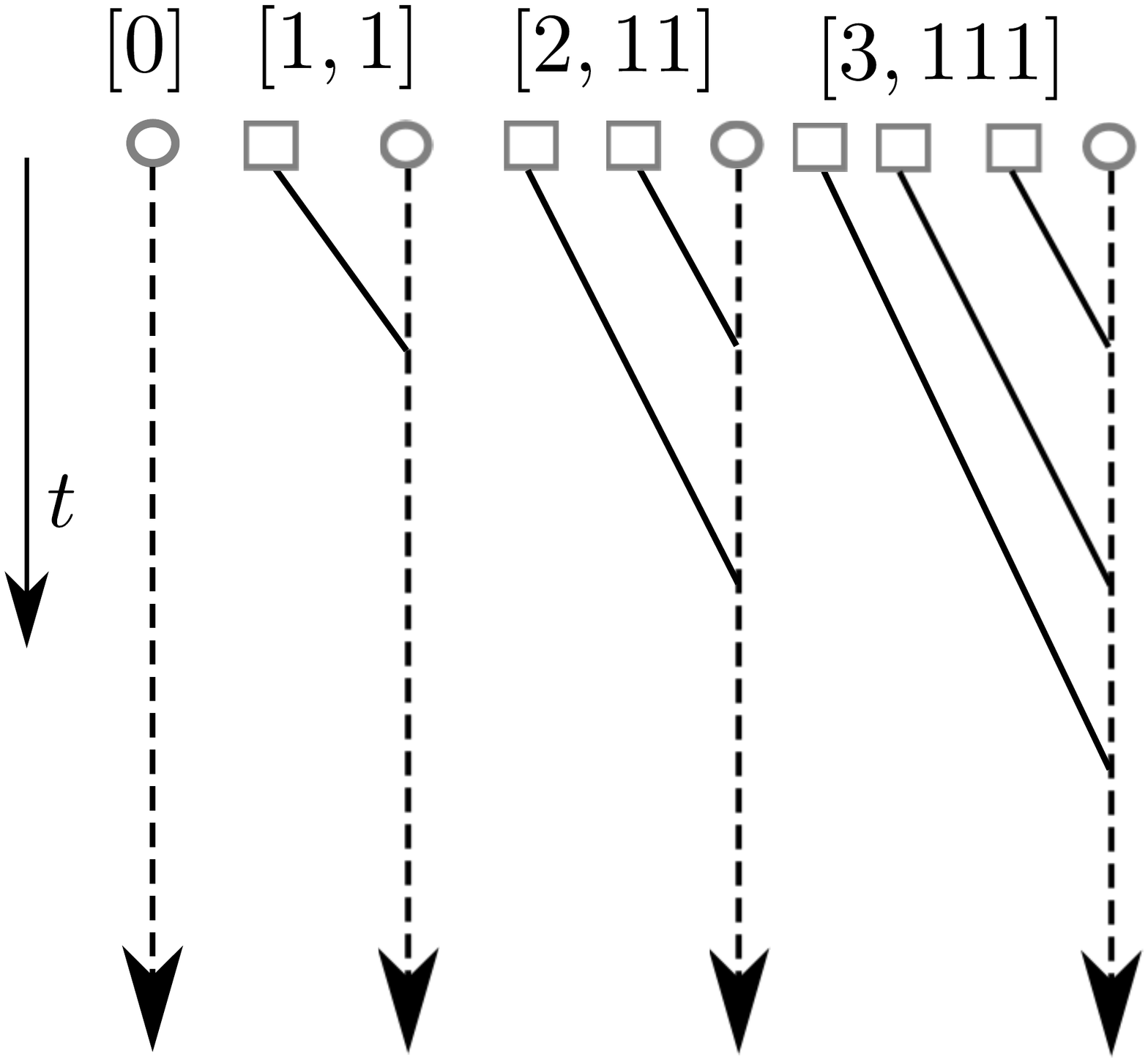}
\caption{\small Perturbative solution for the neutrino distributions. Dashed arrows represent the neutrinos, while continuous line represent the dark matter field. The circle represents the background initial neutrino distribution $f^{[0]}$, while the boxes represent the initial dark matter fields, of which we take correlations to create power-spectrum diagrams later on. All these diagrams involve the linearly evolved dark matter field.}\label{fig:pertrubative_nu_1}
\end{figure}

\begin{figure}[htb!]
\centering
\includegraphics[trim={0cm 5cm 3cm 5cm},width=5.8cm]{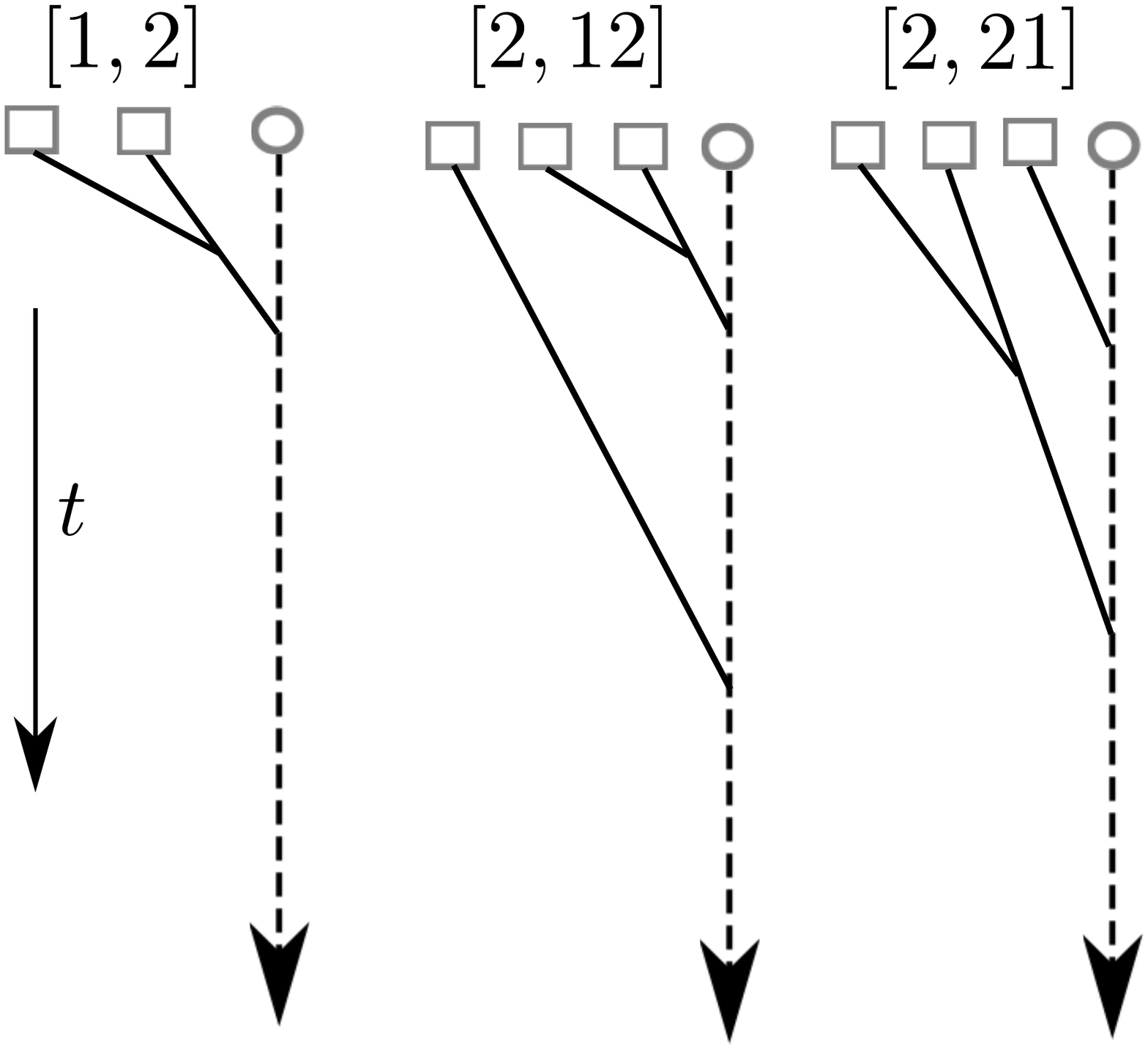}\hspace{2.1cm}
\includegraphics[trim={0cm 5cm 2.8cm 5cm},width=5.8cm]{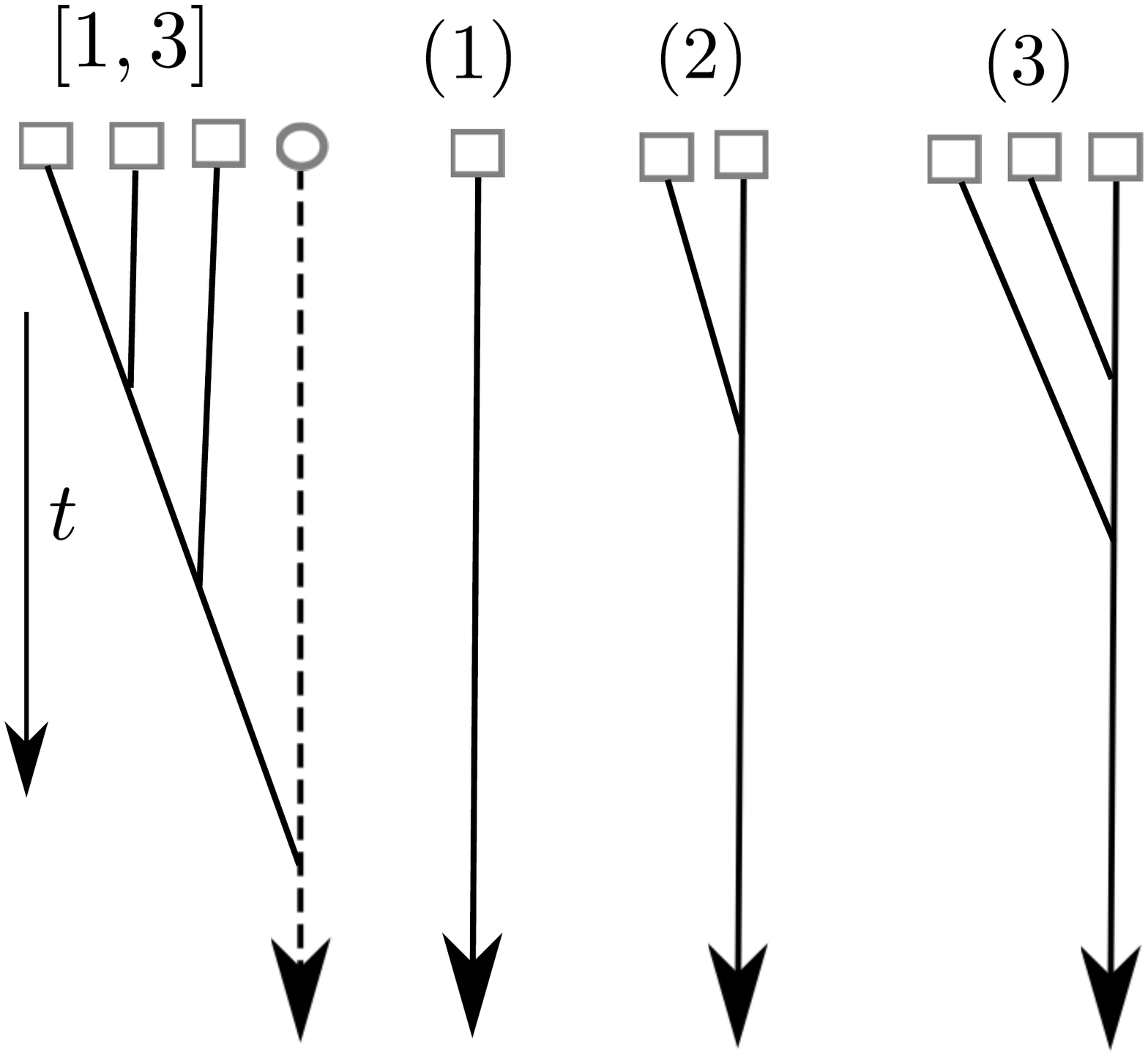}
\caption{\small Perturbative solution for the neutrino distributions that use non-linear dark matter fields at second order. First on the right, we have $f^{[1,2]}$, where a second order solution is obtained by using the second order dark matter field on the $f^{[1]}$ solution for neutrinos.  Similarly, $f^{[2,12]}$ and $f^{[2,21]}$ represent third order solutions obtained using $f^{[2]}$ and the second order dark matter potential. The fourth diagram from the left represents a perturbative solution for the neutrino distributions that use non-linear dark matter fields at third order, and that we denote by $f^{[1,3]}$. For later convenience, we also represent the perturbative solutions for dark matter, $\delta_{\rm dm}^{\{(1),(2),(3)\}}$. }\label{fig:pertrubative_nu_2}\label{fig:pertrubative_nu_3}
\end{figure}

Trusting a perturbative expansion amounts to assuming that the parameter in which we expand is within the convergence radius of the perturbative series. While we feel this is probably true for fast neutrinos, where the expansion parameter is safely less than one for all wavenumbers, this is not the case for slow neutrinos, where the expansion parameter is order one for $k\gtrsim \knl$, beyond  of what we expect is the radius of convergence of the perturbative series. In both cases, however, the solution is determined by the gravitational field, for which we have no knowledge at wavenumbers $k\gtrsim\knl$. These wavenumbers however affect also the solution at $k\ll\knl$. In fact, one can see that a long wavelength perturbation can be sourced by a product of fields some of which have  high-wavenumber $ q_i\gtrsim \knl$. Since these modes are not under perturbative control their $q$- and time-dependence is unknown. Because  we cannot perform the time and momentum integrals correctly in general the perturbative solution will inherit mistakes from the contribution of this part of the phase space. Sensitivity to high wavenumbers is mapped in real space to sensitivity to products of fields evaluated at the same location, which shows in another way that they are sensitive to the uncontrolled short-distance physics. If we are interested in $k\ll\knl$, we can redefine the products of fields as renormalized products, which can correctly account for the contribution from  short distance fluctuations by the addition of suitable counterterms, in a perturbative expansion in $\delta_{\rm dm}\ll1$ and $k/\knl\ll1$. We can  write
\bea\label{eq:product_renorm}
&&[\delta(\vec x_1,\tau_1)\,\delta(\vec x_2,\tau_2)]_\R=[\delta(\vec x_1,\tau_1)]_\R\,[\delta(\vec x_2,\tau_2)]_\R
+[\delta(\vec x_1,\tau_1)\,\delta(\vec x_2,\tau_2)]_\C\ ,
\eea
for compactness and whenever it does not create confusion,  we leave out the $_{\rm dm}$ subscript.  $[\delta(\vec x,\tau)]_\R$ represents the renormalized  $\delta(\vec x,\tau)$, that is the result of the computation in perturbation theory of $\delta(\vec x,\tau)$ with the inclusion of the counterterms to the solution for $\delta(\vec x,\tau)$ (these counterterms originate from the stress-tensor of the effective equations that describe the long-wavelength dark matter dynamics). Instead $[\delta(\vec x_1,\tau_1)\,\delta(\vec x_2,\tau_2)]_\C$ represents the contribution of the counterterms that renormalize the products of two $\delta$'s at the same location. This term is given by 
\bea\label{eq:approx}
&&[\delta(\vec x_1,\tau_1)\delta(\vec x_2,\tau_2)]_\C= \int^{{\rm Max}(\tau_1,\tau_2)} d\tau'\; \left[C^{(2)}_1(\tau_1,\tau_2,\tau')+\epsilon^{(2)}_{\rm stoch, 1}(\tau_1,\tau_2,\xfl(\vec x_1,\tau_1,\tau'),\tau')+\right. \nn \\ 
&&\left.\quad \d^i\d^j\Phi(\xfl(\vec x_1, \tau,\tau'),\tau')\left(C^{(2)}_2(\tau,\tau') \delta^{ij}+C^{(2)}_3(\tau,\tau')\frac{\d_i\d_j}{\d^2}\right)+
\right.
\\  &&
\left. \quad\epsilon_{{\rm stoch, 2}}^{(2), ij}(\tau_1,\tau_2,\xfl(\vec x_1,\tau_1,\tau'),\tau') \frac{\d_i\d_j}{\d^2}+\ldots\right]\ \    \frac{1}{\knl^3}\;\delta^{(3)}_D\left(\xfl(\vec x_1,\tau_1,\tau')-\xfl(\vec x_2,\tau_2,\tau')\right)\ . \nn
\eea
These counterterms have the function of canceling the residual contribution from short distance physics that is present in $[\delta(\vec x_1,\tau_1)]_\R\,[\delta(\vec x_2,\tau_2)]_\R$, and, in the limit $k/\knl\ll1$, to give the correct contribution. Let us define the terms appearing in~(\ref{eq:approx}). The counterterms for the products of fields have support only when the two fields are at the same locations, and have support on the past light cone of the resulting point, as the theory is non-local in time~\cite{Carrasco:2013mua,Senatore:2014eva}.  In particular, 
\bea\label{eq:deltafluid}
&&\delta^{(3)}_D\left(\xfl(\vec x_1,\tau_1,\tau')-\xfl(\vec x_2,\tau_2,\tau')\right)=\\ \nn
&&\qquad  \delta^{(3)}_D(\vec x_1-\vec x_2)+ \d_{x^i_1}\delta^{(3)}_D(\vec x_1-\vec x_2) \left[ \int_{\tau(t')}^{\tau(t_1)} d\tau''_1\;  {v_{\rm dm}}^i(\tau''_1,\vec x_1)- \int_{\tau(t')}^{\tau(t_2)} d\tau''_2\; {v_{\rm dm}^i}(\tau''_2,\vec x_2)\right]+\ldots\ . 
\eea
$\epsilon_{\rm stoch, 1}(\tau_1,\tau_2,\xfl(\vec x_1,\tau_1,\tau'),\tau')$ represents a stochastic term, which has zero expectation value and Poisson statistics
\be
\langle\epsilon_{\rm stoch, 1}(\tau_1,\tau_2,\vec x_1,\tau'_1)\;\epsilon_{{\rm stoch, 2}}(\tau_3,\tau_4,\vec x_2,\tau'_2)\rangle=\delta^{(3)}_D\left(\vec x-\vec x'\right) g^{\rm stoch}_{(1,2)}(\tau_1,\tau_2,\tau_3,\tau_4,\tau_1',\tau_2')\ ,
\ee
with $g^{\rm stoch}_{(1,2)}$ being a function characterized by time scales of order Hubble. Similar expressions hold for all correlation functions of stochastic terms. 

Some intuition for (\ref{eq:approx}) can be developed by noticing that the term in  $C_1^{(2)}$ corrects for the expectation value of the correlation function.  The term in $\epsilon_{\rm stoch, 1}$ accounts for the difference that in a given realization occurs between the expectation value and the actual value. The term in $\d_i\d_j\Phi$ encodes the effect of the deformation of the fields due to the presence of a tidal tensor fields. In general, the $\ldots$ in~(\ref{eq:approx}) represent all the possible terms that are allowed to be there by diff. invariance, hierarchically organized in an expansion in the smallness of the fluctuating fields and in $k/\knl$.

Similar expressions to~(\ref{eq:approx}) hold for higher products of fluctuating fields. For example, for the product of three fields we can write 
\bea\label{eq:three_point_renorm}
&&[\delta(\vec x_1,\tau_1)\,\delta(\vec x_2,\tau_2)\delta(\vec x_3,\tau_3)]_\R\\ \nn
&&\qquad =\left\{[\delta(\vec x_1,\tau_1)\,\delta(\vec x_2,\tau_2)]_\R\,[\delta(\vec x_3,\tau_3)]_\R+{\rm permutations}\right\}
+[\delta(\vec x_1,\tau_1)\,\delta(\vec x_2,\tau_2)\,\delta(\vec x_3,\tau_3)]_\C\ ,
\eea
where
\bea\label{eq:counter3}
&&[\delta(\vec x_1,\tau_1)\,\delta(\vec x_2,\tau_2)\,\delta(\vec x_3,\tau_3)]_\C=\\ \nn
&&\qquad \int^{{\rm Max}(\tau_1,\tau_2,\tau_3)} d\tau'\;  \left[C^{(3)}_1(\tau_1,\tau_2,\tau_3,\tau')+\epsilon^{(3)}_{\rm stoch, 1}(\tau_1,\tau_2,\tau_3,\xfl(\vec x_1,\tau_1,\tau'),\tau')\right. \\ \nn
&&\left.\qquad +\left(\d^i\d^j\Phi(\xfl(\vec x_1, \tau,\tau'),\tau')\left(C^{(3)}_2(\tau,\tau') \delta^{ij}+C^{(3)}_3(\tau,\tau')\frac{\d_{x_1^i}\d_{x_1^j}}{\d_{\vec x_1}^2}\right)+{\rm permutations} \right)+\ldots\right]\\ \nn
&&\qquad\qquad   \frac{1}{\knl^6}\cdot\delta^{(3)}_D\left(\xfl(\vec x_1,\tau_1,\tau')-\xfl(\vec x_2,\tau_2,\tau'_{2})\right)\delta^{(3)}_D\left(\xfl(\vec x_1,\tau_1,\tau')-\xfl(\vec x_3,\tau_3,\tau')\right)\ .
\eea 
We will see that these counterterms will allow us to renormalize the perturbative expressions we obtain next in Sec.~\ref{sec:nocounter} for the case of fast neutrinos. We do this in Sec.~\ref{sec:fastcounterterms}. For slow neutrinos, where the perturbative series does not hold when we insert interactions with $k\gtrsim \knl$, we will need to add additional counterterms to renormalize the same perturbative expressions. We discuss this subtlety later on in Sec.~\ref{sec:slowcounterterms}.

\section{Diagrams with no counterterms\label{sec:nocounter}}

We are now ready to compute correlation functions for neutrinos. As mentioned earlier, we will focus on the one-loop total-matter power spectrum, which requires us to compute $P_{\rm dm,\,dm}$ and $P_{\rm diff,\,dm}$. Notice also that $P_{\rm diff,\,dm}(k)$ can be computed to zeroth order in~$f_\nu$, while we need to compute~$P_{\rm dm,\,dm}(k)$ to first order in $f_\nu$.
In this section, we compute the diagrams that do not involve the counterterms whose contribution we compute in the next section.

\subsection{Contribution to $P_{\rm diff,\,dm}$\label{sec:diagramnum}}

Let us start with $P_{\rm diff,\,dm}(k)$. Given the perturbative solutions $f^{[1]},\; f^{[2]}\,\; f^{[3]}$ in (\ref{eq:first_order_sol}), (\ref{eq:second_order_sol}), and~(\ref{eq:third_order_sol}) we can form loop diagrams by evaluating the gravitation potential to first, second, or third order: $\Phi^{(1)},\;\Phi^{(2)},\; \Phi^{(3)}$. We will use the notation $P_{\rm diff,\,dm,(a,b)}^{[i,j_1\ldots j_i]\;|\;(b)}$ to indicate a diagram where the neutrino side (i.e. the $\delta_{\rm diff.}$ side) is evaluated at order $a$  using the $f^{[i,j_1\ldots j_i]}$ perturbative solution, and the dark matter side to order $b$. We have $j_1+\ldots+j_i=a$. At one-loop, there are six diagrams.  We will provide the expressions for all of them. 

Let us start from the three diagrams involving~$f^{[1]}$. The first one is when we take the $\Phi$ in $f^{[1]}$ to be $\Phi^{(1)}$ and contract it with dark matter evolved to third order (see Fig.~\ref{fig:perturbative_nu_diagram_1}). Denoting by $\calH=a H$ and by $D(a)$ the growth factor, we have
\bea\nn
&&P_{\rm diff,\,dm,\,(1,3)}^{[1,1]\;|\; (3)}(k,a_0)=\\ \nn
&&\qquad=\left\langle\int d^3v \; \int^{a_0}\frac{da_1}{\calH(a_1)} e^{- i \vec k\cdot \vec v (s-s(a_1))}\left(\frac{3}{2}\right) \Omega_{\rm dm}(a_1)\calH(a_1)^2 \frac{i k^{i_1}}{(-k^2)} D(a_1) \delta^{(1)}(\vec k,a_0) \frac{\d f_0(v)}{\d v^{i_1}}\right. \\ 
&&\qquad\left. \times\quad  D(a_0)^3 \int \frac{d^3 q_1}{(2\pi)^3}\int \frac{d^3 q_2}{(2\pi)^3}\int \frac{d^3 q_3}{(2\pi)^3} \; \delta_D^{(3)}\left(\vec q_1+\vec q_2+\vec q_3\right) \; F_3(\vec q_1,\vec q_2,\vec q_3) \right. \; \\ \nn
&&\qquad\left.\times\; \quad\delta^{(1)}(\vec q_1,a_0)\delta^{(1)}(\vec q_2,a_0)\delta^{(1)}(\vec q_3,a_0) \right\rangle \ ,
\eea
where we have approximated the time dependence of the fields with the EdS one~\footnote{To keep the algebra simpler in  the calculation of $P_{\rm diff,\, dm}$ we will use the approximate EdS dependence for the time dependence of dark matter perturbations at zeroth order in $f_\nu$.  It amounts to assuming that the relative growth of the perturbations in $\Lambda{\rm CDM}$ cosmology is the same as in EdS cosmology. This approximation has been verified to be quite good for $\Lambda$CDM. This approximation however alters the cancellation of a term that goes as $k^0P(k)$ for $k\to 0$ both in some of the ordinary diagrams (specifically, between $P_{\rm diff,\,dm,\,(3,1)}^{f^{[2,12]\;|\;(1)}}+P_{\rm diff,\,dm,\,(3,1)}^{f^{[2,21]\;|\;(1)}}$ and $P_{\rm diff,\,dm,\,(3,1)}^{f^{[3,111]\;|\;(1)}}$) and in the analogous counterterms diagrams, that should vanish on symmetry principles. Though we have checked that the amount of the non-cancellation is actually small we will remove the term that goes as  $k^0P(k)$ for $k\to 0$ from the mentioned diagrams for which this cancellation should have happened. In this way, no functional form that would be forbidden by symmetries is introduced by the approximate treatment of the time dependence.}. Integrating by parts the $\d/\d v^{i}$ derivative and performing the contractions we obtain:
\bea\nn
&&P_{\rm diff,\,dm,\,(1,3)}^{[1,1]\;|\; (3)}=3 \;P_{{\rm dm},11}(k,a_0)\; \int^{a_0} da_1\;  \left(\frac{3}{2}\right) \Omega_{\rm dm}(a_1)\calH(a_1) D(a_0)^3 D(a_1)\;(s-s(a_1))  \; 
 \\ 
&&\qquad\qquad\qquad\qquad \tilde f^{[0]}(\vec k (s-s(a_1))) \int \frac{d^3q}{(2\pi)^3}\;  P_{{\rm dm},11}(q,a_0)  \; F_3(-\vec k,\vec q,-\vec q)   \ ,
\eea
where we remind the reader  that $\tilde f^{[0]}$ is the Fourier transform of the relative distribution of the difference of neutrinos and dark matter.

\begin{figure}[htb!]
\centering
\includegraphics[trim={0cm 5cm 3cm 5cm},width=6.5cm]{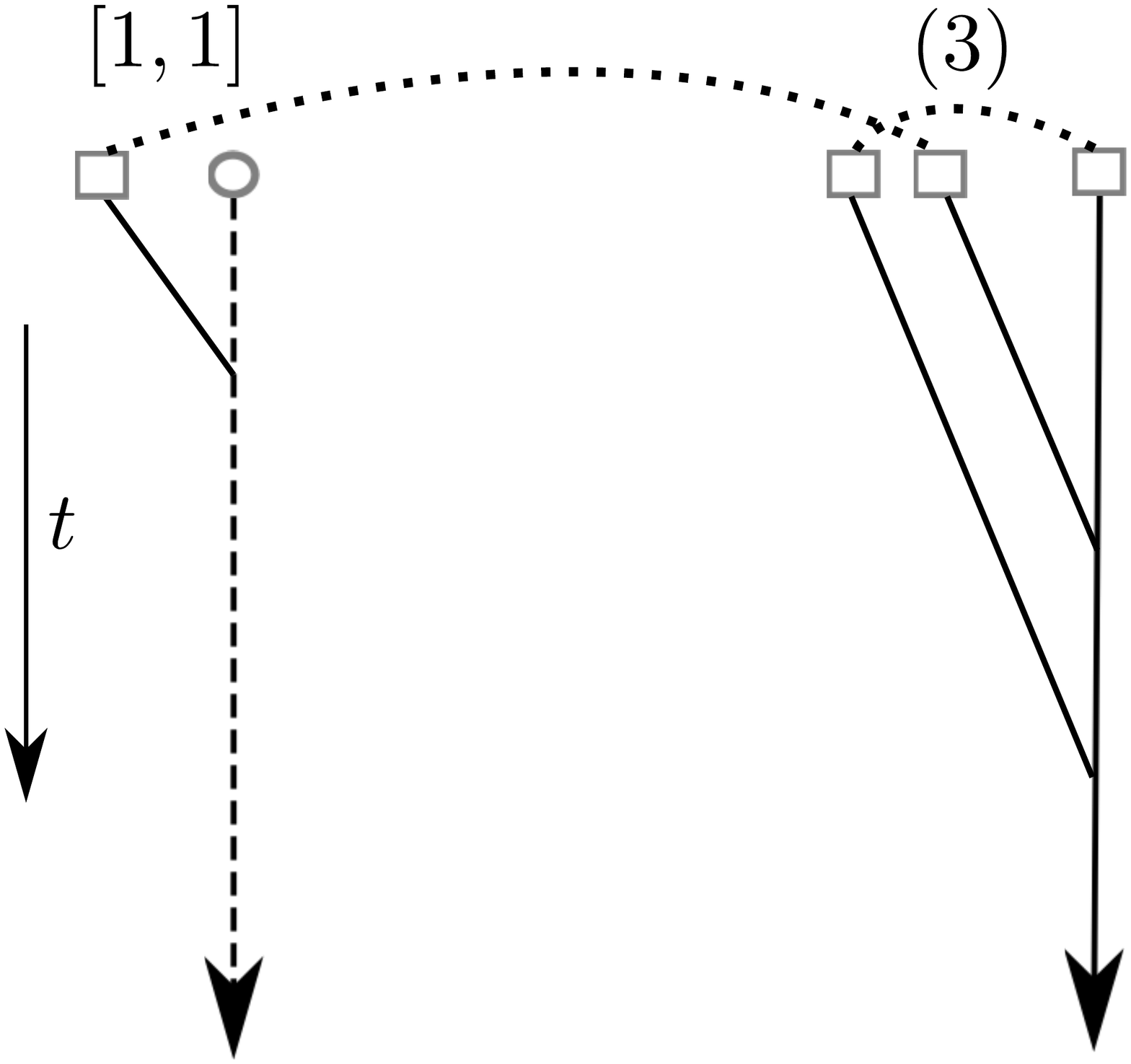}\hspace{3.5cm}
\includegraphics[trim={0cm 5cm 3cm 5cm},width=6.5cm]{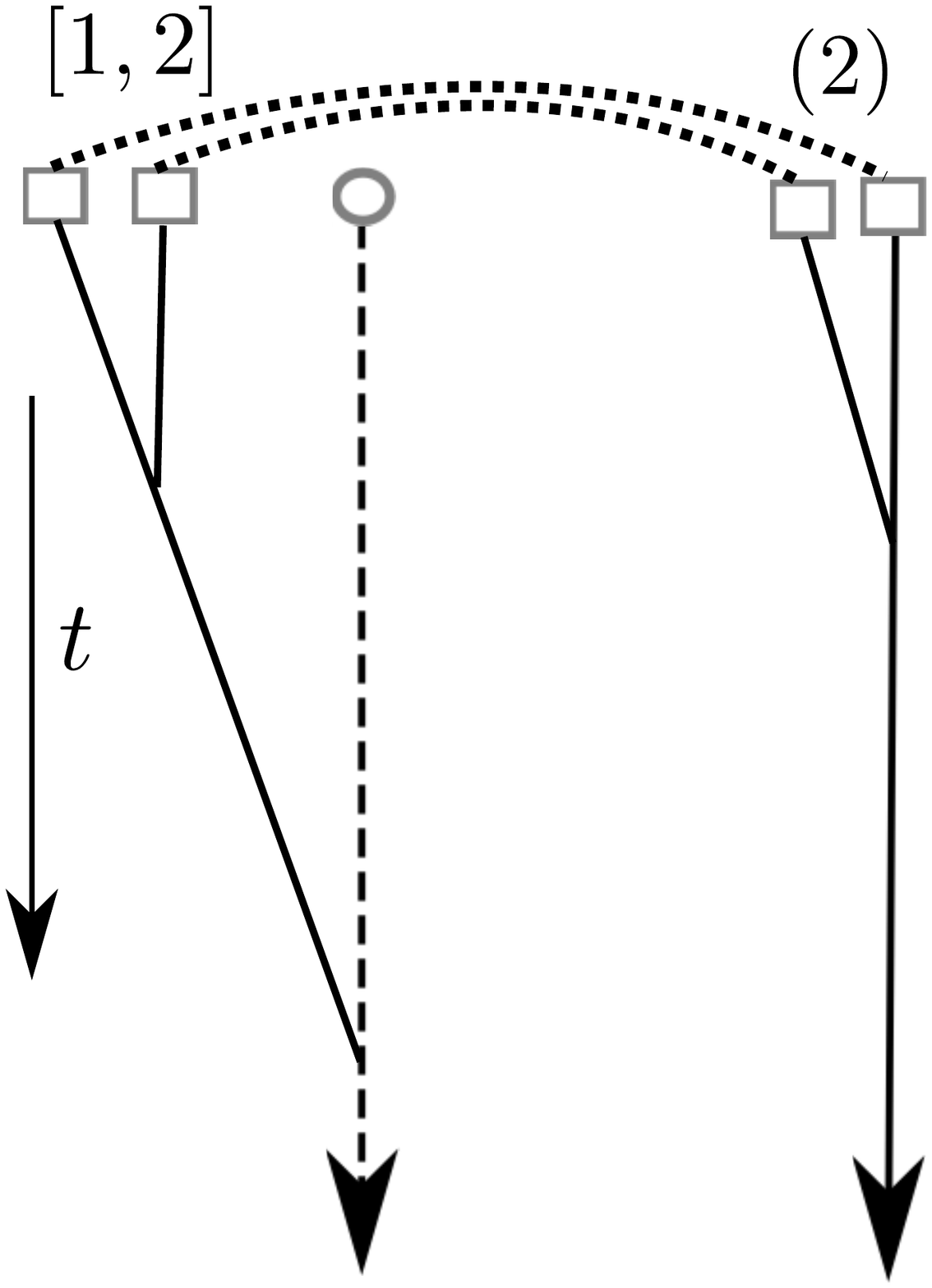}
\caption{\small {\it Left:} One-loop diagram obtained contracting $f^{[1,1]}$ with $\delta_{\rm dm}^{(3)}$. Dotted line means that we take the expectation value over the initial conditions. {\it Right:} One-loop diagram obtained contracting $f^{[1,2]}$ with $\delta_{\rm dm}^{(2)}$.}\label{fig:perturbative_nu_diagram_1}\label{fig:perturbative_nu_diagram_2}
\end{figure}

Next, we take in $f^{[1]}$ the second-order potential $\Phi^{(2)}$ (see Fig.~\ref{fig:perturbative_nu_diagram_2}). By taking dark matter at second order, we can form a 22 diagram. Skipping the algebra which is similar to the former case, we obtain:
\bea\nn
&&P_{\rm diff,\,dm,\,(2,2)}^{[1,2]\;|\; (2)}(k,a_0)=2 \; \int^{a_0} da_1\;  \left(\frac{3}{2}\right)\Omega_{\rm dm}(a_1) \calH(a_1) D(a_0)^2 D(a_1)^2\;(s-s(a_1)) \tilde f^{[0]}(\vec k (s-s(a_1)))
 \\ 
&&\qquad\qquad\qquad \int \frac{d^3q}{(2\pi)^3}\;  \;P_{{\rm dm},11}(|\vec k-\vec q|,a_0)P_{{\rm dm},11}(q,a_0)  \; F_2(-\vec q,-\vec k+\vec q)  \; F_2(\vec q,\vec k-\vec q)  \ .
\eea

The last diagram involving $f^{[1]}$ is when we use $\Phi^{(3)}$ (see Fig.~\ref{fig:perturbative_nu_diagram_3}). By contracting with a linear dark matter field, we obtain a 31 diagram given by:
\bea\nn
&&P_{\rm diff,\,dm,\,(3,1)}^{[1,3]\;|\;(1)}(k,a_0)=\\ \nn
&&\qquad 3 \;P_{{\rm dm},11}(k,a_0)\; \int^{a_0} da_1\;  \left(\frac{3}{2}\right)\Omega_{\rm dm}(a_1) \calH(a_1) D(a_0) D(a_1)^3\;(s-s(a_1))  \; \tilde f^{[0]}(\vec k (s-s(a_1)))
 \\ 
&&\qquad\qquad\qquad\qquad \int \frac{d^3q}{(2\pi)^3}\;  P_{{\rm dm},11}(q,a_0)  \; F_3(\vec k,\vec q,-\vec q)   \ .
\eea

\begin{figure}[htb!]
\centering
\includegraphics[trim={0cm 5cm 3cm 2cm},width=6.5cm]{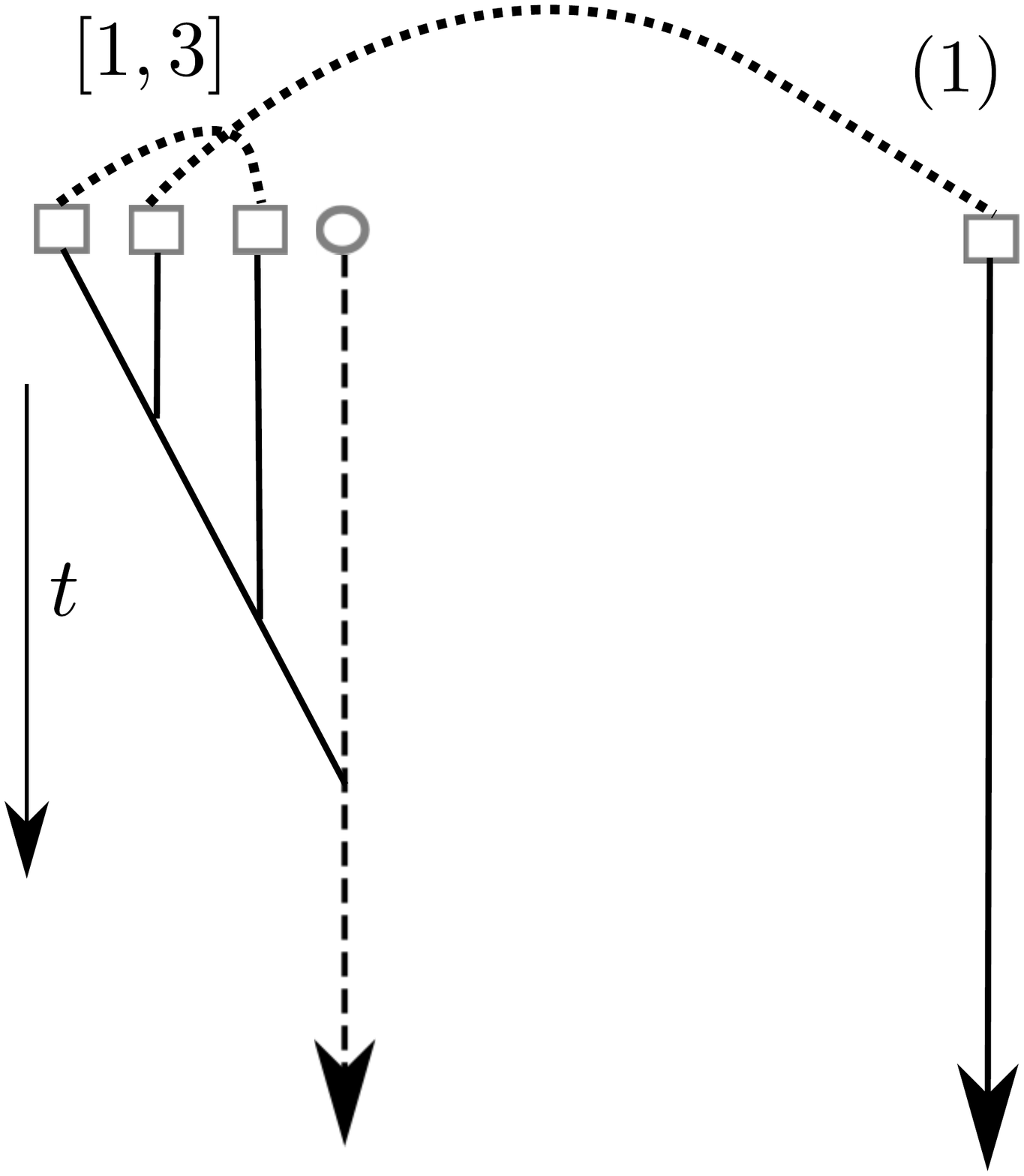}\hspace{2.5cm}
\includegraphics[trim={0cm 5cm 3cm 4cm},width=6.5cm]{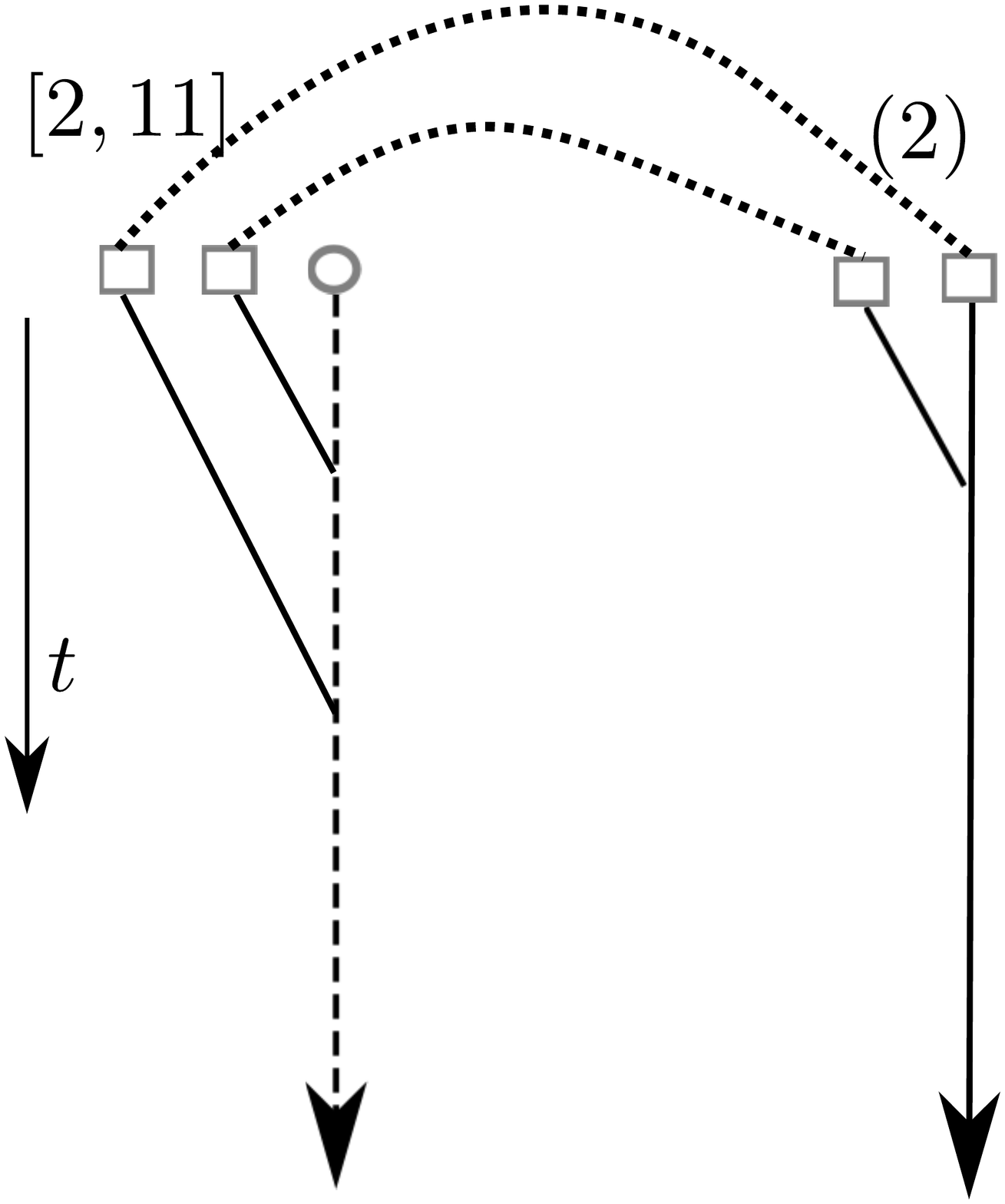}
\caption{\small {\it Left:} One-loop diagram obtained contracting $f^{[1,3]}$ with $\delta_{\rm dm}^{(1)}$. {\it Right:} One-loop diagram obtained contracting $f^{[2,11]}$ with $\delta_{\rm dm}^{(2)}$.}\label{fig:perturbative_nu_diagram_3}\label{fig:perturbative_nu_diagram_4}
\end{figure}

Next we consider the two diagrams we can build using $f^{[2]}$. The first is when we use the two $\Phi$'s to linear order, and we contract with dark matter evolved to second order (see Fig.~\ref{fig:perturbative_nu_diagram_4}). We obtain a 22 diagram of the form:
\bea\nn
&&P_{\rm diff,\,dm,\,(2,2)}^{[2,11]\;|\;(2)}(k,a_0)=2 \; \int^{a_0} da_1\;\int^{a_1} da_2\;  \left(\frac{3}{2}\right)^2 \calH(a_1)\Omega_{\rm dm}(a_1)\calH(a_2) \Omega_{\rm dm}(a_2) D(a_0)^2 D(a_1) D(a_2)\,\\ \nn
&&\qquad \int \frac{d^3q}{(2\pi)^3}\; \left[\left(\frac{\vec q\cdot\vec k }{q^2}(s-s(a_1))\right) \left(\frac{\vec q_2\cdot\vec k }{q_2^2}(s-s(a_1))+(s(a_1)-s(a_2))\right) \right.
 \\ 
&&\qquad  \left.\tilde f^{[0]}(\vec k (s-s(a_1))+\vec q_2 (s(a_1)-s(a_2))) \;P_{{\rm dm},11}(q_2)P_{{\rm dm},11}(q,a_0)  \; F_2(-\vec q,-\vec q_2) \right]_{\vec q_2=\vec k-\vec q} \ .
\eea

The other diagram that we can build out of $f^{[2]}$ is a 31 diagram obtained when one of the two $\Phi$'s is second order and the other first order and we contract with linear matter~(see Fig.~\ref{fig:perturbative_nu_diagram_5}). We obtain the following expression:
\bea\nn\label{P31fromf2}
&&P_{\rm diff,\,dm,\,(3,1)}^{[2,21]\;|\;(1)}(k,a_0)+P_{\rm diff,\,dm,\,(3,1)}^{[2,12]\;|\;(1)}(k,a_0)=\\ \nn
&&=\quad 2 \;P_{{\rm dm},11}(k,a_0)\; \int^{a_0} da_1\; \int^{a_1} da_2  \left(\frac{3}{2}\right)^2 \calH(a_1)\Omega_{\rm dm}(a_1)\calH(a_2) \Omega_{\rm dm}(a_2) D(a_0) D(a_1) D(a_2) \\ \nn
&&\quad \int \frac{d^3q}{(2\pi)^3}\;  P_{{\rm dm},11}(q,a_0)  \; F_2(\vec k,-\vec q) \times \\ \nn
&&\left\{ \left[\left(\frac{\vec q_1\cdot\vec k }{q_1^2}(s-s(a_1))\right) \left(\frac{\vec q_2\cdot\vec k }{q_2^2}(s-s(a_1))+(s(a_1)-s(a_2))\right)\; \tilde f^{[0]}(\vec k (s-s(a_1))+\vec q_2 (s(a_1)-s(a_2)))\right.\right.\\ \nn
&&\qquad\qquad\times\quad \left.\left. D(a_1)\right]_{\vec q_1=\vec k-\vec q, \; \vec q_2=\vec q}+ \right.
 \\  \nn
&&\ \left[\left(\frac{\vec q_1\cdot\vec k }{q_1^2}(s-s(a_1))\right) \left(\frac{\vec q_2\cdot\vec k }{q_2^2}(s-s(a_1))+(s(a_1)-s(a_2))\right)\;  \tilde f^{[0]}(\vec k (s-s(a_1))+\vec q_2 (s(a_1)-s(a_2)))\right.\\ 
&&\qquad\qquad\times\quad \left.\left. D(a_2) \right]_{\vec q_1=\vec q, \; \vec q_2=\vec k-\vec q} \right\}   \ .
\eea

\begin{figure}[htb!]
\centering
\includegraphics[trim={0cm 5cm 3cm 4cm},width=6.5cm]{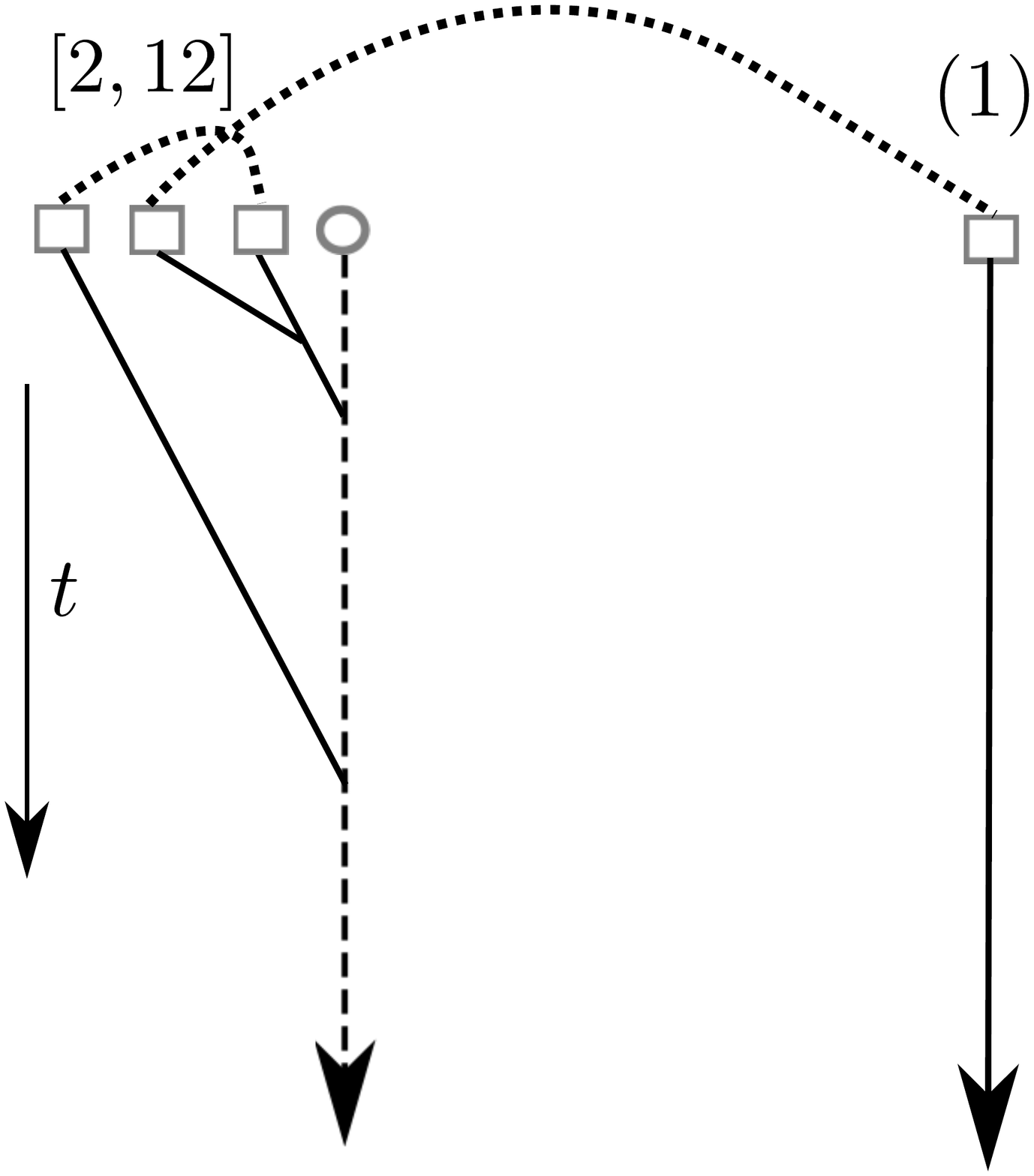}\hspace{2.5cm}
\includegraphics[trim={0cm 5cm 3cm 4cm},width=6.5cm]{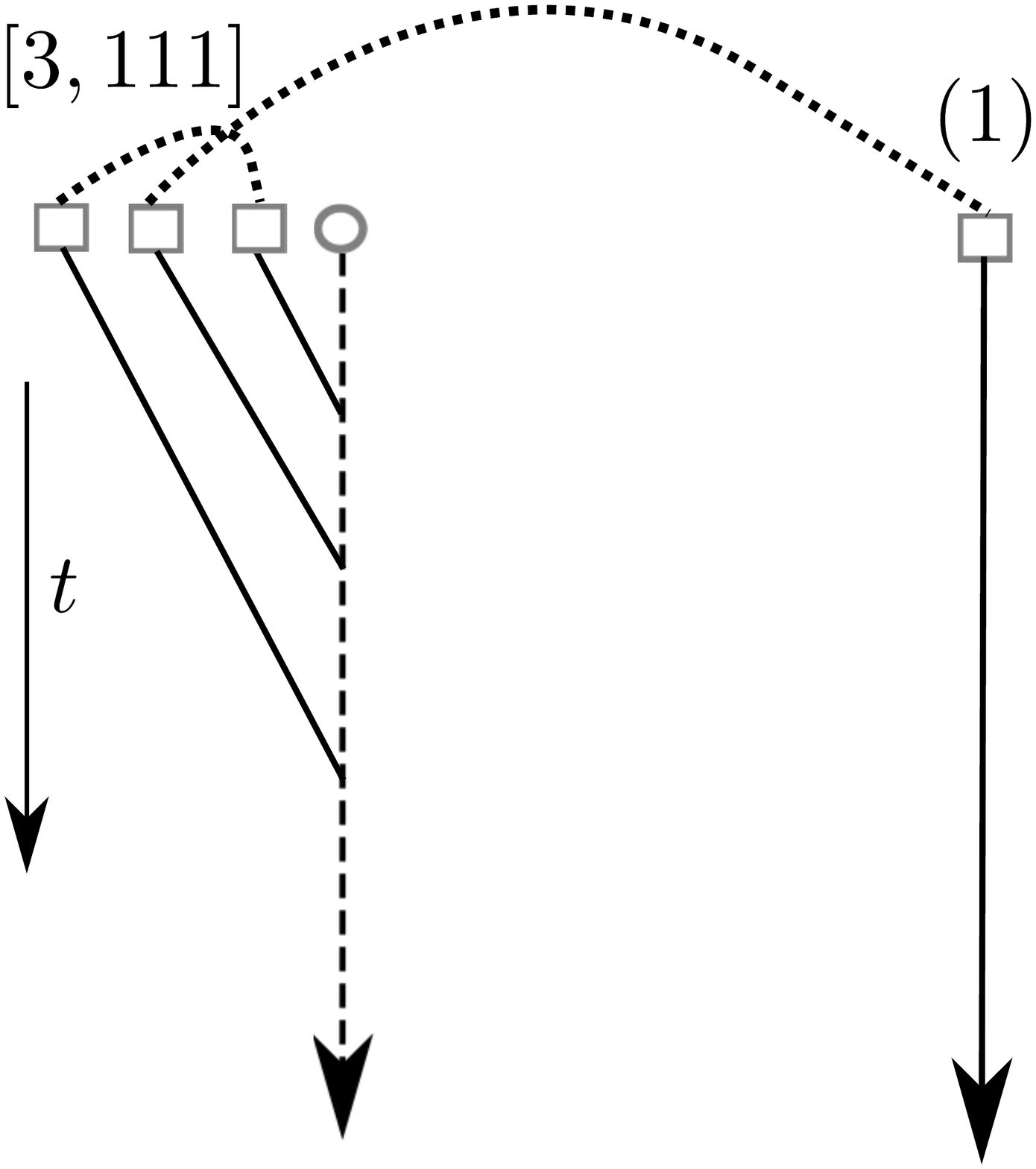}
\caption{\small {\it Left:} One-loop diagram obtained contracting $f^{[2,12]}$ with $\delta_{\rm dm}^{(1)}$. We do not show the analogous diagram build using $f^{[2,21]}$. {\it Right:} One-loop diagram obtained contracting $f^{[3,111]}$ with $\delta_{\rm dm}^{(1)}$.}\label{fig:perturbative_nu_diagram_5}\label{fig:perturbative_nu_diagram_6}
\end{figure}

Finally we can form one diagram by using $f^{[3]}$, with all the $\Phi$'s taken at linear order (see Fig.~\ref{fig:perturbative_nu_diagram_6}). We obtain a 31 diagram that can be written in the following form:
\bea\nn\label{eq:P13integrand}
&&P_{\rm diff,\,dm,\,(3,1),\,{\rm integrand}}^{[3,111]\;|\; (1)}(k,a_0,\vec q; \{\vec q_1,\vec q_2,\vec q_3\})
= \;P_{{\rm dm},11}(k,a_0)\; \int^{a_0} da_1\; \int^{a_1} da_2\int^{a_2} da_3\\ \nn
&&\qquad\qquad  \left(\frac{3}{2}\right)^3 \calH(a_1)\calH(a_2)\calH(a_3)\Omega_{\rm dm}(a_1) \Omega_{\rm dm}(a_2) \Omega_{\rm dm}(a_3) D(a_0) D(a_1) D(a_2)D(a_3) \\ \nn
&&\qquad\qquad \left(\frac{\vec q_1\cdot\vec k }{q_1^2}(s-s(a_1))\right) \left(\frac{\vec q_2\cdot\vec k }{q_2^2}(s-s(a_1))+\frac{\vec q_2\cdot (\vec q_2+\vec q_3)}{q_2^2}(s(a_1)-s(a_2))\right) \\ \nn
&&\qquad\qquad\left(\frac{\vec q_3\cdot\vec k }{q_3^2}(s-s(a_1))+\frac{\vec q_3\cdot (\vec q_2+\vec q_3)}{q_3^2}(s(a_1)-s(a_2))+(s(a_2)-s(a_3))\right)\; \\ \nn
&&\qquad\qquad\tilde f^{[0]}(\vec k (s-s(a_1))+(\vec q_2+\vec q_3) (s(a_1)-s(a_2))+\vec q_3 (s(a_2)-s(a_3))) \  P_{{\rm dm},11}(q,a_0)\ .\\
\eea
so that
\bea\label{eq:P13fromf3}
&&P_{\rm diff,\,dm,\,(3,1)}^{[3,111]\;|\;(1)}(k,a_0)= \int \frac{d^3 q}{(2\pi)^3} \left[P_{\rm diff,\,dm,\,(3,1),\,{\rm integrand}}^{[3,111]\;|\; (1)}(k,a_0,\vec q; \{\vec k,\vec q,-\vec q\})\right.\\ \nn
&&\qquad \left.+P_{\rm diff,\,dm,\,(3,1),\,{\rm integrand}}^{[3,111]\;|\; (1)}(k,a_0,\vec q; \{\vec q,\vec k,-\vec q\})+P_{\rm diff,\,dm,\,(3,1),\,{\rm integrand}}^{[3,111]\;|\; (1)}(k,a_0,\vec q; \{\vec q,-\vec q,\vec k\})\right]\ .
\eea

This completes our listing of the diagrams to evaluate $P_{\rm diff,\, dm}$. We consider the diagrams for $P_{\rm dm,\,dm}$ in the next section.

\subsection{Contribution to $P_{\rm dm,\,dm}$}

We now move to consider the contributions to $P_{\rm dm,\,dm}$ originating from the neutrinos perturbations. We call this correction $\Delta P_{\rm dm,\,dm}$. In fact, the overdensities in neutrinos affect the gravitational potential that in turn  affects the dark matter overdensity. Let us analyze this effect.

In this part we follow the notation of~\cite{Lewandowski:2016yce}. By writing $\pi_{\rm dm}^i(\vec x,t)=\rho(\vec x,t) v_{\rm dm}^i(\vec x,t)$, and by defining the usual $\theta_{\rm dm}=\d_i v^i_{\rm dm}$, the equations of motions of dark matter in the presence of neutrinos become:
\bea\label{eq:poisson3} \label{nonlincont3}\label{nonlineuler3} \label{eq:newdmequations3}
&&\frac{\d^2}{a^2}\Phi= \frac{3}{2}H_0^2 \frac{a_0^3}{a^3} \left(\Omega_{\rm dm,0} \delta_{\rm dm}+\Omega_{\nu, 0} \delta_\nu\right)\simeq  \frac{3}{2}H_0^2 \frac{a_0^3}{a^3} \Omega_{\rm NR,0} \left(\delta_{\rm dm}+f_\nu \left(\delta_\nu-\delta_{\rm dm}\right)\right)\\ \nn
&&\frac{1}{a} \frac{\d \delta_{\rm dm}}{\d a} + \frac{1}{a}  \theta_{\rm dm} + \frac{1}{a} \partial_i \left( \delta_{\rm dm} v^i_{\rm dm} \right)=0  \ , \\ \nn
&&\frac{1}{a} \frac{\d \theta_{\rm dm}}{\d a} + H \theta_{\rm dm} +\frac{1}{a} \d^2\Phi + \frac{1}{a} \partial_i \left( v^j_{\rm dm} \partial_j v^i_{\rm dm} \right)  =0
 \ .
\eea
We remind the reader that we have neglected all counterterms,. They will be discussed later in Sec.~\ref{sec:fastcounterterms}. We have also neglected the velocity vorticity that is generated only by the counterterms, and so is negligible at leading orders~\cite{Carrasco:2013sva,Mercolli:2013bsa}.
We first define 
\be
\Theta_{\rm dm}=-\frac{\theta_{\rm dm}}{f_g {\cal H}} \ ,
\ee
where $f_g$ is the standard growth factor $f_g=d\log D_0/d\log a$, where $D_0$ is the growth factor in the absence of neutrinos. We drop the subscript ${}_{\rm dm}$ whenever doing this does not create confusion.  In terms of $\Theta$,  equations~(\ref{eq:newdmequations3}) read as
\begin{align}\label{eq:perturbative_explicit}
&a\frac{\d\delta_{\vk}}{\d a}-f_g\Theta_{\vk}=f_g\int \frac{d^3q_1}{(2\pi)^{3}}\int \frac{d^3q_2}{(2\pi)^{3}} (2\pi)^{3}\delta^{(3)}_{D}(\vk-\vq_1-\vq_2)\alpha(\vq_1,\vq_2)\Theta_{\vq_1}\delta_{\vq_2},\\ \nn
&a\frac{\d\Theta_{\vk}}{\d a}-f_g\Theta_{\vk}+\frac{3}{2}\frac{\Omega_{\rm dm}(a)}{f_g}(\Theta_{\vk}-\delta_{\vk}-f_\nu \delta_{\nu,\vec k})=\\ \nn
&\qquad\qquad=f_g\int \frac{d^3q_1}{(2\pi)^{3}}\int \frac{d^3q_2}{(2\pi)^{3}}(2\pi)^{3}\delta^{(3)}_{D}(\vk-\vq_1-\vq_2)\beta(\vq_1,\vq_2)\Theta_{\vq_1}\Theta_{\vq_2} \nonumber \ , 
\end{align}
where
\begin{equation} 
\alpha ( \qvec_1 , \qvec_2 )  = 1 + \frac{\qvec_1 \cdot \qvec_2}{q_1^2} \label{alphadef}\ ,\qquad \beta( \qvec_1 , \qvec_2 ) = \frac{ | \qvec_1 + \qvec_2 |^2 \qvec_1 \cdot \qvec_2}{2 q_1^2 q_2^2}    \ .
 \end{equation}
 
At zeroth order in $f_\nu$, one can solve the above equations iteratively: one can write $\delta^{(n)}_{\vec k}(a)$ and~$\Theta^{(n)}_{\vec k}(a)$ at any perturbative order as~\cite{Lewandowski:2016yce} 
 \begin{eqnarray}  \label{dtGreen}
 &&\delta^{(n)}_{\vec k}=\int^a_0 d\tilde a \bigg(G^{\delta}_{1}(a,\tilde{a})S^{(n)}_1(\tilde{a},\vec k)+G^{\delta}_{2}(a,\tilde{a})S^{(n)}_2(\tilde{a},\vec k)\bigg) \ ,\\ \nn
  &&\Theta^{(n)}_{\vec k}=\int^a_0 d\tilde{a} \bigg(G^{\Theta}_{1}(a,\tilde{a})S^{(n)}_1(\tilde{a},\vec k)+G^{\Theta}_{2}(a,\tilde{a})S^{(n)}_2(\tilde{a},\vec k)\bigg) \ , \label{dtGreen2}
  \end{eqnarray}
where $G^{\delta}_{1}$, $G^{\delta}_{2}$ are the density Green's functions, $G^{\Theta}_{1}$, $G^{\Theta}_{2}$ are the velocity Green's functions, and $S^{(n)}_1(\tilde{a},\vec k)$ and $S^{(n)}_2(\tilde{a},\vec k)$ are the source terms of the continuity and Euler equations at the $n$-th order respectively
\begin{align}
&S_1^{(n)}(a,\vk)=f_g(a)\sum\limits_{m=1}^{n-1}\int \frac{d^3q}{(2\pi)^3}\;  \alpha(\vq,\vk-\vq) \, \tT^{(m)}_{\vq}(a)\td^{(n-m)}_{\vk-\vq}(a),   \label{source}   \\
&S_2^{(n)}(a,\vk)=f_g(a)\sum\limits_{m=1}^{n-1}\int \frac{d^3q}{(2\pi)^3}\; \beta(\vq,\vk-\vq)\, \tT^{(m)}_{\vq}(a)\tT^{(n-m)}_{\vk-\vq}(a).
\end{align}
Expressions for the Green's functions are given in App.~C of~\cite{Lewandowski:2016yce}.

At first order in $f_\nu$, the contribution of $\delta_\nu$ to the evolution of dark matter is related to the influence it has through the gravitational potential. $\delta_\nu$ induces therefore a perturbation in $\delta_{\rm dm}$ and~$\Theta_{\rm dm}$ that are given by (see Fig.~\ref{fig:pertrubative_nu_dm_1}):
\bea\label{eq:inducedmatterpt}
&&\delta_{{\rm dm}\leftarrow \nu}^{(n)}(\vec k,a)= f_\nu \int^a d\tilde a\; G^\delta_2(a,\tilde a) \; \frac{3}{2} \frac{\Omega_{\rm dm}(\tilde a)}{f_g(\tilde a)} \delta_\nu^{(n)}(\vec k,\tilde a)\ , \\\nn
&&\Theta_{{\rm dm}\leftarrow \nu}^{(n)}(\vec k,a)= f_\nu\int^a d\tilde a\; G^\theta_2(a,\tilde a)  \; \frac{3}{2} \frac{\Omega_{\rm dm}(\tilde a)}{f_g(\tilde a)} \delta_\nu^{(n)}(\vec k,\tilde a)\ .
\eea

\begin{figure}[htb!]
\centering
\includegraphics[trim={0cm 5cm 3cm 5cm},width=6cm]{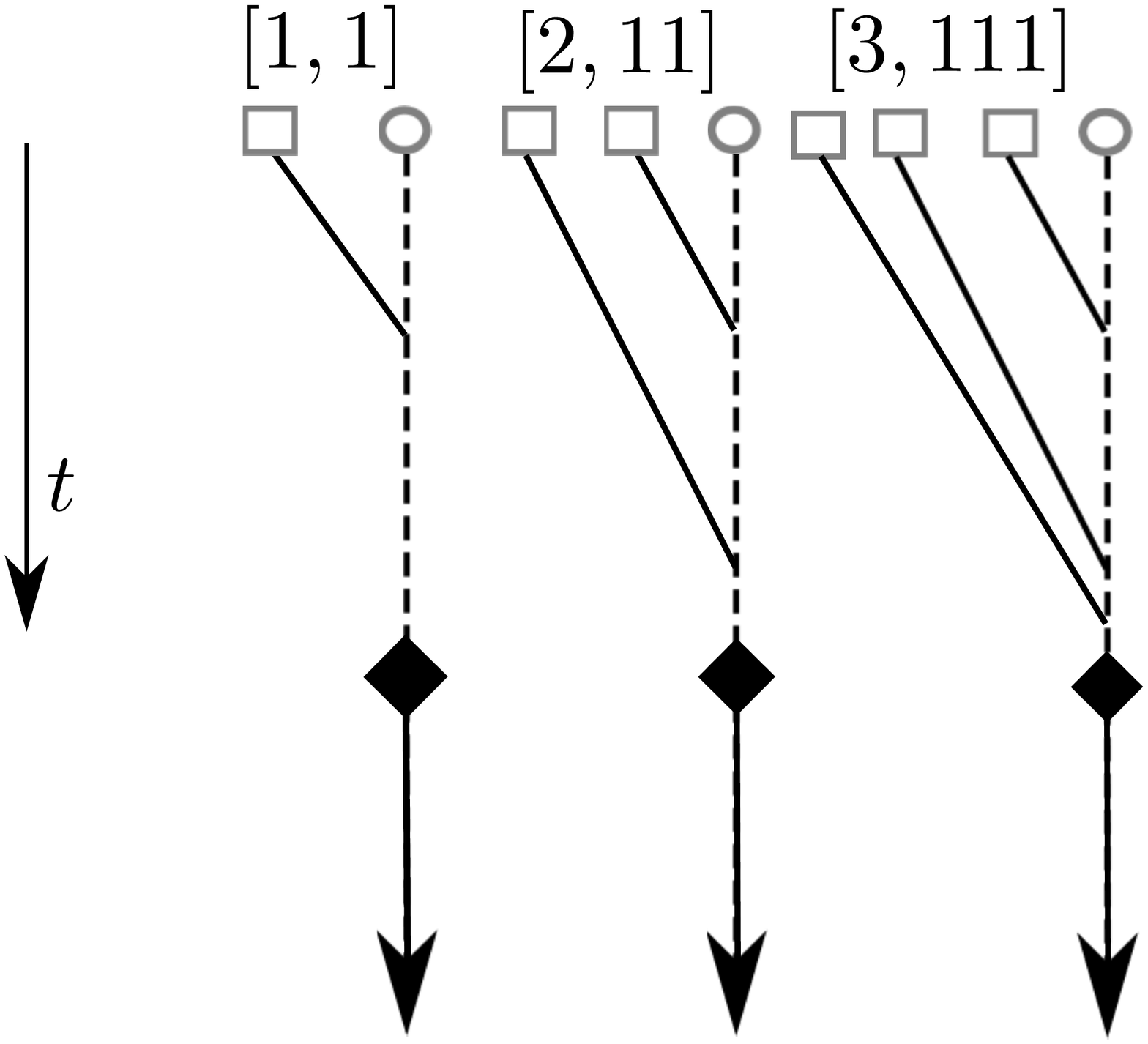}
\caption{\small Perturbative solution for the linearly-induced dark matter perturbations from the neutrino perturbations, $\delta_{{\rm dm}\leftarrow \nu}^{(n)}$.  The filled diamond represents the insertion of the mixing vertex $ \frac{3}{2} \frac{\Omega_{\rm dm}}{f_g}$ that converts a neutrino field into a dark matter field. We do not show the analogous diagrams for $\Theta_{{\rm dm}\leftarrow \nu}$, which differs from these ones simply by the choice of the Green's function to be used after the mixing vertex.}\label{fig:pertrubative_nu_dm_1}
\end{figure}

At this point, one can form diagrams either contracting directly $\delta_{{\rm dm}\leftarrow \nu}^{(n)}$ and $\Theta_{{\rm dm}\leftarrow \nu}^{(n)}$, or by using these solutions in the perturbative expressions resulting from (\ref{eq:perturbative_explicit}). In other words, we should insert~(\ref{eq:inducedmatterpt}) in all possible ways into~(\ref{dtGreen}) to obtain a perturbation of the relevant order. Then we perform the necessary contractions. This gives rise to two different kinds of diagrams that we discuss separately.

\subsubsection*{Integral diagrams}

If we form diagrams by directly contracting the fluctuations in  $\delta_{{\rm dm}\leftarrow \nu}^{(n)}$ and $\Theta_{{\rm dm}\leftarrow \nu}^{(n)}$ on one side with fields on the other side, we obtain diagrams which are related to ones we already computed to obtain $P_{\rm diff,\,dm}$ by a rather simple integration. In formulas we have:
\be\label{eq:integral_diagrams}
\Delta P_{\rm dm,\,dm}^{[n,i_1\ldots i_n]\;|\; (m)}(k,a_0)=\int^{a_0} d\tilde a\; G^\delta_2(a_0,\tilde a)\;\frac{3}{2}\frac{\Omega_{\rm dm}(\tilde a)}{f_g(\tilde a)}P_{\rm diff,\,dm}^{[n,i_1\ldots i_n]\;|\; (m)}(k,\tilde a,a_0)\ ,
\ee
where $P_{\rm diff,\,dm}^{[n,i_1\ldots i_n]\;|\; (m))}$ is the correspondent diagram that contributes to $\langle\delta_{\rm diff}^{(n)}(\vec k,\tilde a)\,\delta^{(m)}(\vec k',a)\rangle$ and that was computed in the former subsection~(\ref{sec:diagramnum}). See Fig.~\ref{fig:perturbative_nu_dm_diagram_1} for a diagrammatic representation of two of the six diagrams of this kind.

\begin{figure}[htb!]
\centering
\includegraphics[trim={0cm 5cm 3cm 5cm},width=6.5cm]{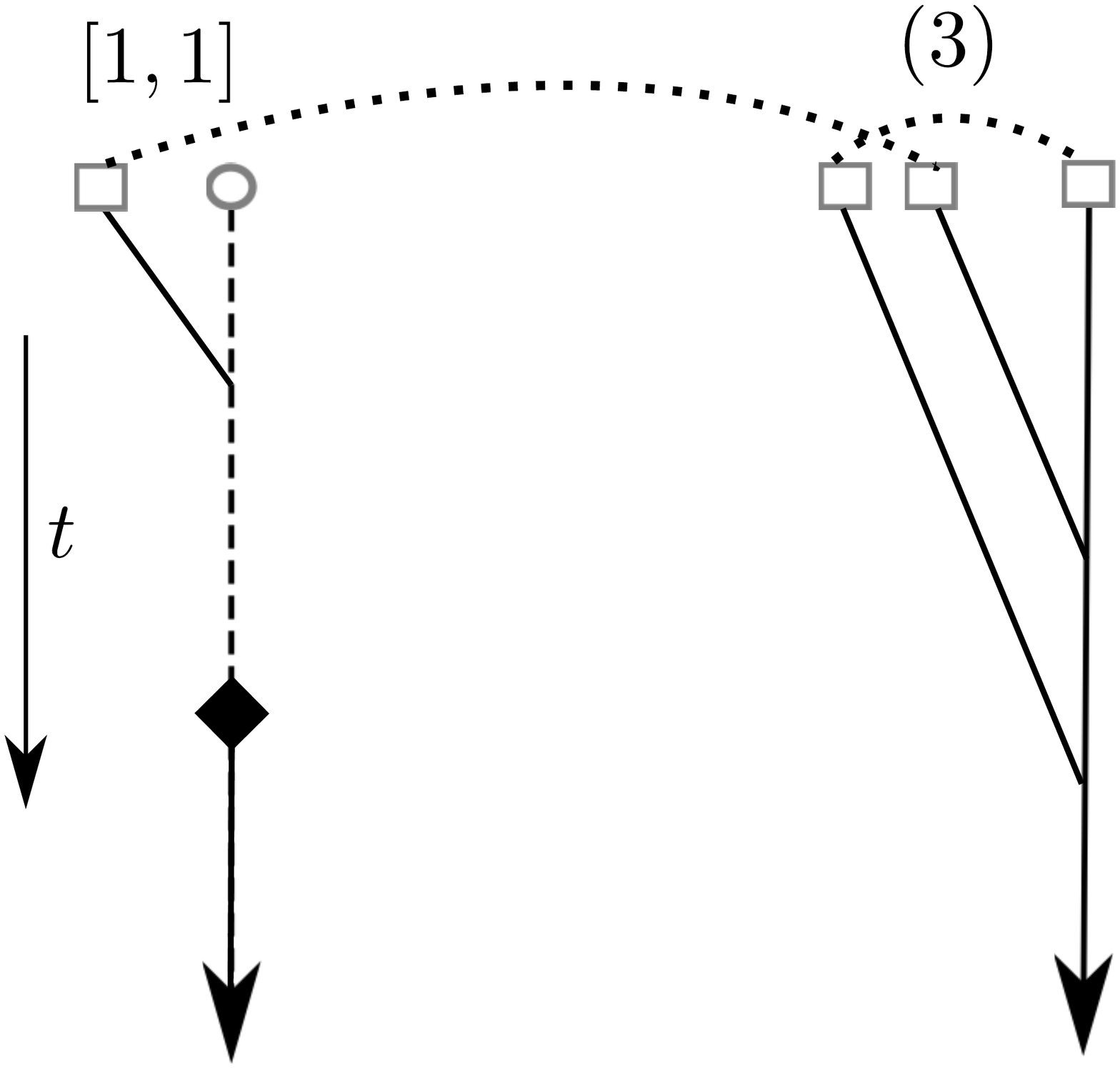}\hspace{3.5cm}
\includegraphics[trim={0cm 5cm 3cm 5cm},width=6.5cm]{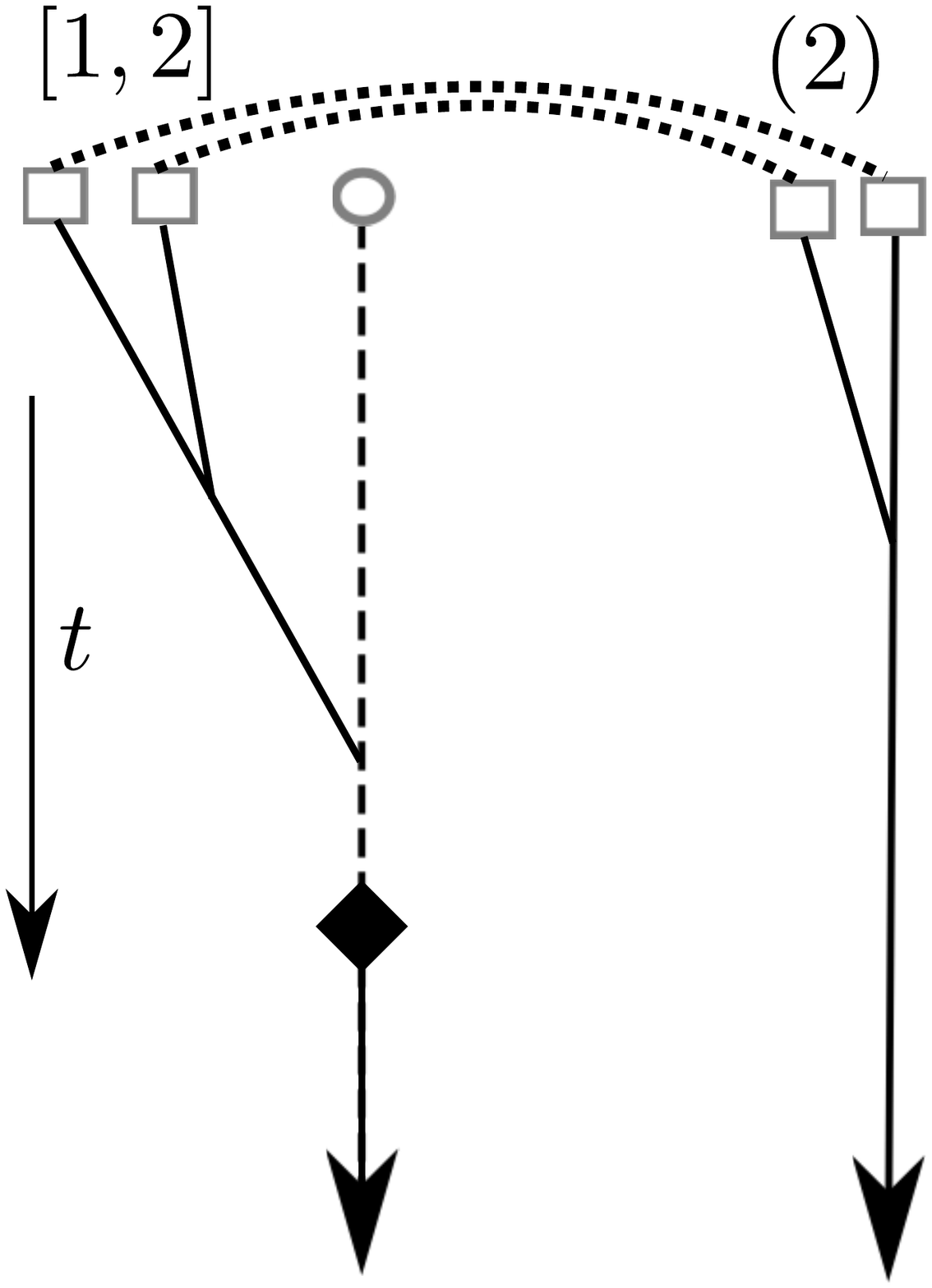}
\caption{\small {\it Left:} One-loop diagram obtained contracting $\delta^{(1)}_{\rm dm\leftarrow\nu}$ with $\delta_{\rm dm}^{(3)}$. {\it Right:} One-loop diagram obtained contracting $\delta^{(2)}_{\rm dm\leftarrow\nu}$ induced by $f^{[1,2]}$ with $\delta_{\rm dm}^{(2)}$. We do not show the analogous diagrams that can simply be obtained by applying the same manipulations as here to the diagrams of Fig.~\ref{fig:perturbative_nu_diagram_3} and~\ref{fig:perturbative_nu_diagram_5}. These are {\it integral} diagrams, as they are obtained by simple integration of the analogous diagrams for $\langle\delta_{\rm diff}\delta_{\rm dm}\rangle$.}\label{fig:perturbative_nu_dm_diagram_1}\label{fig:perturbative_nu_dm_diagram_2}
\end{figure}

\subsubsection*{Non-Integral diagrams}

The other way on top of (\ref{eq:integral_diagrams}) to obtain contributing diagrams  is to use  $\delta_{{\rm dm}\leftarrow \nu}^{(n)}$ or $\Theta_{{\rm dm}\leftarrow \nu}^{(n)}$ in some non-linear terms in the equations of motion for dark matter in (\ref{eq:perturbative_explicit}), to obtain an higher order perturbation to be contracted with the relevant one. This leads to several diagrams, that we list here. We denote this solutions with the notation $\delta_{\rm dm}^{[q,i_1\ldots i_q](m)}$, with $i_1+\ldots+i_q=n$, which means that we have constructed an $n+m$ order solution by considering  a $\delta_{{\rm dm}\leftarrow \nu}^{(n)}$ or $\Theta_{{\rm dm}\leftarrow \nu}^{(n)}$ induced by $f^{[q,i_q\ldots i_q]}$, , with $i_1+\ldots+i_q=n$, and then using dark matter interaction to order $m$. The  diagram obtained by contracting this solution with a $\delta_{\rm dm}^{(p)}$ solution is denoted as $P_{\rm dm,\,dm,\,(m+n,p)}^{[q,i_1\ldots i_q](m)\;|\;(p)}$.  

First, there are two solutions, $\delta^{[1,1](1)}_{\rm dm}$ and $\delta^{[1,1](2)}_{\rm dm}$ respectively, where we use $\delta_{{\rm dm}\leftarrow \nu}^{[1,1]}(\vec k,a)$ and $\Theta_{{\rm dm}\leftarrow \nu}^{[1,1]}(\vec k,a)$ (see Fig.~\ref{fig:perturbative_nu_dm_diagram_3}). The simplest way we find to evaluate the resulting diagrams obtained by contracting with $\delta_{\rm dm}^{(1)}$ and $\delta_{\rm dm}^{(2)}$ respectively is to use the following trick.  We   solve~(\ref{eq:newdmequations3}) by using~(\ref{dtGreen}) and by defining first a linear dark matter perturbation which is given by the sum of the linear perturbation in a cosmology in absence of neutrino perturbations, {\it plus}  $\delta_{{\rm dm}\leftarrow \nu}^{[1,1]}(\vec k,a)$. To the outcome of the resulting diagram, we subtract the evaluation of the same diagram where we just take the linear perturbation to be the one in absence of neutrino perturbations. This procedure leads to two diagrams that we call $\Delta P_{\rm dm,\,dm,\,(3,1)}^{[1,1](2)\;|\;(1)}$ and $\Delta P_{\rm dm,\,dm,\,(2,2)}^{[1,1](1)\;|\;(2)}$. The expression for each of the diagrams of which we need to take the difference can be read-off from App. C of~\cite{Lewandowski:2016yce}, with the important difference that one has to be careful in upgrading the expression to account for not just the correct time-dependence, but also for the correct $k$-dependence for $\delta^{(1)}_{\rm dm}$ and $\Theta^{(1)}_{\rm dm}$, as in the presence of neutrinos the linear growth factor is $k$-dependent~\footnote{One can find the details in the publicly available Mathematica code at \website.}.

\begin{figure}[htb!]
\centering
\includegraphics[trim={0cm 5cm 3cm 5cm},width=6.5cm]{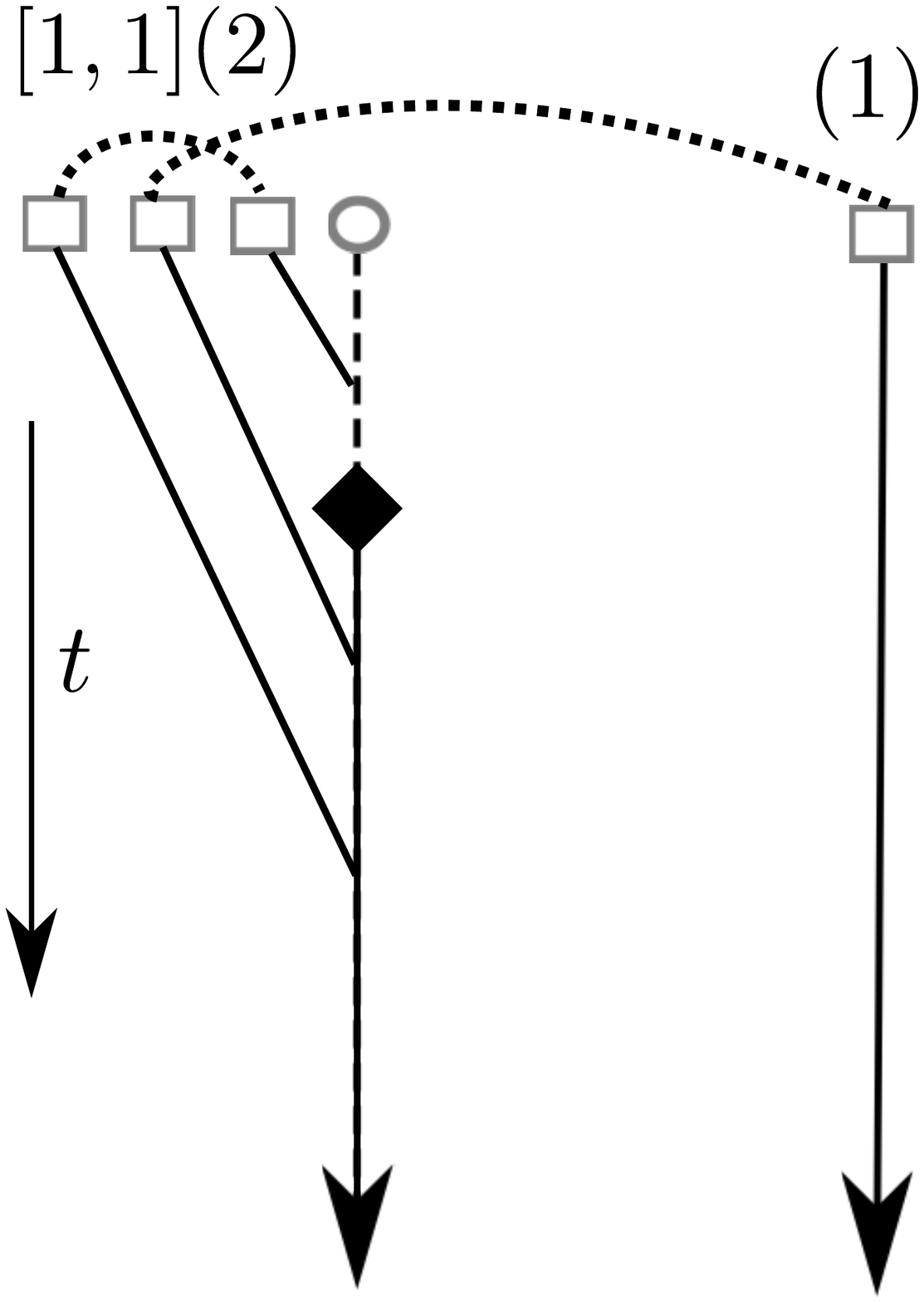}\hspace{2.5cm}
\includegraphics[trim={0cm 5cm 3cm 5cm},width=6.5cm]{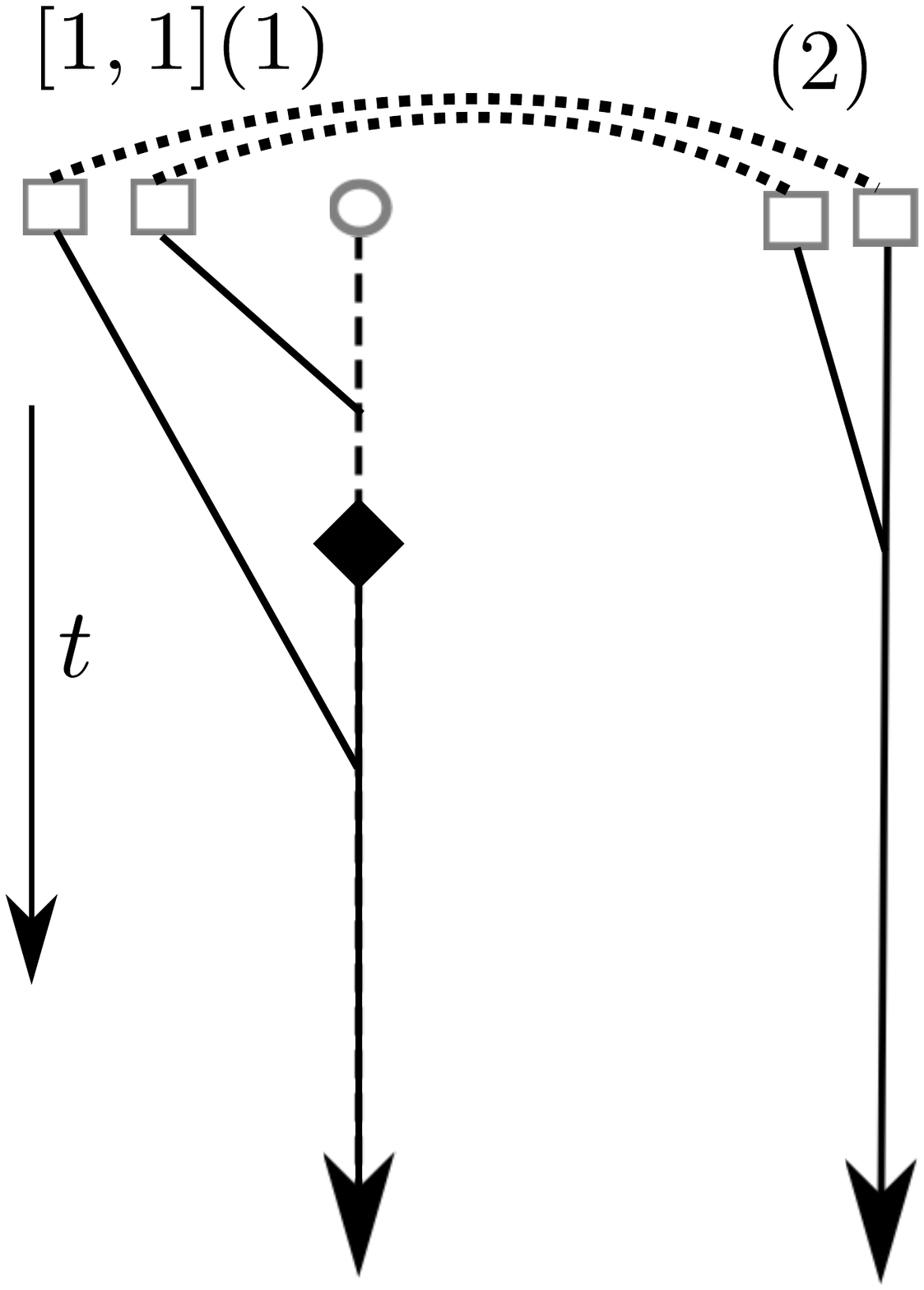}
\caption{\small {\it Left:} One-loop diagram $\Delta P_{\rm dm,\,dm,\,(3,1)}^{[1,1](2)\;|\;(1)}$, obtained contracting   $\delta^{[1,1](1)}_{\rm dm}$ with $\delta_{\rm dm}^{(1)}$. {\it Right:}   One-loop diagram $\Delta P_{\rm dm,\,dm,\,(2,2)}^{[1,1](1)\;|\;(2)}$, obtained contracting a $\delta^{[1,1](2)}_{\rm dm}$ with $\delta_{\rm dm}^{(2)}$. In these plots, it is left implicit that we should consider the sum over all possible ordering of the insertion of a linear $\delta_{{\rm dm}\leftarrow \nu}^{[1,1]}$ and $\Theta_{{\rm dm}\leftarrow \nu}^{[1,1]}$. We have only represented the case where we insert $\delta_{{\rm dm}\leftarrow \nu}^{[1,1]}$ at the first interaction, but it could happen at the second interaction as well, and also it could involve $\Theta_{{\rm dm}\leftarrow \nu}^{[1,1]}$.}\label{fig:perturbative_nu_dm_diagram_3}\label{fig:perturbative_nu_dm_diagram_4}
\end{figure}

Next, there are two additional 31 diagrams (see Fig.~\ref{fig:perturbative_nu_dm_diagram_5}). They are obtained by plugging  $\delta_{{\rm dm}\leftarrow \nu}^{(2)}$ or $\Theta_{{\rm dm}\leftarrow \nu}^{(2)}$, as obtained either by using $f^{[2,11]}$ or by using $f^{[1,2]}$, into a $\delta^{(2)}$ in (\ref{dtGreen}), obtaining in this way two $\delta^{(3)}$'s that can be contracted with a linear fluctuation obtaining what we denote as $\Delta P_{\rm dm,\,dm,\,(3,1)}^{[2,11](1)\;|\;(1)}$ and $\Delta P_{\rm dm,\,dm,\,(3,1)}^{[1,2](1)\:|\;(1)}$. In formulas, we have
\bea\label{eq:delta3f2nonintegral}
&&\delta_{\rm dm}^{[2,11](1)}(\vec k, a_0)=\int^a_0 d\tilde a \bigg(G^{\delta}_{1}(a,\tilde{a})S^{{[2,11](1)}}_1(\tilde{a},\vec k)+G^{\delta}_{2}(a,\tilde{a})S^{{[2,11](1)}}_2(\tilde{a},\vec k)\bigg)\ ,
\eea
where
\begin{align}
&S_1^{{[2,11](1)}}(a,\vk)=\\ \nn
&\quad\qquad =f_g(a)\int \frac{d^3q}{(2\pi)^3}\;  \alpha(\vq,\vk-\vq) \, \left[\tT^{{[2,11]}}_{{\rm dm}\leftarrow\nu}(a,\vq)\,\td^{(1)}(a,\vk-\vq)+\tT^{(1)}(a,\vq)\,\td^{{[2,11]}}_{{\rm dm}\leftarrow \nu}(a,\vk-\vq)\right],   \label{source}   \\
&S_2^{{[2,11](1)}}(a,\vk)=2 f_g(a)\int \frac{d^3q}{(2\pi)^3}\; \beta(\vq,\vk-\vq)\, \tT^{{[2,11]}}_{{\rm dm}\leftarrow\nu}(a,\vec q)\,\tT^{(1)}(a,\vk-\vq)\ ,
\end{align}
and
\bea\nn\label{eq:delta3f12nonintegral}
&&\delta_{\rm dm}^{[1,2](1)}(\vec k, a_0)=\int^a_0 d\tilde a \bigg(G^{\delta}_{1}(a,\tilde{a})S^{[1,2](1)}_1(\tilde{a},\vec k)+G^{\delta}_{2}(a,\tilde{a})S^{[1,2](1)}_2(\tilde{a},\vec k)\bigg)\ ,\\
\eea
where
\begin{align}
&S_1^{[1,2](1)}(a,\vk)=\\ \nn
&\qquad\qquad f_g(a)\int \frac{d^3q}{(2\pi)^3}\;  \alpha(\vq,\vk-\vq) \, \left[\tT^{[1,2]}_{{\rm dm}\leftarrow\nu}(a,\vq)\,\td^{(1)}(a,\vk-\vq)+\tT^{(1)}(a,\vq)\,\td^{[1,2]}_{{\rm dm}\leftarrow \nu}(a,\vk-\vq)\right],   \label{source}   \\
&S_2^{[1,2](1)}(a,\vk)=2 f_g(a)\int \frac{d^3q}{(2\pi)^3}\; \beta(\vq,\vk-\vq)\, \tT^{[1,2]}_{{\rm dm}\leftarrow\nu}(a,\vec q)\,\tT^{(1)}(a,\vk-\vq)\ .
\end{align}
By performing the contractions with $\delta_{\rm dm}^{(1)}$, one obtains $\Delta P_{\rm dm,\,dm,\,(3,1)}^{[2,11](1)\;|\;(1)}(k,a_0)$ and $\Delta P_{\rm dm,\,dm,\,(3,1)}^{[1,2](1)\:|\;(1)}(k,a_0)$. 

\begin{figure}[htb!]
\centering
\includegraphics[trim={0cm 5cm 3cm 5cm},width=6.5cm]{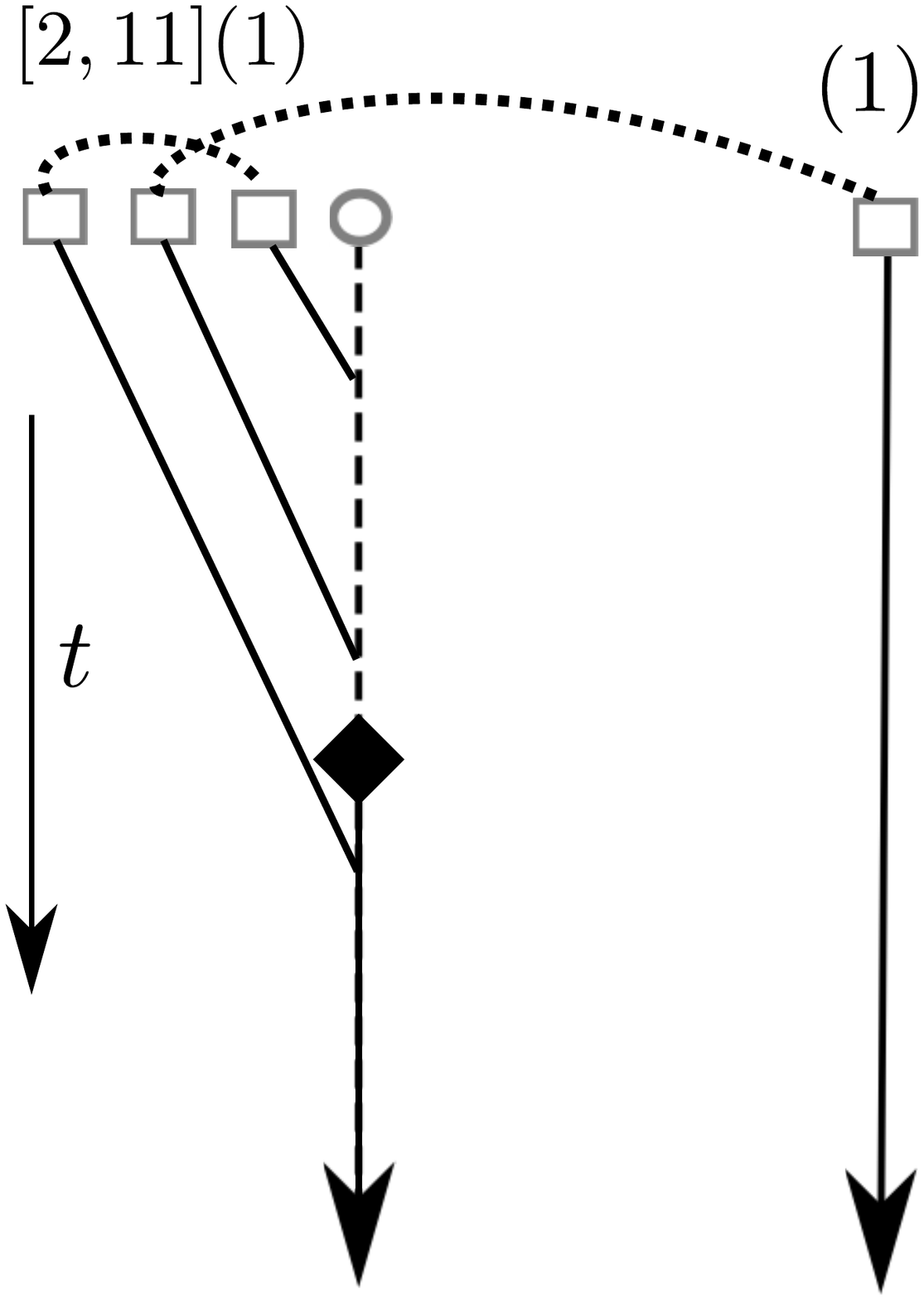}\hspace{2.5cm}
\includegraphics[trim={0cm 5cm 3cm 5cm},width=6.5cm]{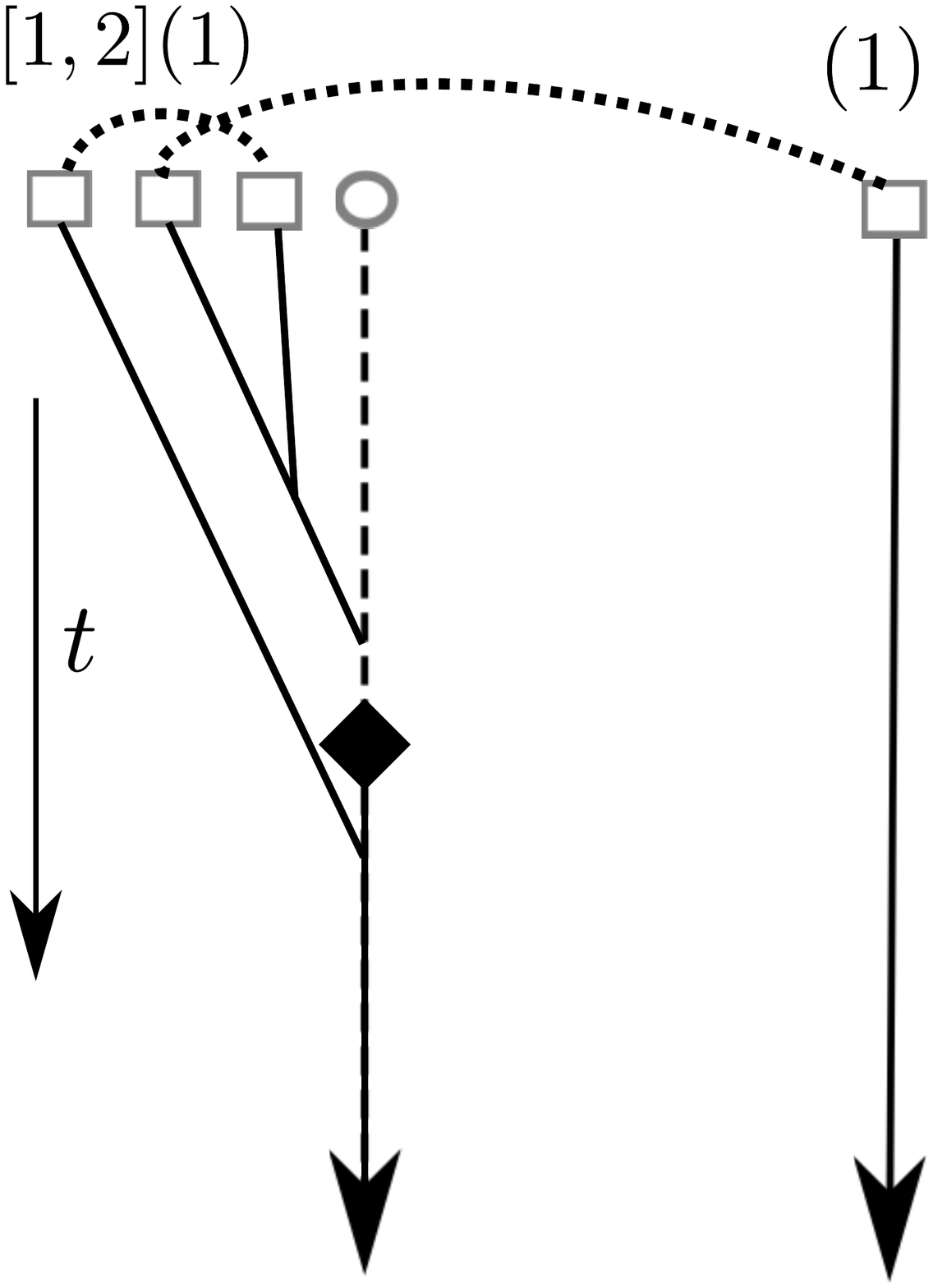}
\caption{\small {\it Left:} One-loop diagram $\Delta P_{\rm dm,\,dm,\,(3,1)}^{[2,11](1)\;|\;(1)}$ obtained contracting  $\delta_{\rm dm}^{[2,11](1)}$ with $\delta_{\rm dm}^{(1)}$. {\it Right:}~One-loop diagram $\Delta P_{\rm dm,\,dm,\,(3,1)}^{[1,2](1)\:|\;(1)}$, obtained contracting $\delta_{\rm dm}^{[1,2](1)}$ with $\delta_{\rm dm}^{(1)}$. In these plots, it is left implicit that we should consider the analogous diagram obtained using $\Theta_{\rm dm}^{[2,11](1)}$ in place of~$\delta_{\rm dm}^{[2,11](1)}$ and $\Theta_{\rm dm}^{[1,2](1)}$ in place of~$\delta_{\rm dm}^{[1,2](1)}$.}\label{fig:perturbative_nu_dm_diagram_5}\label{fig:perturbative_nu_dm_diagram_6}
\end{figure}

This concludes the evaluation of the ordinary diagrams that do not involve the counterterms. We now give two technical (and at some level physical) details that concern the numerical evaluation of these diagrams. 

{\bf Early times:} The first is the following. We pointed out earlier that by evaluating the diagrams using $\tilde f^{[0]}$ in~(\ref{eq:newftilde}), we evaluate automatically the effect induced by $\delta_{\rm diff}=\delta_\nu-\delta_{\rm dm}$. However, some of the Feynman diagrams that are obtained using $\tilde f^{[0]}$ in~(\ref{eq:newftilde}) become quite sensitive to the initial time of integration. This is in contrast with the fact that we wish not to use our neutrino equations at times earlier than $a\sim 1/100$, as for simplicity we wish to work with our non-relativistic expressions. This effect is due to the fact that perturbations at a given $k$ do not start to grow while the free streaming length is longer than the corresponding wavelength, making $\delta_{\rm diff}$ non-vanishing at early times. Therefore, in all the diagrams $P_{\rm diff, dm}$, $P_{\rm dm,\,dm}$ and also in the counterterm diagrams that we discuss next,  instead of using $\tilde f^{[0]}$ we use $\tilde f^{[0]}_\nu$. This corresponds to evaluating the diagrams just with $\delta_{\rm diff}\to\delta_\nu$. The full expression is therefore obtained by subtracting to each diagram, an analogous diagram where we replace $\delta_{\nu}^{(n)}$ with $\delta_{\rm dm}^{(n)}$, evaluated using the standard perturbative expressions, rather than the Boltzmann solution with $\tilde f^{[0]}=-1$, which are evaluated rather easely.

{\bf Log-enhancement:} A related technical complication comes from using the expressions for the linear dark matter fields induced at linear level by $\delta_{\rm diff}$: $\delta_{{\rm dm}\leftarrow \nu}^{[1,1]}(\vec k,a)$ and $\Theta_{{\rm dm}\leftarrow \nu}^{[1,1]}(\vec k,a)$ in (\ref{dtGreen}).  In fact, the Green's function at early times goes as $\sim 1/a^2$, and therefore the resulting expression is logarithmically sensitive to the initial time of the evaluation. This is a physical effect that can be understood as follows. In the presence of neutrinos, dark matter grows in a universe where a fraction of non-relativistic matter does not cluster. This leads to a correction to the exponent of the growth factor proportional to $f_\nu$, which therefore, after Taylor expansion, appears as a logarithmic correction. Such a phenomenon is not unusual whenever we treat perturbatively a linear term in the equations of motion: usually the resulting solution breaks  down after a long enough time. In our universe, the logarithmic divergence is  regulated at the epoch of matter radiation equality,  where however our non-relativistic equations do not apply anymore. Luckily, there is an easy fix to this problem. We can construct $\delta_{{\rm dm}\leftarrow {\rm diff}}^{(1)}(\vec k,a)$ and $\Theta_{{\rm dm}\leftarrow {\rm diff}}^{(1)}(\vec k,a)$, which are the dark matter fields induced by $\delta_{\rm diff}$, without using (\ref{dtGreen}) but rather using the difference of two exact linear solutions obtained from a relativistic Boltzmann solver such as CMBFAST~\cite{Seljak:1996is}, CAMB~\cite{Lewis:1999bs} or CLASS~\cite{Blas:2011rf}. One solution obtained running with massive neutrinos, and the other with massless neutrinos (and the same overall $\Omega_{\rm NR}$). Therefore, in every diagram involving $\delta_{{\rm dm}\leftarrow {\rm diff}}^{(1)}(\vec k,a)$ and $\Theta_{{\rm dm}\leftarrow {\rm diff}}^{(1)}(\vec k,a)$ we should keep in mind that that is not constructed with (\ref{eq:inducedmatterpt}), but rather with the relativistic Boltzmann solver.

Finally, in order to ease the numerical evaluation, we manipulate some of the diagrams to make them UV and IR safe, as described in~\cite{Carrasco:2013sva,Lewandowski:2017kes}. We subtract from each diagram the spurious UV and IR divergencies that cancel from the overall answer. We are now ready to pass to the evaluation of the diagrams involving the counterterms.

\section{Diagrams involving counterterms}

\subsection{Fast neutrinos\label{sec:fastcounterterms}}

Let us proceed to evaluate the contribution of the counterterms, which arise by using in the perturbative expressions the terms with subscript $[\ldots]_\C$ that appear in products of fields such as (\ref{eq:product_renorm}).  We will see that the inclusions of the counterterms corrects, in a perturbative way, the uncontrolled UV contribution of the perturbative diagrams that could affect the result at the perturbative order at which we work. Therefore,  we will limit to counterterms to diagrams of the 13 kind, as they are the leading ones, and we will work only up to linear order in the fields.
As we discussed, the structure of the perturbative expansion is different if we consider `fast' or `slow' neutrinos, where `fast' neutrinos are those whose free-streaming length $1/\kfs\sim v/H\gtrsim \knl$. 
We therefore treat them separately, and start with the fast neutrinos.

Before moving on, it is worthwhile to point out that the UV contribution of the perturbative diagrams can be corrected not just by using the renormalized products of fields, as in~(\ref{eq:product_renorm}), in the perturbative diagrams, as we use in the main part of this paper. An alternative approach is to construct an Effective Boltzmann equation by adding to the normal Boltzmann equations suitable new terms that act as counterterms, correcting the UV dependence of the perturbative calculations. These terms are the analogous of the effective stress tensor, $\tau_{ij}$, that is added to the fluid equations for dark matter. However, unlike for dark matter, the presence of fast neutrinos makes it impossible to construct an effective equation that is local in space, unless we also expand in $\vnl/v\ll1$. We perform the explicit construction of the Effective Boltzmann equation in App.~\ref{app:EffectiveBoltzmann}, where we show that, in an expansion in $\vnl/v\ll1$, we are able to add to the ordinary Boltzmann equation new terms that correct the UV dependence in perturbation theory. Here, instead, we proceed to renormalize the perturbative diagrams by using~(\ref{eq:product_renorm}) directly. This will allow us not to expand in $\vnl/v\ll1$. However  the same spatial non-locality due to the fast neutrinos will be mirrored in both formalisms. When  using (\ref{eq:product_renorm}) directly in the perturbative diagrams it will lead to a particular sensitivity to the time dependence of the counterterms.

\subsubsection{Contributions to $P_{\rm diff,\, {\rm dm}}$}

We start with the contribution to $P_{\rm diff,\, {\rm dm}}$. The simplest way to evaluate the contribution of the counterterms is to use the renomarlized products of fields that we discussed around eq.~(\ref{eq:product_renorm}). When computing the loops without counterterms, we encountered products of fields, $\delta(\vec q_1)\delta(\vec q_2)\ldots$, and we evaluated them using the perturbative expressions (using the notation of (\ref{eq:product_renorm}), $[\delta(\vec q_1)]_\R[\delta(\vec q_2)]_\R\ldots$~(\footnote{We actually have not yet computed $[\delta(\vec q_1)]_\R[\delta(\vec q_2)]_\R\ldots$, because we have not yet used the counterterms of the dark matter equations. We are going to include them as well shortly.})). Now, we need to include the contribution from the terms $[\delta(\vec q_1)\delta(\vec q_2)\ldots]_\C$. Let us start with the diagram obtained from $f^{[3,111]}$, and replace the products of the three dark matter fields with~(\ref{eq:three_point_renorm}). The leading term is when we use the term in $C_1^{(2)}$ in~(\ref{eq:approx}). We notice that
\bea\label{eq:temphelp}
&&[\delta(\vec q,\tau)\,\delta(\vec q',\tau')]_\C=\int d^3x \int d^3x' \;e^{i\,\vec q\cdot \vec x}\,e^{i\,\vec q'\cdot \vec x'}\;[\delta(\vec x,\tau)\,\delta(\vec x',\tau')]_\C\;\\ \nn
&&\qquad\quad\qquad\qquad \supset\; (2\pi)^3   \delta_D^{(3)}(\vec q+\vec q')\;\frac{1}{\knl^3} \, \int^{{\rm Max}(\tau,\tau')} d\tau''\; C^{(2)}_1(\tau,\tau',\tau'')\ .
\eea
Using this contribution from the counterterms for the products of fields $\delta(\vec q_2)\delta(\vec q_3)$ in $f^{[3,111]}$, we obtain a 31 diagram of the form (see Fig.~\ref{fig:perturbative_nu_diagram_counter_1}):
\bea\nn\label{eq:cunterhelptemp}
&&P_{\rm diff,\,dm,\,(3,1),{\rm\;counter}}^{{[3,111]\;|\;(1)}}(k,a_0)\supset\;P_{{\rm dm},11}(k,a_0)\; \int \frac{d^3q}{(2\pi)^3} \int^{a_0} da_1\; \int^{a_1} da_2\int^{a_2} da_3\\ \nn
&&\qquad\qquad   \left(\frac{3}{2}\right)^3 \calH(a_1)\calH(a_2)\calH(a_3) \Omega_{\rm dm}(a_1) \Omega_{\rm dm}(a_2) \Omega_{\rm dm}(a_3)D(a_0) D(a_1) D(a_2)D(a_3) \\ \nn
&&\qquad\qquad \left.\left(\frac{\vec q_1\cdot\vec k }{q_1^2}(s-s(a_1))\right) \left(\frac{\vec q_2\cdot\vec k }{q_2^2}(s-s(a_1))+\frac{\vec q_2\cdot (\vec q_2+\vec q_3)}{q_2^2}(s(a_1)-s(a_2))\right)\right. \\ \nn
&&\qquad\qquad\left(\frac{\vec q_3\cdot\vec k }{q_3^2}(s-s(a_1))+\frac{\vec q_3\cdot (\vec q_2+\vec q_3)}{q_3^2}(s(a_1)-s(a_2))+(s(a_2)-s(a_3))\right)\; \\ \nn
&&\qquad\qquad\left.\tilde f^{[0]}(\vec k (s-s(a_1))+(\vec q_2+\vec q_3) (s(a_1)-s(a_2))+\vec q_3 (s(a_2)-s(a_3)))\right]_{\vec q_1=\vec k,\,\vec q_2=\vec q,\,\vec q_3=-\vec q}\\ 
&&\qquad\qquad  \frac{1}{D(a_2)D(a_3)}\int^{a_2} d\tilde a\; C^{(2)}_1(a_2,a_3,\tilde a)\ .
\eea
As we have been trying to emphasize with our way of writing this term, the integrand in this diagram is extremely similar to the perturbative diagram we just computed in (\ref{eq:P13fromf3}) and indicated with $P_{\rm diff,\,dm,\,(3,1),\,{\rm integrand}}^{f^{[3,111]\;|\;(1)}}(k,a_0,\vec q; \{\vec k,\vec q,-\vec q\})$. The difference is that there is an additional time integration over an unknown function of time,~$C^{(2)}_1(a_2,a_3,\tilde a)$. It is worthwhile to discuss this further.

The time dependence associated to the factor in~$C^{(2)}_1(a_2,a_3,\tilde a)$ comes from two different sources. First, the theory is not time translational invariant, which means that the counterterm parameters that appear in~(\ref{eq:temphelp}) will be time dependent. This accounts for the dependence on the final times, $a_2$ and $a_3$ in this case. Furthermore, since the theory is non-local in time~\cite{Carrasco:2013mua,Carroll:2013oxa}, the counterterms can be expressed in terms of integral over the whole past trajectory of the fluid element under consideration. Such a phenomenon did not affect the computations involving dark matter  too much because in that case the Green's function were space independent and therefore the time integrals could be formally evaluated for each order of perturbation theory. This lead  to the presence in the final answer of a different time dependent constant  for each order in the perturbative expansion a given counterterm was evaluated at~\footnote{Equivalently, one had to add higher convective derivative terms according to the perturbative order in the calculation~\cite{Senatore:2014eva,Mirbabayi:2014zca} (see also~\cite{Bertolini:2016bmt}).}. Schematically
\be
\int^t dt'\; G_R(t,t')\; c(t')\ D(t')^n= K_{(n)}(t)\ ,
\ee
where $G_R$ is the $k$-independent retarded Green's function, and $K_{(n)}(t)$ is a generic function of time. In the case of fast neutrinos, both the absence of time-translation invariance and of time locality affects the evaluation of the counterterms in a more harmful way. In fact in this case the Green's function is $k$-dependent and therefore the integration over time leads to a generic function of $k$. Again, very schematically, we can write
\be
\int^t dt'\; G_R(k,t,t')\; c(t')\ D(t')^n= K_{(n)}(k,t)\ ,
\ee  
where $K_{(n)}(k,t)$ is a generic function of $t$ {\it and} $k$.
Luckily, the situation is not so desperate as it might seem. In fact the counterterms are expected to vary on a time scale of order Hubble. Therefore, one can imagine to expand these function in some basis of functions $g_i$ varying on time scales of order $H$. Schematically $C^{(2)}_1(\tau_1,\tau_2,\tau')=\sum_i c_i \, g_i(\tau_1,\tau_2,\tau')$. The integrals for each term can be evaluated numerically. Only a few functions $g_i$ should be sufficient for our required precision as the contributions from different terms should become  effectively degenerate. 

In this paper we will take a simpler approach. We will assume that the time dependence associated to the counterterms is such that in each diagram they induce a time dependence which is identical to the one that is produced by the perturbative expansion. This choice guarantees that we are able to remove any uncontrolled contribution from perturbation theory and it is expected to be also quite accurate to describe the contribution of the finite counterterms.  We leave a detailed study of its accuracy to future work.

With our choice of the time dependence for the counterterms, we can write (\ref{eq:cunterhelptemp}) as
\bea\nn
&&P_{\rm diff,\,dm,\,(3,1),{\rm\;counter}}^{f^{[3,111]\;|\;(1)}}(k,a_0)\supset  \frac{C_{\rm vev}(a_0)}{\knl^3} \;P_{{\rm dm},11}(k,a_0)\; \int \frac{d^3q}{(2\pi)^3} \int^{a_0} da_1\; \int^{a_1} da_2\int^{a_2} da_3 \\ \nn
&&\qquad\qquad  \left(\frac{3}{2}\right)^3 \calH(a_1)\calH(a_2)\calH(a_3)\Omega_{\rm dm}(a_1) \Omega_{\rm dm}(a_2) \Omega_{\rm dm}(a_3) D(a_0) D(a_1) D(a_2)D(a_3) \\ \nn
&&\qquad\qquad\left[ \left(\frac{\vec q_1\cdot\vec k }{q_1^2}(s-s(a_1))\right) \left(\frac{\vec q_2\cdot\vec k }{q_2^2}(s-s(a_1))+\frac{\vec q_2\cdot (\vec q_2+\vec q_3)}{q_2^2}(s(a_1)-s(a_2))\right) \right.\\ \nn
&&\qquad\qquad\left(\frac{\vec q_3\cdot\vec k }{q_3^2}(s-s(a_1))+\frac{\vec q_3\cdot (\vec q_2+\vec q_3)}{q_3^2}(s(a_1)-s(a_2))+(s(a_2)-s(a_3))\right)\; \\ \nn
&&\qquad\qquad\left.\tilde f^{[0]}(\vec k (s-s(a_1))+(\vec q_2+\vec q_3) (s(a_1)-s(a_2))+\vec q_3 (s(a_2)-s(a_3)))\right]_{\vec q_1=\vec k,\,\vec q_2=\vec q,\,\vec q_3=-\vec q}\ .\\
\eea
The integrand for this term has the same  form as the one in (\ref{eq:P13integrand}) with the replacement $P_{{\rm dm},11}(q,a_0)\to (2\pi)^3C_{\rm vev}(a_0)/\knl^3$. We therefore can write extremely concisely:
\bea\label{eq:P13fromf3counter}
&&P_{\rm diff,\,dm,\,(3,1),{\rm\;counter}}^{f^{[3,111]\;|\;(1)}}(k,a_0)=\frac{C_{\rm vev}(a_0)}{\knl^3} \int \frac{d^3 q}{(2\pi)^3} \left[P_{\rm diff,\,dm,\,(3,1),\,{\rm integrand}}^{{[3,111]\;|\;(1)}}(k,a_0,\vec q; \{\vec k,\vec q,-\vec q\})\right.\\ \nn
&& \left.+P_{\rm diff,\,dm,\,(3,1),\,{\rm integrand}}^{{[3,111]\;|\;(1)}}(k,a_0,\vec q; \{\vec q,\vec k,-\vec q\})+P_{\rm diff,\,dm,\,(3,1),\,{\rm integrand}}^{{[3,111]\;|\;(1)}}(k,a_0,\vec q; \{\vec q,-\vec q,\vec k\})\right]_{P_{{\rm dm},11}(q,a_0)\to (2\pi)^3}.
\eea
Needless to say, this diagram is extremely simple to evaluate once (\ref{eq:P13fromf3}) has been evaluated~\footnote{A further advantage of this approach to evaluate the counterterms, to be compared for example to the result obtained by using the effective Boltzmann equation in App.~\ref{app:EffectiveBoltzmann}, is that we do not need to perform an expansion in $\vnl/v\ll1$ (though we are still assuming that $v\gtrsim \vnl$ as we are assuming that the perturbative expansion converges for all $k's$).}.

\begin{figure}[htb!]
\centering
\includegraphics[trim={0cm 5cm 3cm 4cm},width=6.5cm]{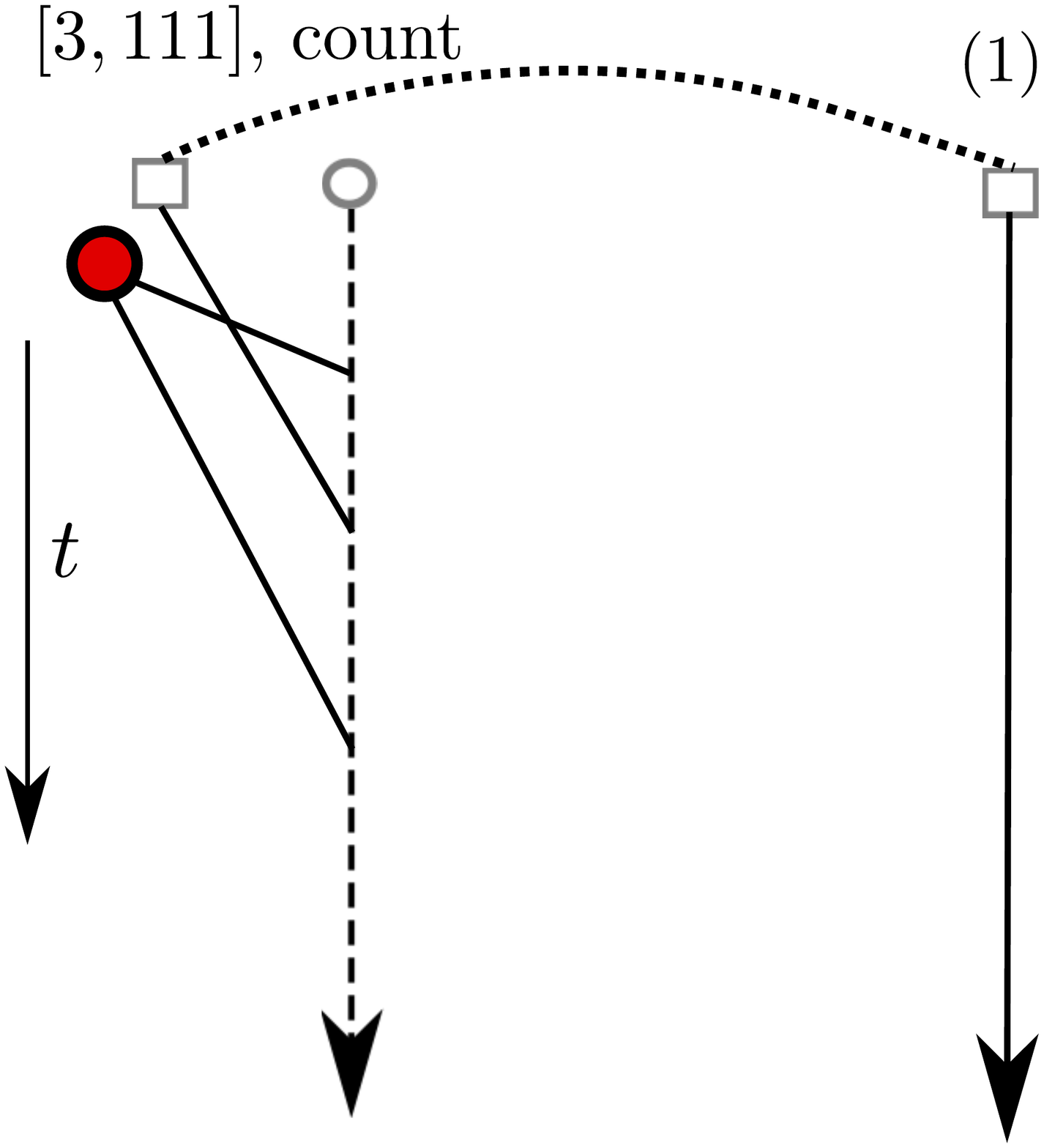}\hspace{2.5cm}
\includegraphics[trim={0cm 5cm 3cm 4cm},width=6.5cm]{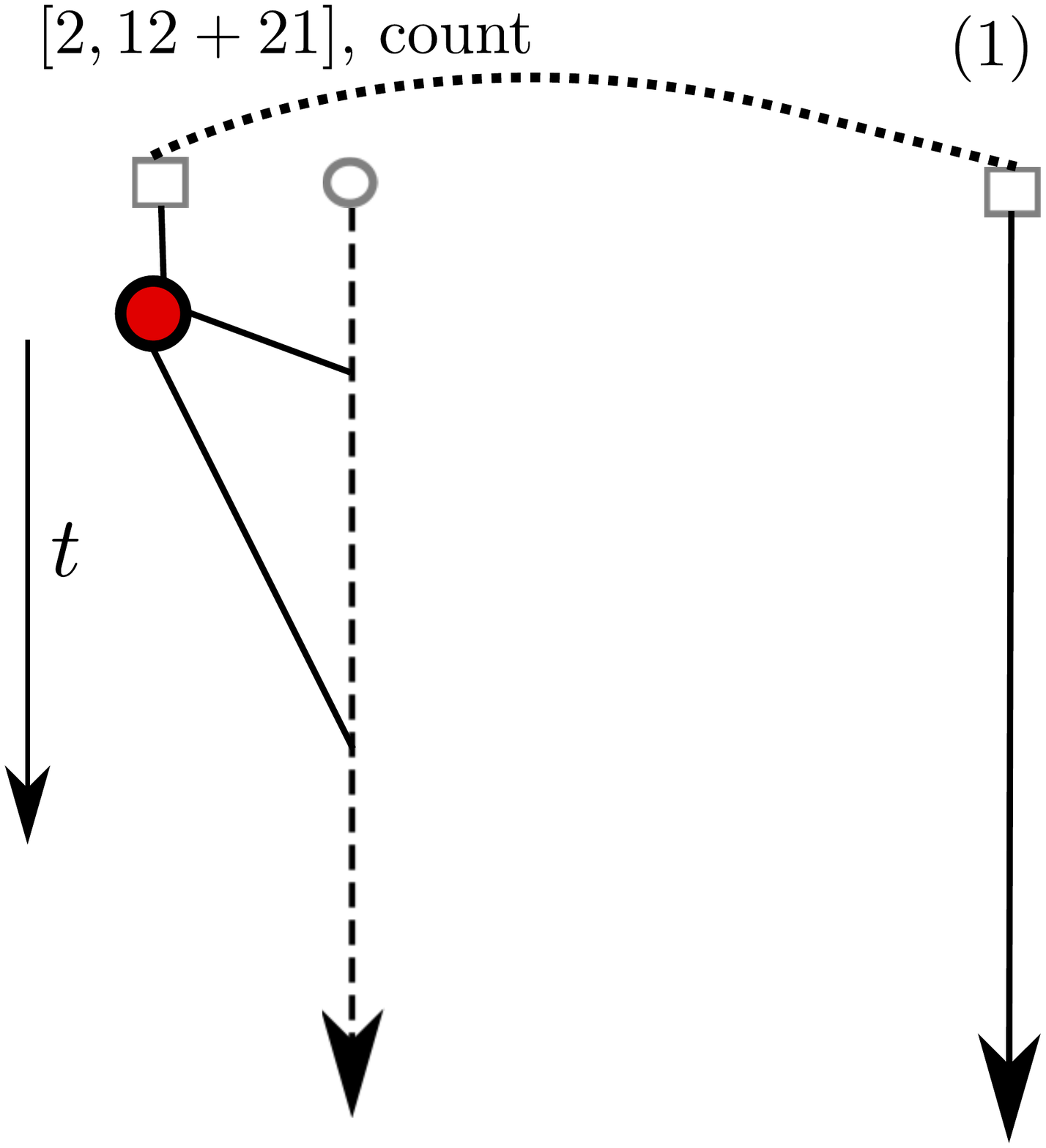}
\caption{\small {\it Left:} $P_{\rm diff,\,dm,\,(3,1),{\rm\;counter}}^{f^{[3,111]\;|\;(1)}}$ counterterm diagram, obtained contracting with $\delta_{\rm dm}^{(1)}$ a $f^{[3,111]}$ where we have renormalized the expectation value of two dark matter fields evaluated at the same location.  The  vacuum expectation value is denoted with a red filled circle with two legs at its bottom. {\it Right:}   $P_{\rm diff,\,dm,\,(3,1),\,{\rm counter,\, \{ flow,\, Iso , \, Ani\}}}^{[2,21+12]\;|\;(1)}$ counterterm diagram, obtained contracting with $\delta_{\rm dm}^{(1)}$ an $f^{[2,21+12]}$ induced by replacing the product of two dark matter fields at the same location in an $f^{[2]}$ with their response to a long wavelength fields. This is represented with a red filled circle with two legs at its bottom and one at its top. Technically, this diagram is obtained by manipulating the non-counterterm diagrams $P_{\rm diff,\,dm,\,(3,1)}^{[2,21]\;|\;(1)}+P_{\rm diff,\,dm,\,(3,1),}^{[2,12]\;|\;(1)}$, a fact that explains the naming of these diagrams.  Keep in mind that there are three kinds of these response diagrams, as there are three kind of responses: `flow', `Iso' and `Ani'.}\label{fig:perturbative_nu_diagram_counter_1}\label{fig:perturbative_nu_diagram_counter_2}
\end{figure}

We now consider the counterterms associated to the diagrams of the 13 kind that arose from the use of $f^{[2]}$, represented in Fig.~\ref{fig:perturbative_nu_diagram_counter_2}. When again considering the contribution given by $[\delta(\vec q)\,\delta(\vec q')]_\C$, we find that here the leading contribution comes from the counterterms linear in the fields. There are three of them. The first and most subtle  is related to the expansion of the terms in $\xfl$ in (\ref{eq:deltafluid}). Using these terms, we obtain counterterms proportional to the velocity of the long modes, and they appear with the same coefficient $C^{(2)}_1$ as the term in the former diagram~\footnote{These are reminiscent of the terms in $v_{\rm dm}$ that upgraded the counterterms in $v$ to $v\r$ in the effective Boltzmann equation in~(\ref{eq:Boltzone_eff4}) in App.~\ref{app:EffectiveBoltzmann}.}. After assuming the same time dependence for the coefficients and performing the same analysis as  before we are lead to a contribution that is indeed very similar to the one of~$P_{\rm diff,\,dm,\,(3,1)}^{[2,21]\;|\;(1)}+P_{\rm diff,\,dm,\,(3,1)}^{[2,12]\;|\;(1)}$ in~(\ref{P31fromf2}), and that takes the form
\bea\nn\label{P31fromf2counterfluid}
&&P_{\rm diff,\,dm,\,(3,1),\,{\rm counter, \,flow}}^{[2,21+12]\;|\;(1)}(k,a_0) =\\ \nn
&&\qquad=\frac{C_{\rm vev}(a_0)}{\knl^3} \left[P_{\rm diff,\,dm,\,(3,1)}^{[2,21]\;|\;(1)}(k,a_0)+P_{\rm diff,\,dm,\,(3,1),}^{[2,12]\;|\;(1)}(k,a_0)\right]_{\{F_2(\vec k,-\vec q)\to \frac{1}{2}  \frac{\vec k\cdot\vec q}{k^2},\; P_{{\rm dm},11}(q,a_0)\to (2\pi)^3\}}\ ,
\eea
where the reader will hopefully pardon our notation:
\be
\left[P_{\rm diff,\,dm,\,(3,1)}^{[2,21]\;|\;(1)}(k,a_0)+P_{\rm diff,\,dm,\,(3,1)}^{[2,12]\;|\;(1)}(k,a_0)\right]_{\{F_2(\vec k,\vec q)\to \frac{1}{2}  \frac{\vec k\cdot\vec q}{k^2},\; P_{{\rm dm},11}(q,a_0)\to (2\pi)^3\}}\ ,
\ee 
means that the substitution of the arguments on the right hand side must per performed {\it before} evaluation of any integral.

A similar reasoning allows us to compute the contribution from the response of the short wavelength fields to the long ones, proportional to $C^{(2)}_2$ and $C^{(2)}_3$ in (\ref{eq:approx}). We obtain
\bea\label{P31fromf2counterresponse}
&&P_{\rm diff,\,dm,\,(3,1),\,{\rm counter, \,Iso}}^{[2,21+12]\;|\;(1)}(k,a_0)(k,a_0)=\\ \nn
&&\qquad=\frac{C_{\rm res,\; Iso}(a_0)}{\knl^3} \left[P_{\rm diff,\,dm,\,(3,1)}^{[2,21]\;|\;(1)}(k,a_0)+P_{\rm diff,\,dm,\,(3,1),}^{[2,12]\;|\;(1)}(k,a_0)\right]_{\{F_2(\vec k,-\vec q)\to 1,\; P_{{\rm dm},11}(q,a_0)\to (2\pi)^3\}}\ , \\ \nn
&&P_{\rm diff,\,dm,\,(3,1),\,{\rm counter, \,Ani}}^{[2,21+12]\;|\;(1)}(k,a_0)=\\ \nn
&&\qquad=\frac{C_{\rm res,\; Ani}(a_0)}{\knl^3} \left[P_{\rm diff,\,dm,\,(3,1)}^{[2,21]\;|\;(1)}(k,a_0)+P_{\rm diff,\,dm,\,(3,1),}^{[2,12]\;|\;(1)}(k,a_0)\right]_{\{F_2(\vec k,-\vec q)\to (\hat q\cdot\hat k)^2,\; P_{{\rm dm},11}(q,a_0)\to (2\pi)^3\}}\ .
\eea

To conclude the contribution of the counterterms to $P_{\rm diff,\, {\rm dm}}$, we need to compute the effect of the counterterms of $\Phi^{(3)}$ that enter in the two diagrams involving $f^{[1]}$: $P_{31}$, which involves $f^{[1,3]}$, and $P_{13}$, which involves $f^{[1,1]}$. These corresponds to using the counterterms contained in the products of the form $[\delta]_\R[\delta]_\R$ (see Fig.~\ref{fig:perturbative_nu_diagram_counter_3}). The counteterm for $\Phi^{(3)}$ is the usual $c_s$ counterterm. In order to make our conventions more uniform  we normalize $c_s$ by writing the ordinary $c_s$-counterterm for purely dark matter in the following way:
\be\label{eq:csdarkmatter}
P_{{\rm dm,\,dm},(1,3),\, c_s}(k,a_1,a_0)=\frac{c_s^2}{\knl^3} P_{{\rm dm},11}(k,a_0) D(a_1)\int^{a_0} d\tilde a\; G^\delta_2(a_0, \tilde a) \; D(\tilde a)^3  \int \frac{d^3q}{(2\pi^3)} 3\, F_3(\vec k,\vec q,-\vec q) \ ,
\ee
where $F_3(\vec k_1,\vec k_2,\vec k_3)$ is the standard kernel for the third order dark matter solution in SPT. This choice of normalization for $c_s$, which is different from what done in the normal context of the EFTofLSS (see for example~\cite{Foreman:2015lca}), makes the $c_s$ diagram depend on the $q$-cutoff $\Lambda$ of the loop, but,  for the value $\Lambda=5\hinvMpc$ that we use and for $\knl\simeq 2\hinvMpc$, it makes it roughly an order of magnitude smaller for the same numerical value of $c_s$. Similarly, our choice of the time dependence for the $c_s$ counterterm has been the same as the analogous perturbative diagram. 
We can now write 
\bea\nn\label{eq:cs_non_corrspoendent}
&&P_{\rm diff,\,dm,\,(1,3),\; {\rm counter,\, c_s}}^{[1,1]\;|\;(3)}(k,a_0) =\\ \nn
&&\quad = \frac{c_s^2}{\knl^3} P_{{\rm dm},11}(k,a_0) \left[\int^{a_0} d\tilde a\; G^\delta_2(a_0, \tilde a) \; D(\tilde a)^3\right] \left[ \int \frac{d^3q}{(2\pi^3)} 3\, F_3(\vec k,\vec q,-\vec q)\right]\\
&&\qquad \int^{a_0} da_1\;  \left(\frac{3}{2}\right) \calH(a_1)\Omega_{\rm dm}(a_1)  D(a_1)\;(s-s(a_1))  \; \tilde f^{[0]}(\vec k (s-s(a_1)))\ ,
\eea
and similarly
\bea\nn
&&P_{\rm diff,\,dm,\,(3,1),\; {\rm counter,\, c_s}}^{[1,3]\;|\;(1)}(k,a_0)=\frac{c_s^2}{\knl^3} P_{{\rm dm},11}(k,a_0) D(a_0)  \int^{a_0} da_1\; \left(\frac{3}{2}\right) \calH(a_1)\Omega_{\rm dm}(a_1) \;(s-s(a_1))  \\ 
&&\qquad\qquad     \left[ \int^{a_1} d\tilde a\; G^\delta_2(a_1, \tilde a) \; D(\tilde a)^3\right]\;\left[\int \frac{d^3q}{(2\pi^3)} 3\, F_3(\vec k,\vec q,-\vec q)\right]\;\tilde f^{[0]}(\vec k (s-s(a_1)))\ .
\eea
In principle, we could use the time dependence for $c_s$ that has been already measured in, for example,~\cite{Foreman:2015lca}. However, for the purpose of this paper (where we do not even compare with simulations data), we will assume that it is time independent once we factor out the time dependence of the perturbative diagram it renormalizes.

\begin{figure}[htb!]
\centering
\includegraphics[trim={0cm 5cm 3cm 4cm},width=6.5cm]{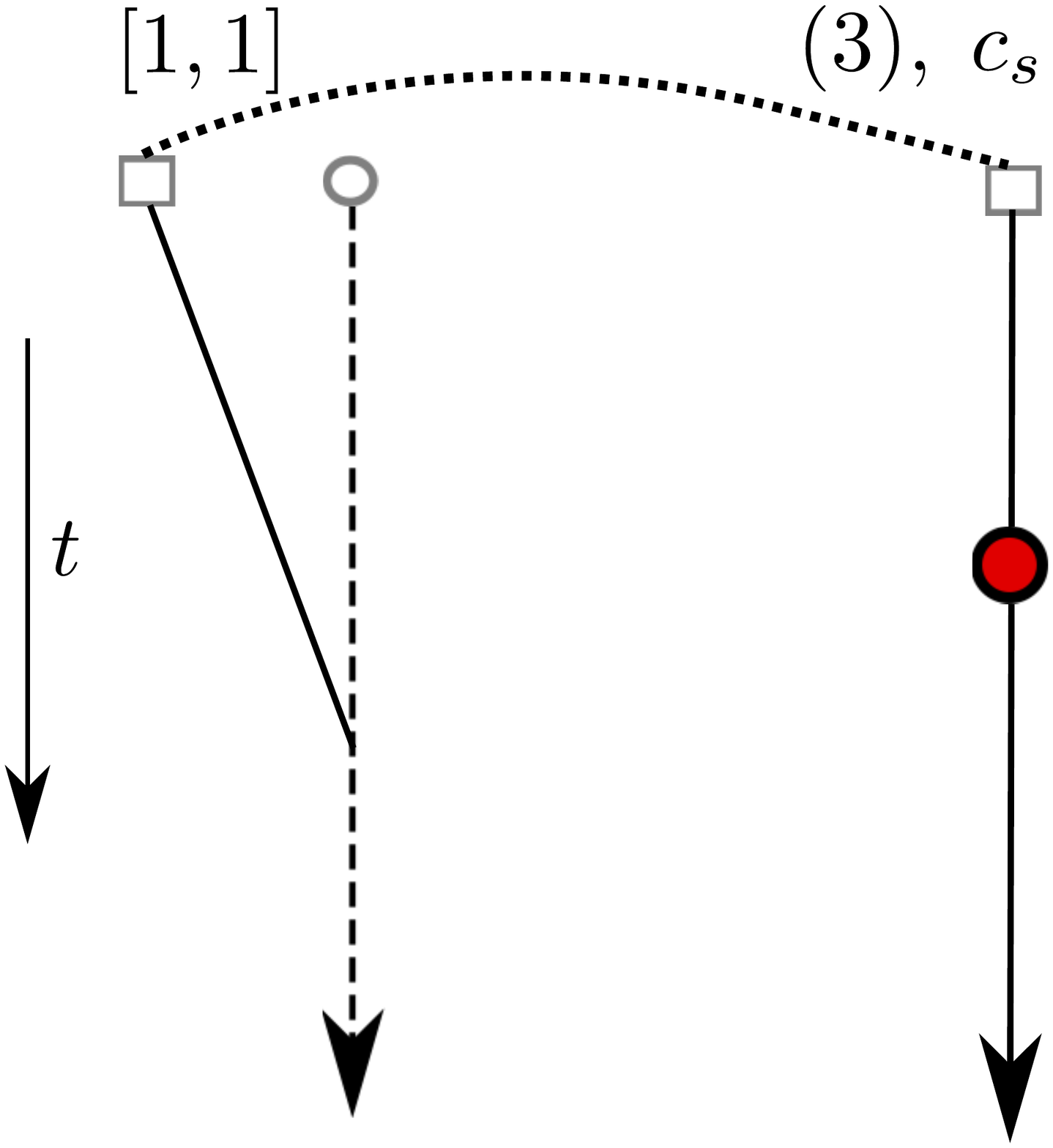}\hspace{2cm}
\includegraphics[trim={0cm 5cm 3cm 4cm},width=6.5cm]{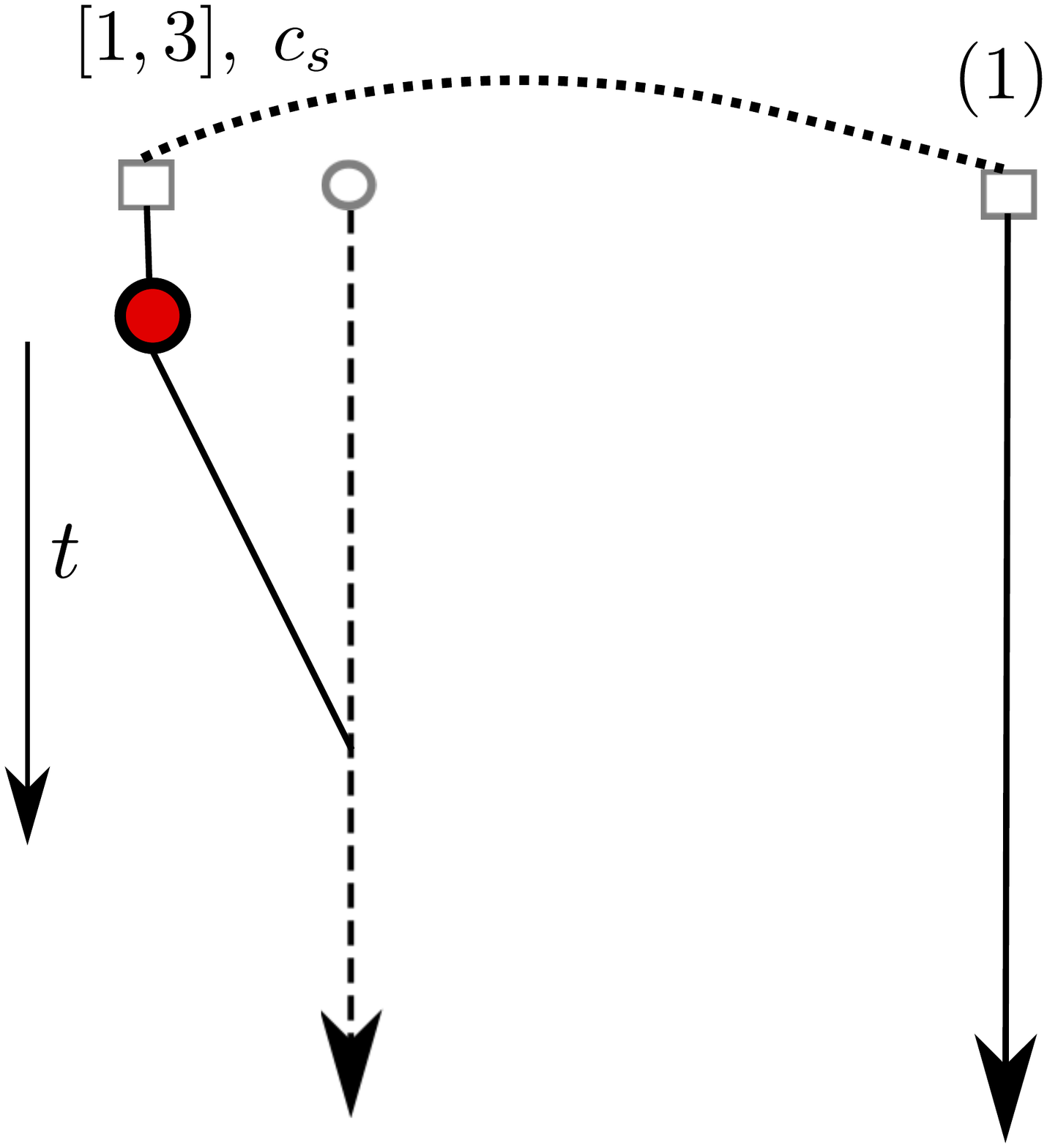}
\caption{\small The contribution to $P_{\rm diff, dm}$ of the two diagrams where we use the $c_s$ counterterm on $\Phi^{(3)}$ for dark matter. {\it Left:} $P_{\rm diff,\,dm,\,(1,3),\; {\rm counter,\, c_s}}^{[1,1]\;|\;(3)}$; {\it Right:} $P_{\rm diff,\,dm,\,(3,1),\; {\rm counter,\, c_s}}^{[1,3]\;|\;(1)}$.}\label{fig:perturbative_nu_diagram_counter_3}\label{fig:perturbative_nu_diagram_counter_3}
\end{figure}

We have now corrected all the relevant uncontrolled UV dependence from the diagrams of the form $P_{\rm diff,\,dm,\,(3,1)}$ associated to fast neutrinos, and we are not going to do the same for the diagrams~$\Delta P_{\rm dm,\,dm,\,(3,1)}$.

\subsubsection{Contributions to $P_{{\rm dm,\,dm}}$\label{sec:Pdmdm_counter}}

We now consider the counterterms that are needed to renormalize the ordinary diagrams $\Delta P_{{\rm dm,\,dm}}$. 
In the presence of neutrinos the dark matter equations of motion in~(\ref{eq:newdmequations}) receive a new non-linear term from the effect of $\delta_{\rm diff}$ on the gravitational potential which requires new counterterms on top of the usual ones for dark matter. 
In fact, in~(\ref{eq:newdmequations}), substituting the Poisson equation into the momentum equation, we can rewrite it in the following way
\bea\label{eq:momentum2}
\dot\pi^i_{\rm dm} + 4 H\pi^i_{\rm dm}+\frac{1}{a}\d_j\left(\frac{\pi^i_{\rm dm}\pi^j_{\rm dm}}{\rho^{(0)}_{\rm dm}(1+\delta_{\rm dm})}\right)+\frac{3}{2}H_0^2 \frac{a_0^3}{a^2} \Omega_{\rm NR,0}\rho^{(0)}_{\rm dm} (1+\delta_{\rm dm}) \frac{\d^i}{\d^2}\left(\delta_{\rm dm}+f_\nu \left(\delta_\nu-\delta_{\rm dm}\right)\right)=0\ .\nn\\
\eea
The terms that involve products of fields at the same location need to be renormalized. For each product of fields, we can therefore write them in the combination of~(\ref{eq:product_renorm}). At zeroth order in $f_\nu$, the counterterms recombine into the usual divergence of the effective stress tensor, which leads to the usual $c_s $counterterm. At linear order in $f_\nu$, we need to write in a renormalized form the products $\delta_{\rm dm} \frac{\d^i}{\d^2} \left(\delta_\nu-\delta_{\rm dm}\right)$, which will induce new counterterm contributions.

\subsubsection*{Non-integral diagrams}

Let us start by considering the additional contribution from renormalizing $\delta_{\rm dm}\d^i\delta_{\nu}\subset \delta_{\rm dm}\d^i(\delta_\nu-\delta_{\rm dm})$. In this expression, $\delta_{\nu}$ can be considered as a known function of the gravitational fields, through the perturbative expressions~(\ref{eq:first_order_sol}) and~(\ref{eq:second_order_sol}). We can use the perturbative expressions because, as we argued, we can trust perturbation theory for fast neutrinos. In particular, this implies that we can stop at the second order solution for this particular term. 

When we insert the linear solution~(\ref{eq:first_order_sol}), we obtain the product at the same location of two dark matter fields that  we can renormalize using the expression~(\ref{eq:product_renorm}) to obtain the contribution of the counterterm (see Fig.~\ref{fig:perturbative_nu_dm_diagram_counter_1}). The expectation value contributes only to the zero mode and we can therefore neglect it. The responses give rise to the following diagrams:
\bea\nn
&&\Delta P_{\rm dm\,,dm,(3,1),\, counter\;\nu, temp}^{{[1,2](1)+[1,1](2)\;|\;(1)}}(k,a_0)= f_\nu P_{{\rm dm},11}(k,a_0) \int \frac{d^3q}{(2\pi)^3}  \int^a_0 da_1 \int^a_1 da_2 \\ \nn
&&\qquad\quad G^\delta_2(a_0, a_1)\cdot \frac{3}{2}\frac{\Omega_{\rm dm}(a_1)}{f_g(a_1)} \cdot 4\;D(a_0)D(a_1)D(a_2) \ F_{\rm temp}(\vec k,\vec q) \\ \nn
&&\qquad\qquad \left[\frac{\vec k\cdot(\vec k-\vec q)}{(\vec k-\vec q)\cdot (\vec k-\vec q)}\left(\left(\frac{3}{2}\right) \Omega_{\rm dm}(a_2) \calH(a_2)  \left(s(a_1)-s(a_2)\right)D(a_2) \right.\right.\\ \nn
 &&\left.\left.\qquad\qquad\qquad\qquad\qquad\qquad\qquad\qquad\qquad\qquad\times\ \tilde f^{[0]}((\vec k-\vec q) (s(a_1)-s(a_2)))\right.\right.\\ 
&&\qquad\qquad+ \frac{\vec k\cdot \vec q}{q^2}D(a_1)\left(\left(\frac{3}{2}\right) \Omega_{\rm dm}(a_2) \calH(a_2)  \left(s(a_1)-s(a_2)\right) \right.\\ \nn
 &&\left.\left.\qquad\qquad\qquad\qquad\qquad\qquad\qquad\qquad\qquad\qquad\times\ \tilde f^{[0]}(\vec q (s(a_1)-s(a_2)))\right)\right] \ .
\eea
Here we introduced a shorthand notation: $\Delta P_{\ldots}^{[1,2](1)+[1,1](2)\;|\;(1)}=\Delta P_{\ldots}^{[1,2](1)\;|\;(1)}+\Delta P_{\ldots}^{[1,1](2)\;|\;(1)}$.
The flow term, the response to $\d^2\Phi$, and to $\d_i\d_j\Phi$ give rise to respectively the following terms
\bea\nn\label{eq:dmdmcounternuresponse}
&&\Delta  P_{\rm dm\,,dm,(3,1),\, counter\;\nu,\, flow}^{{[1,2](1)+[1,1](2)\;|\;(1)}}(k,a_0)=(2\pi)^3 \frac{C_{\rm vev}(a_0)}{\knl^3}\left[P_{\rm dm\,,dm,(3,1),\, counter\;\nu, temp}^{{{[1,2](1)+[1,1](2)\;|\;(1)}}}(k,a_0)\right]_{F_{\rm temp}\to -\frac{1}{2}\frac{\vec k\cdot\vec q}{k^2}}\ ,\\ \nn
&&\Delta  P_{\rm dm\,,dm,(3,1),\, counter\;\nu, \,Iso}^{{[1,2](1)+[1,1](2)\;|\;(1)}}(k,a_0)=(2\pi)^3 \frac{C_{\rm res,\; Iso}(a_0)}{\knl^3}\left[P_{\rm dm\,,dm,(3,1),\, counter\;\nu, temp}^{{[1,2](1)+[1,1](2)\;|\;(1)}}(k,a_0)\right]_{F_{\rm temp}\to 1}\ ,\\ \nn
&&\Delta  P_{\rm dm\,,dm,(3,1),\, counter\;\nu, \,Ani}^{{[1,2](1)+[1,1](2)\;|\;(1)}}(k,a_0)=(2\pi)^3 \frac{C_{\rm res,\; Ani}(a_0)}{\knl^3}\left[P_{\rm dm\,,dm,(3,1),\, counter\;\nu, temp}^{{[1,2](1)+[1,1](2)\;|\;(1)}}(k,a_0)\right]_{F_{\rm temp}\to(\hat k\cdot\hat q)^2}\ .\\ 
\eea

\begin{figure}[htb!]
\centering
\includegraphics[trim={0cm 5cm 3cm 4cm},width=6.5cm]{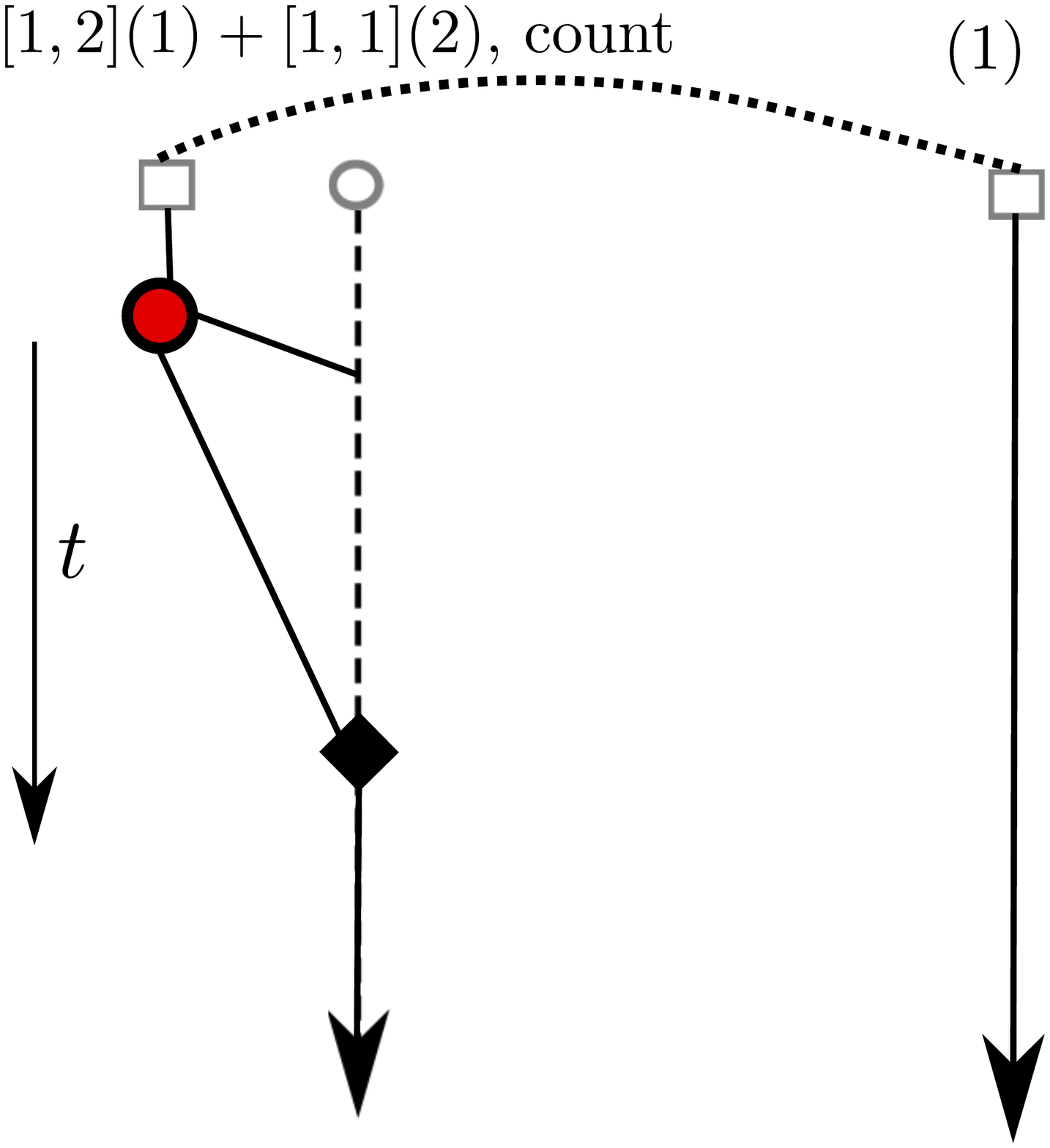}\hspace{2cm}
\includegraphics[trim={0cm 5cm 3cm 4cm},width=6.5cm]{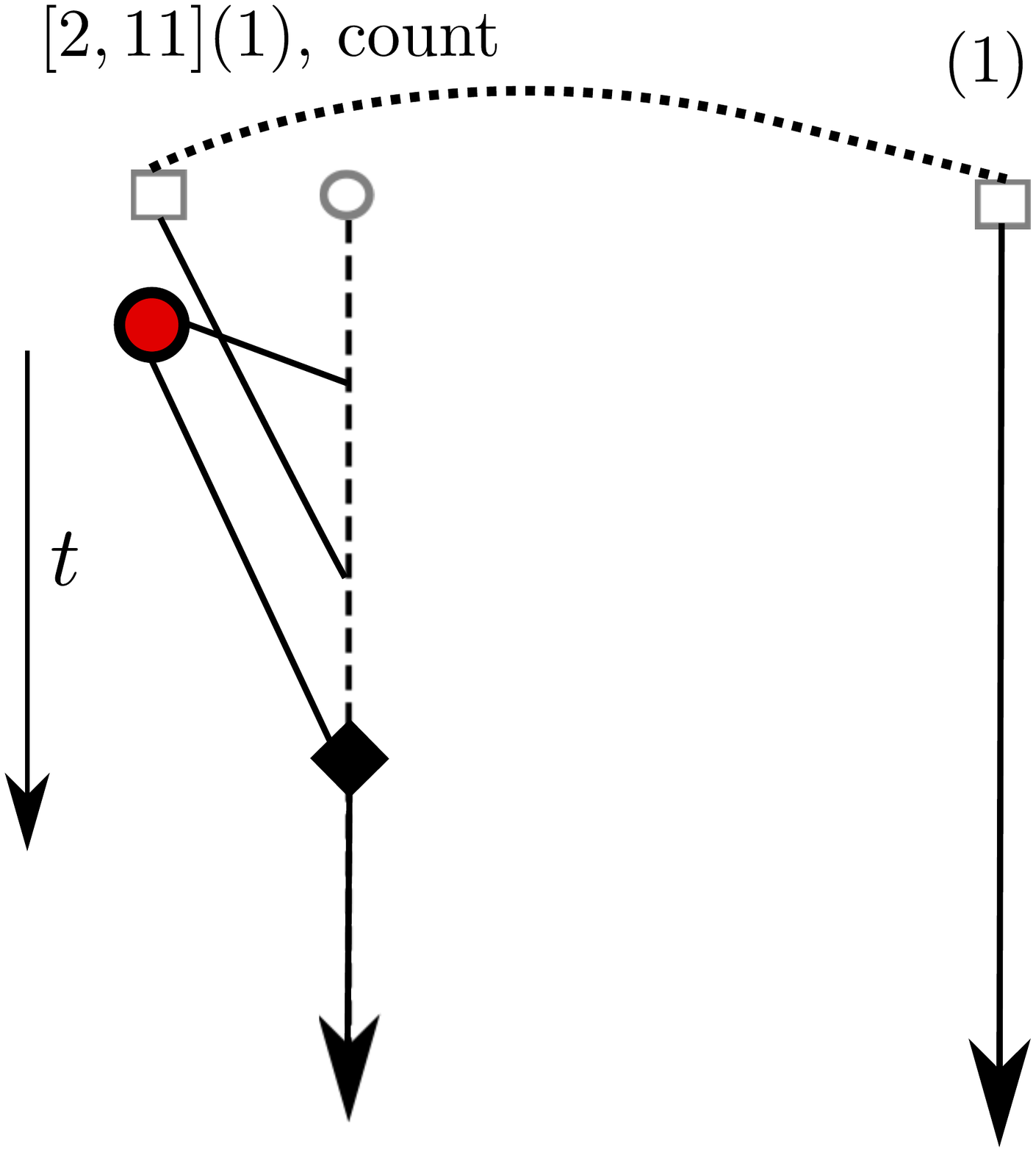}
\caption{\small The contribution to $\Delta P_{\rm dm, dm}$ of the counterterms that arise from the renormalization of the $\delta_{\rm dm}\d_i\delta_{\nu}$ vertex. {\it Left:} $\Delta  P_{\rm dm\,,dm,(3,1),\, counter\;\nu,\,  \{ flow,\, Iso , \, Ani\}}^{{[1,2](1)+[1,1](2)\;|\;(1)}}$ is the contribution of the response of the products of two dark matter fields at the same location. {\it Right:} $\Delta P_{\rm dm\,,dm,(3,1),\, counter\;\nu}^{{[2,11](1)\;|\;(1)}}$ is the contribution of the vev of these fields.}\label{fig:perturbative_nu_dm_diagram_counter_1}\label{fig:perturbative_nu_dm_diagram_counter_2}
\end{figure}

We now  consider the counterterm diagram that we obtain  when we insert the quadratic solution for $\delta_\nu$,~(\ref{eq:second_order_sol}) into the momentum equation and take the counterterm for the expectation value (see Fig.~\ref{fig:perturbative_nu_dm_diagram_counter_2}). At the level of the counterterms we obtain a diagram of the following form 
\bea\nn
&&\Delta  P_{\rm dm\,,dm,(3,1),\, counter\;\nu}^{{[2,11](1)\;|\;(1)}}(k,a_0)= f_\nu  (2\pi)^3 \frac{C_{\rm vev}(a_0)}{\knl^3} \;P_{{\rm dm},11}(k,a_0) \int \frac{d^3q}{(2\pi)^3}  \int^a_0 da_1 \int^a_1 da_2\int^a_2 da_3 \\ \nn
&&\qquad G^\delta_2(a_0, a_1)\cdot \frac{3}{2}\frac{\Omega_{\rm dm}(a_1)}{f_g(a_1)} \cdot 2\, D(a_0)D(a_1)D(a_2) D(a_3) \left(\frac{3}{2}\right)^2 \Omega_{\rm dm}(a_2) \calH(a_2)   \Omega_{\rm dm}(a_3) \calH(a_3) \\ \nn
&&\qquad\qquad\left[-\frac{(\vec k-\vec q)\cdot \vec q}{q^2} \left(s(a_1)-s(a_2)\right) \left( \frac{(\vec k-\vec q)\cdot \vec k}{k^2} \left(s(a_1)-s(a_2)\right)+ \left(s(a_2)-s(a_3)\right)\right)\right. \\ \nn 
&&\qquad \qquad\qquad \qquad \qquad\qquad \qquad \times\ \tilde f^{[0]}((\vec k-\vec q) (s(a_1)-s(a_2))+\vec k(s(a_2)-s(a_3)))\\\nn
&&\left.\qquad\qquad+\frac{(\vec k-\vec q)\cdot \vec k}{k^2} \left(s(a_1)-s(a_2)\right) \left( -\frac{(\vec k-\vec q)\cdot \vec q}{q^2} \left(s(a_1)-s(a_2)\right)+ \left(s(a_2)-s(a_3)\right)\right)\right. \\ 
&&\qquad \qquad\qquad \qquad \qquad\qquad \qquad \left. \times\ \tilde f^{[0]}((\vec k-\vec q) (s(a_1)-s(a_2))-\vec q(s(a_2)-s(a_3)))\right]\ . 
\eea

We now consider the renormalization of the remaining contribution $-\delta_{\rm dm}\d^i\delta_{\rm dm}\subset \delta_{\rm dm}\d^i(\delta_\nu-\delta_{\rm dm})$. Strictly speaking, these terms take the form of a stress tensor at leading order in $f_\nu$ so that they can be included by readjusting the $c_s$ in the ordinary diagrams. However, in order to keep the symmetry of the calculation it is worthwhile to include them. The construction of the resulting diagrams are very similar to the one we wrote in (\ref{eq:dmdmcounternuresponse}), with the replacement of the linear solution for neutrinos with the one for dark matter. We therefore keep a similar, though somewhat redundant, notation. Explicitly we have the simple expression:
\bea\nn
&& \Delta  P_{\rm dm\,,dm,(3,1),\, counter\; dm,\, temp}^{{[1,2](1)+[1,1](2)\;|\;(1)}}(k,a_0)= -f_\nu P_{{\rm dm},11}(k,a_0) \int \frac{d^3q}{(2\pi)^3}  \int^a_0 da_1 \int^a_1 da_2 \\ \nn
&&\qquad G^\delta_2(a_0, a_1)\cdot \frac{3}{2}\frac{\Omega_{\rm dm}(a_1)}{f_g(a_1)} \cdot 4\;D(a_0)D(a_1)^3 \ F_{\rm temp}(\vec k,\vec q) \ \left[\frac{\vec k\cdot(\vec k-\vec q)}{(\vec k-\vec q)\cdot (\vec k-\vec q)} + \frac{\vec k\cdot \vec q}{q^2}\right] \ .
\eea
and
\bea\nn\label{eq:dmdmcounternuresponsedm}
&&\Delta  P_{\rm dm\,,dm,(3,1),\, counter\; dm,\, flow}^{{[1,2](1)+[1,1](2)\;|\;(1)}}(k,a_0)=(2\pi)^3 \frac{C_{\rm vev}(a_0)}{\knl^3}\left[P_{\rm dm\,,dm,(3,1),\, counter\; dm,\, temp}^{{[1,2](1)+[1,1](2)\;|\;(1)}}\right]_{F_{\rm temp}\to -\frac{1}{2}\frac{\vec k\cdot\vec q}{k^2}}\ ,\\ \nn
&&\Delta  P_{\rm dm\,,dm,(3,1),\, counter\; dm,\, Iso}^{{[1,2](1)+[1,1](2)\;|\;(1)}}(k,a_0)=(2\pi)^3 \frac{C_{\rm res,\; Iso}(a_0)}{\knl^3}\left[P_{\rm dm\,,dm,(3,1),\, counter\; dm,\, temp}^{{[1,2](1)+[1,1](2)\;|\;(1)}}\right]_{F_{\rm temp}\to 1}\ ,\\ \nn
&&\Delta  P_{\rm dm\,,dm,(3,1),\, counter\; dm,\, Ani}^{{[1,2](1)+[1,1](2)\;|\;(1)}}(k,a_0)=(2\pi)^3 \frac{C_{\rm res,\; Ani}(a_0)}{\knl^3}\left[P_{\rm dm\,,dm,(3,1),\, counter\; dm,\, temp}^{{[1,2](1)+[1,1](2)\;|\;(1)}}\right]_{F_{\rm temp}\to(\hat k\cdot\hat q)^2}\ .\\ 
\eea

There is an additional `non-integral' diagram that we should consider (see Fig.~\ref{fig:perturbative_nu_dm_diagram_counter_4}). This is the one where we apply the ordinary dark matter $c_s$ counterterm to the  dark matter solution induced at linear level by the linear $\delta_{\rm diff}, \delta_{\rm diff}^{(1)}$.  We get:
\bea\nn
&&\Delta  P_{\rm dm\,,dm,(3,1),\; {\rm counter,\, c_s}}^{{[1,1],(2)\;|\;(1)}}(k,a_0) =\\ \nn
&&\quad = \frac{c_s^2}{\knl^3} P_{{\rm dm},11}(k,a_0) \left[\int^{a_0} d\tilde a\; G^\delta_2(a_0, \tilde a) \; D(\tilde a)^2 \Delta_{\rm dm\leftarrow \,diff}(k,\tilde a)\right] \left[ \int \frac{d^3q}{(2\pi^3)} 3\, F_3(\vec k,\vec q,-\vec q)\right]\ , 
\eea
where $\Delta_{\rm dm\leftarrow \,diff}(k,\tilde a)$ is the linear transfer function for dark matter as induced by $\delta_{\rm diff}$. $\Delta_{\rm dm\leftarrow \,diff}(k,\tilde a)$ can be obtained as the difference of the transfer functions of two cosmologies with the same amount of $\Omega_{\rm NR}$, but with massless or massive neutrinos, using  any relativistic Boltzmann code.

\subsubsection*{Integral diagrams}

We should not forget that the counterterms were used also to renormalize the expressions for $P_{\rm diff,\, dm}$. Each of these diagrams induces a diagram for $\Delta P_{\rm dm,\,dm}$ by simply replacing $P_{\rm diff,\,dm}^{(n,m)}$ used in (\ref{eq:integral_diagrams}) with the relevant counterterm diagram (see Fig.~\ref{fig:perturbative_nu_dm_diagram_counter_3}):
\be\label{eq:integral_diagrams_counter}
\Delta P_{\rm dm,\,dm,\,counter}^{[n,i_1\ldots i_n]\;|\; (m)}(k,a_0)=\int^{a_0} d\tilde a\; G^\delta_2(a_0,\tilde a)\;\frac{3}{2}\frac{\Omega_{\rm dm}(\tilde a)}{f_g(\tilde a)}P_{\rm diff,\,dm\,\,counter}^{[n,i_1\ldots i_n]\;|\; (m)}(k,\tilde a,a_0)\ ,
\ee
Notice that the assumption that the time-dependence of the counterterm diagrams is exactly equal to the one in perturbation theory helps us not only in evaluating $P_{\rm diff,\,dm,\; counter}$ but also in performing the last time integration in $P_{\rm dm,\,dm,\; counter}^{{\rm integral}, (n,m)}$. As mentioned in the context of the discussion of the diagrams involving $P_{\rm diff,\,dm,\; counter}$ one could try to employ the described additional methods to account for a more general time dependence.

\begin{figure}[htb!]
\centering
\includegraphics[trim={0cm 5cm 3cm 4cm},width=6.5cm]{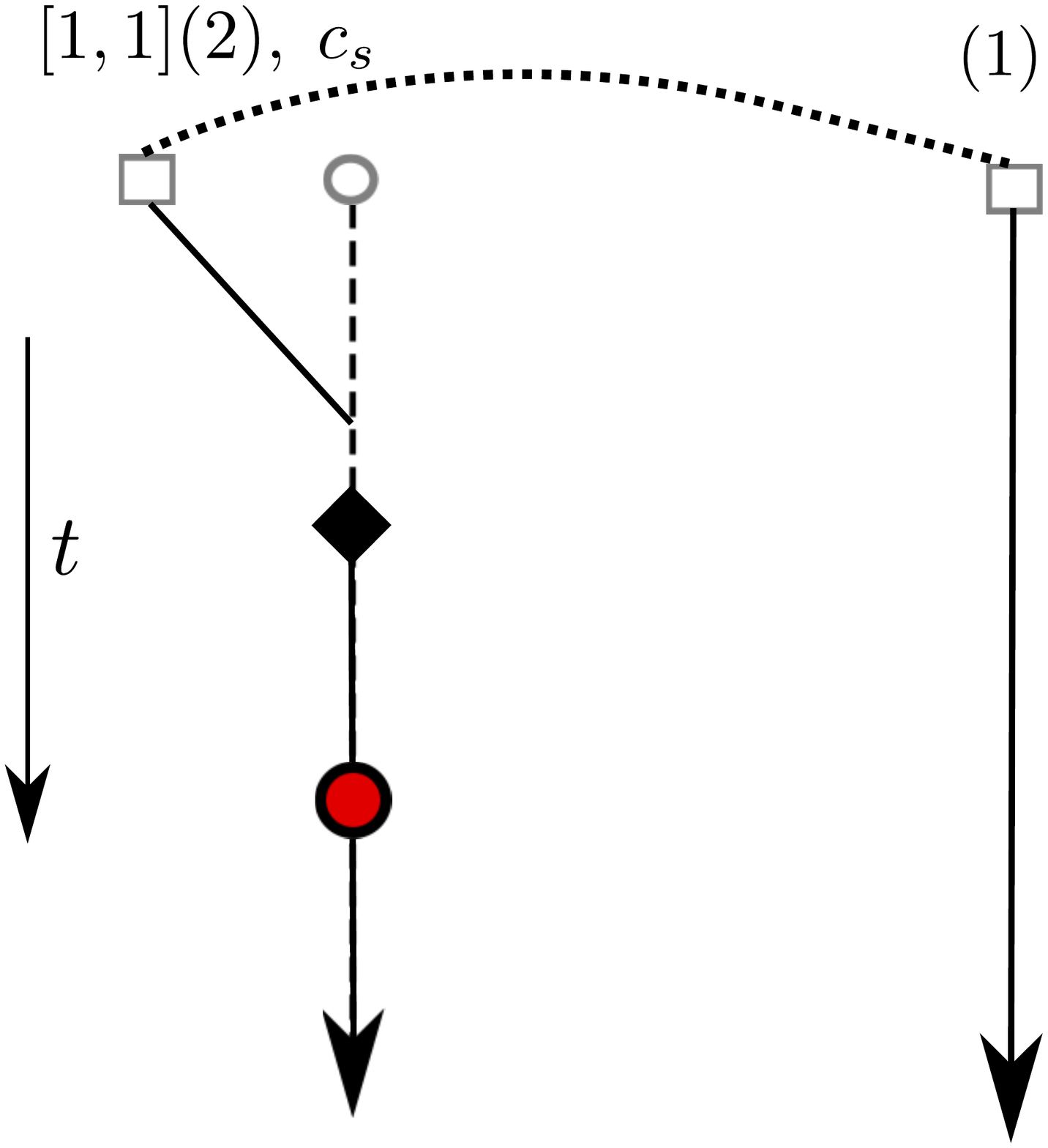}\hspace{2cm}
\includegraphics[trim={0cm 5cm 3cm 4cm},width=6.5cm]{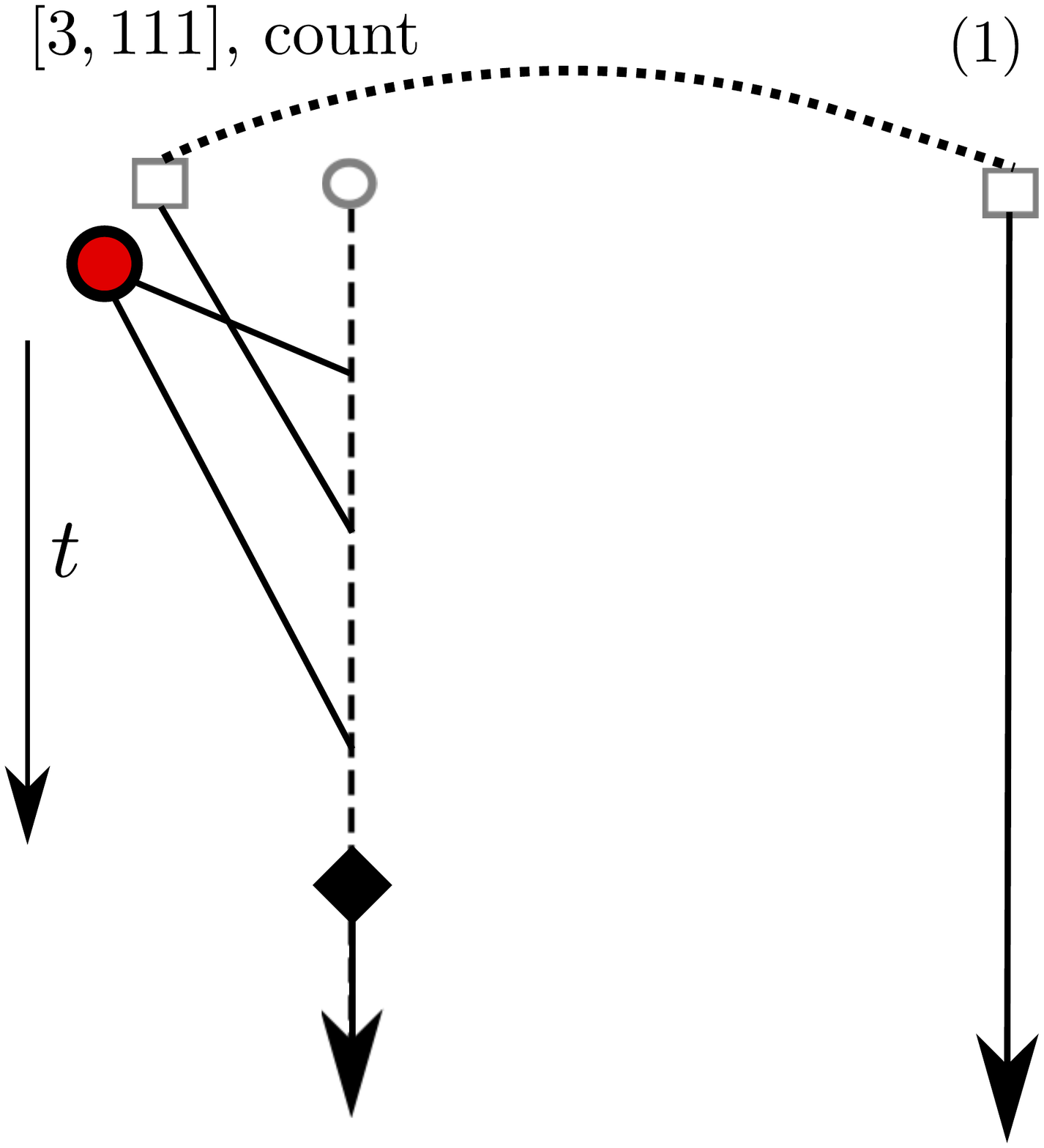}
\caption{\small  {\it Left:} The counterterm diagram that is obtained after converting a $\delta_\nu$ into $\delta_{\rm dm}$, and then applying the $c_s$ counterterm to the resulting fluctuation: $\Delta  P_{\rm dm\,,dm,(3,1),\; {\rm counter, c_s}}^{{[1,1],(2)\;|\;(1)}}$. {\it Right:} A representative of the 
$\Delta P_{\rm dm,\,dm,\,counter}^{[n,i_1\ldots i_n]\;|\; (m)}$ that are obtained by simple integration of the counterterm diagrams for $P_{\rm diff,\;dm}$, after inserting at the end of the diagram, the mixing vertex and the resulting Green's function. There is a total a six {\it integral} diagrams like similar to this one. Here we represent~$\Delta P_{\rm dm,\,dm,\,counter}^{[3,111]\;|\; (1)}$. 
}\label{fig:perturbative_nu_dm_diagram_counter_3}\label{fig:perturbative_nu_dm_diagram_counter_4}
\end{figure}

\vspace{0.3cm}

A final comment: strictly speaking, in all the expressions in this section, we have considered only the contribution associated to the `fast' neutrinos, so that we should really have used a sort of 
\be
\tilde f^{[0]}_{\rm fast}(q)=\int_{v\gtrsim a \vnl} d^3 v \; e^{-i\,\vec q\cdot\vec v} \; \left(f^{[0]}(v)\; \Theta(v-a\vnl)-1\right)\ . 
\ee
However, as we discuss in the context of `slow' neutrinos later on, there is no harm in adding the contributions of these diagrams also for  the slow neutrinos, because anything that we add here for them can be corrected by the counterterm we add later on, so we can just use the full  $\tilde  f^{[0]}(q)$ in the counterterms for fast neutrinos.\vspace{0.3cm}

This concludes the list of all the diagrams involving fast neutrinos, where, manifestly, all the relevant uncontrolled UV dependence from the diagrams of the form $P_{\rm diff,\,dm,\,(3,1)}$ and $\Delta P_{\rm dm,\,dm,\,(3,1)}$ due to fast neutrinos has been removed.

\subsection{Slow neutrinos\label{sec:slowcounterterms}}

We now discuss the slow neutrinos. At first we could try to simply treat them as we did for the fast neutrinos, that is: use perturbation theory and renormalize any UV-sensitive expression with the appropriate counterterms. This is indeed what we did in the former section where, even though we named the section `fast neutrinos', we performed the integration over velocity including the `slow' neutrinos as well. The problem in trusting the contribution we obtained in Sec.~\ref{sec:fastcounterterms} also for the slow neutrinos is that if we did so we would be trusting the perturbative expressions for slow neutrinos. This should be done with care. While perturbations are small for fast neutrinos, they are of order one for `slow' neutrinos everytime we insert a vertex with a fluctuation with $k\gtrsim\knl$, and so, each insertion of such a vertex is naively unsuppressed. In this case, the perturbative expansion would converge to the right answer only if the radius of convergence of the perturbative series went beyond where the expansion parameter is order one. This is something we do not feel confident we can do.

We therefore need to correct the expressions we obtain in Sec.~\ref{sec:fastcounterterms} by adding a suitable counterterm that corrects the contribution for `slow' neutrinos. This appears rather straightforward to do. Indeed, `slow' neutrinos behave exactly as dark matter (which indeed, after a Hubble time, acquires velocity of order $\vnl$) and their free streaming length is not longer than the one of dark matter, of order $1/\knl$. Therefore, for $k\lesssim\knl$, we can simply use the equations for an effective fluid-like system with which we describe dark matter~\footnote{The fact that `slow' neutrinos can be described as an effective fluid-like system confirms  that we can actually trust the perturbative solution of the Boltzmann equation for slow neutrinos as long as all modes are longer than the non-linear scale. However, as we mentioned, we cannot trust the counterterms we derived earlier.}.

There is an important additional subtlety that occurs at the level of the counterterms. The number and momentum of slow neutrinos is not conserved. This means that, naively, we should add the effective fluid-like equations terms that are absent in the presence of a single fluid-like system.  Let us elaborate more on this. Concerning the  non-conservation of the number of slow neutrinos, we know that the overall number of fast and slow neutrinos is conserved. Therefore, since, in dealing with the fast neutrinos, we did not include the fact that some of them can become slow (in practice, we used the renormalized local products of fields and integrated over {\it all} velocities), and since we are only interested in the dynamics of the overall set of neutrinos, in the effective slow neutrino equations we do not need to add counterterms to the continuity equation that would violate number conservation~\footnote{Since the splitting between slow and fast neutrinos is dictated by the velocity relative to dark matter, if we were to explicitly write the counterterms to the continuity equation, we would do so by including terms proportional to $v_{\rm \nu, slow}^i(\xfl(\vec x,\tau,\tau'),\tau)-v_{\rm dm}^i(\xfl(\vec x,\tau,\tau'),\tau)$.}. Instead, the situation is different for the momentum, as the momentum of neutrinos is non-conserved. Therefore, the counterterm to the Euler equation takes the form of an effective force, rather than of the divergence of an effective stress tensor. We therefore write:
\bea\label{eq:slowneutrinoseq}
&&\dot \delta_{\rm \nu, slow} + \frac{1}{a}  \theta_{\rm \nu, slow} + \frac{1}{a} \partial_i \left( \delta_{\rm \nu, slow} v^i_{\rm \nu, slow} \right)=0  \label{nonlincontslow}\ , \\  \nn
&&\dot \theta_{\rm \nu, slow} + H \theta_{\rm \nu,slow} +\frac{1}{a} \d^2\Phi + \frac{1}{a} \partial_i \left( v^j_{\rm \nu, slow} \partial_j v^i_{\rm \nu, slow} \right)  =-\d_i\frac{1}{\rho_{\nu,\rm slow} a}\gamma^i_{\nu,\rm slow}
\label{nonlineulerslow} \ ,
\eea
where $\theta_{\nu,\rm slow}=\d_i v^i_{\nu,\rm slow}$, and where again we stress that the absence of counterterm for the continuity equation is due to the fact that we are studying only the properties of all neutrinos. 
Similarly to the case where we describe baryons and dark matter as a an effective two fluid-like system~\cite{Lewandowski:2014rca}, here $\gamma^i$ represents an external effective force that can be expressed in terms of the long wavelength fields in the usual way, following the same rules as for the effective stress tensor. An important difference is that  in the case of two fluids $\gamma^i$ can depend on the relative velocity between dark matter and neutrinos~\footnote{More precisely, it is the neutrino velocity in the inertial (or Fermi) frame where $\d\Phi$ has been set to zero at all times. At leading order in $f_\nu$, in this frame the dark matter velocity is equal to zero.}. Therefore, symmetries allow us to write a term like
\be
\gamma^i(\vec x,\tau)\supset\int^\tau d\tau'\; K_{\gamma,1}(\tau,\tau')  \left(v_{\rm \nu, slow}^i(\xfl(\vec x,\tau,\tau'),\tau)-v_{\rm dm}^i(\xfl(\vec x,\tau,\tau'),\tau)\right)\ .
\ee
However, at long wavelengths, $k\lesssim \knl$, the solution for the velocity of dark matter, $v^i_{\rm dm}$ is the same as the one for slow neutrinos, $v_{\nu,\rm slow}$, so that the relative velocity vanishes~\footnote{This statement would change if we were to consider isocurvature modes or the initial conditions for neutrinos. However, as discussed around~(\ref{eq:first_order_sol}), their effect is negligible.}. If we now consider all the other long wavelength fields that  $\gamma^i$ can depend upon, such as $\d^i \d^2\Phi,\, \d_i\theta,\,\ldots$, we quickly realize that all terms  scale as $k^2\delta$. We therefore can use (\ref{eq:csdarkmatter}) to write the contribution to $P_{\nu,\rm dm}$ from the additional counterterms associated to the slow neutrinos  as 
\be\label{eq:slowadditonalcounter}
\Delta P_{\nu,\rm dm, counter-slow}(k,a)= f_\nu  f_{\rm \nu,slow}(a)\;P_{{\rm dm,\,dm},(1,3),\, c_s}(k,a_1,a_0)
\ee
where  
\be\label{eq:fslowdef}
f_{\rm \nu,slow}(a)\sim \int^{a\vnl}d^3v\; f^{[0]}(v)\ ,
\ee 
is the fraction of neutrinos that are considered `slow'. We will treat $f_{\rm \nu,slow}$ as a free parameter to be fitted to data, though the definition in~(\ref{eq:fslowdef}) can be used to give a prior on it. In practice, the `slow' neutrinos force us to add an additional counterterm. An equivalent derivation of this result, using the Boltzmann equation, is given in App.~\ref{app:slowneutrinos1} and~\ref{app:slowneutrinos2}.

Similarly to what we had for fast neutrinos the counterterm that gives rise to (\ref{eq:slowadditonalcounter}) also affects the purely matter power spectrum as
\be\label{eq:integral_diagrams_counter_slow}
P_{\rm dm,\,dm,\; counter-slow}^{}(k,a_0)=\int^{a_0} d\tilde a\; G^\delta_2(a_0,\tilde a)\;\frac{3}{2}\frac{\Omega_{\rm dm}(\tilde a)}{f_g(\tilde a)}P_{\rm diff,\,dm,\; counter-slow}(k,\tilde a,a_0)\ .
\ee

On top of this, there is the renormalization of the analogous term that  we had for fast neutrinos in the Euler equation: $\delta_{\rm dm}\d^i\delta_{\nu,\rm slow}$. This terms just gives a correction of order $f_\nu f_{\nu,\,\rm slow}$ to the $c_s$ in the ordinary dark matter counterterm.

\section{Results\label{sec:results}}

Here we illustrate the result of the numerical integration of the loop diagrams that we described earlier, presenting the full result of the correction to the power spectrum of the total matter at one loop order. The cosmological parameters that we use are the following
\bea\nn
&&\Omega_{\rm dm}=0.2430308\,,\quad \Omega_b = 0.0468\,,\quad \Omega_{\Lambda}=0.704948\,,\quad \Omega_\nu=0.0051692\,,\quad H_0=68.8\,, \\ 
&&P_\zeta = 2.187 \cdot 10^{-9}\,,\quad n_s=0.9676\,,\quad k_{\rm pivot}=0.05 \hinvMpc \ .
\eea
We choose the following masses for the neutrinos, which saturate the Planck $2\sigma$ upper bound
\be
m_{\nu_1}=0.0712909 {\rm eV}, \quad m_{\nu_2}=0.0718178 {\rm eV}, \quad m_{\nu_3}=0.0868913 {\rm eV}, \quad f_\nu=0.0178352 \ ,
\ee
which corresponds to a normal hierarchy.

For clarity we first summarize the overall result in Fig.~\ref{fig:Summary}. Here we plot the one-loop correction to the total matter power spectrum, plus the largest contribution from the counterterms.  At high wavenumbers, higher than the typical $\kfs$,  the  contribution not coming from the counterterms are quite well approximated by $-16 f_\nu P_{\rm dm,\,dm}$. At lower wavenumbers it is smaller, but just by an order one fraction (about half of that). Concerning the counterterms, we have just plotted the leading ones, which come in four curves. Three of them have the same functional form of the usual $c_s$ dark-matter counterterm, but different prefactors: respectively $c_s^2$, $c_s^2f_{\nu,\,\rm slow}$, $\Delta c_s^2$, where $\Delta c_s^2$ is a known function of $C_{\rm vev}, \; C_{\rm res, \; Iso}$ and $C_{\rm res,\; Ani}$. The fourth curve is the contribution due to the inclusion of the $c_s$ dark-matter counterterm in the neutrino diagrams: it does not have the same functional form of the ordinary $c_s$ counterterm, but it has the same prefactor, $c_s^2$. We are now going  to explain in some detail the main features of these plots.

\begin{figure}[htb!]
\centering
\includegraphics[width=12.5cm]{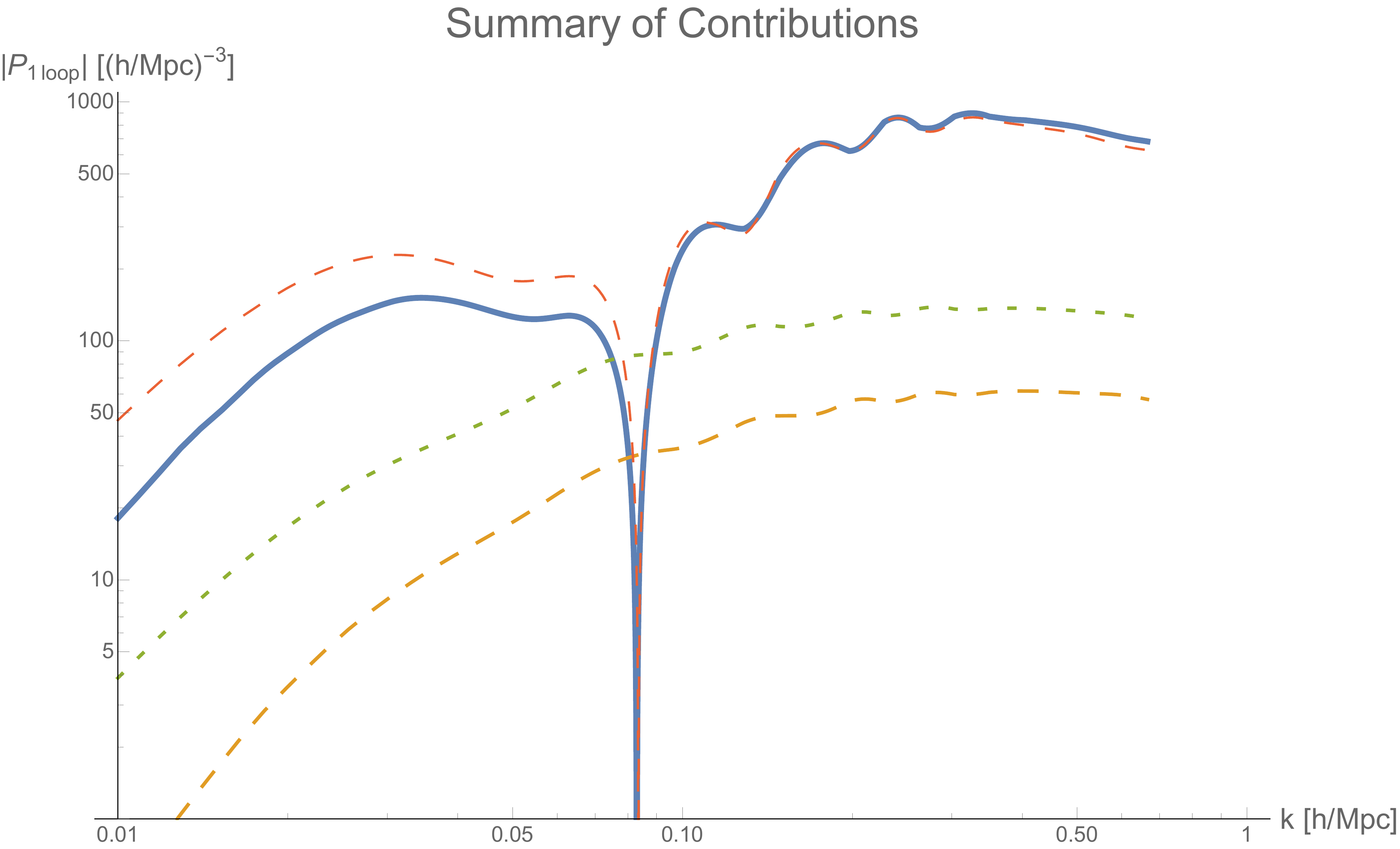}\hspace{0.01cm}
\includegraphics[width=4.cm]{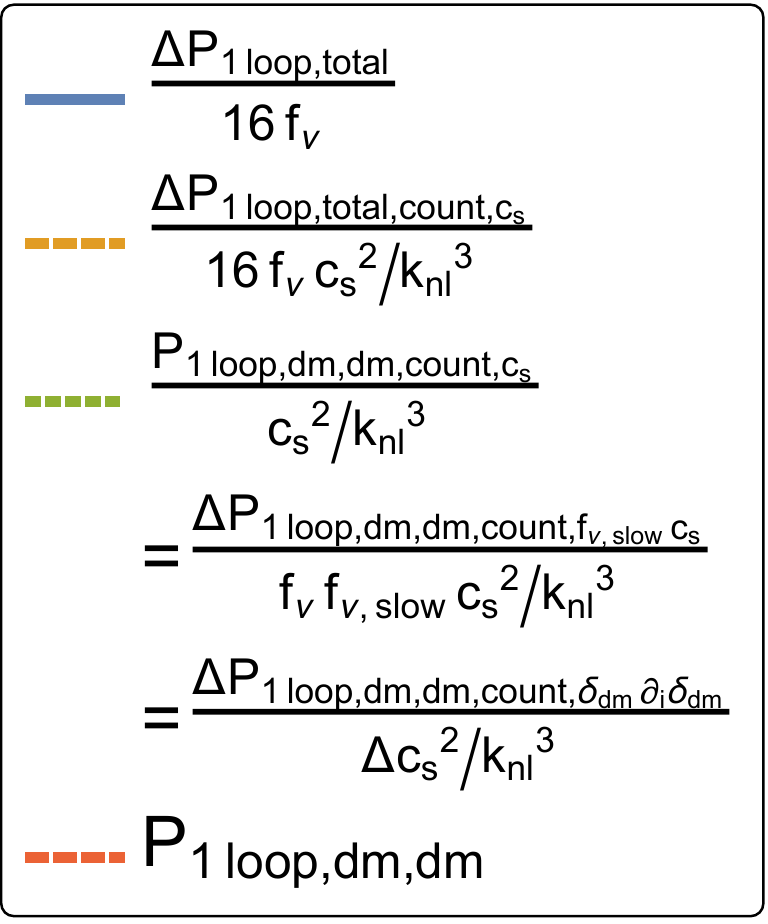}

\caption{Summary plot of the contribution to the total-matter one-loop power spectrum. We plot the contribution from diagrams without counterterms and the contribution due to the leading counterterms, which come in four curves. Three of them have the same functional form as the usual dark-matter $c_s$ counterterm, but different prefactors: respectively $c_s^2$, $c_s^2f_{\nu,\,\rm slow}$, $\Delta c_s^2$, where $\Delta c_s^2$ is a known function of $C_{\rm vev}, \; C_{\rm res, \; Iso}$ and $C_{\rm res,\; Ani}$. The fourth curve is the contribution due to the inclusion of the $c_s$ dark-matter counterterm in the neutrino diagrams: it does not have the same functional form of the ordinary $c_s$ counterterm, but it has the same prefactor, $c_s^2$.}\label{fig:Summary}
\end{figure}

We start by presenting, in Fig.~\ref{fig:linear_nu_transfer}, the ratio  of the linear neutrinos power spectrum and the dark matter one. We clearly see the well-known suppression of power at high wavenumbers due to neutrino  free streaming. At lower wavenumbers neutrinos behave as cold dark matter. For our parameters we see that the break occurs at about $k\sim 10^{-2}\hinvMpc$, though it should be noticed that the break is not extremely sharp. This implies that $\delta_{\rm diff}=\delta_\nu-\delta_{\rm dm}$ at linear level goes to zero at low wavenumbers, while it approaches $-\delta_{\rm dm}$ at higher ones. This features of the linear power spectrum will  be important for interpreting the qualitative features of the one-loop results that we present below.

\begin{figure}[htb!]
\centering
\includegraphics[width=9cm]{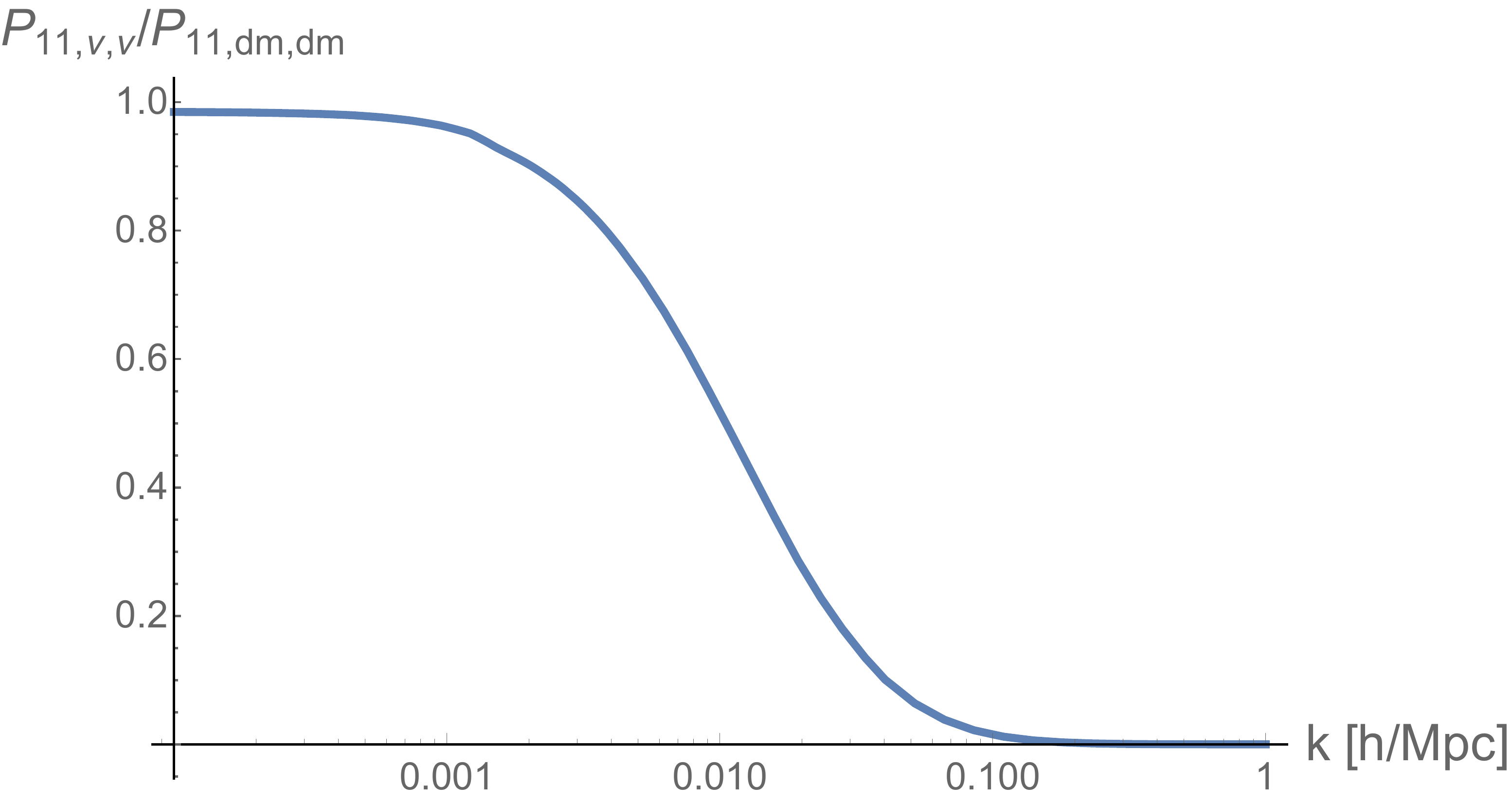}
\caption{\small Ratio of the linear power spectra for neutrinos and dark matter. We see that the neutrino linear power spectrum is suppressed at high wavenumber due to the free streaming of their fast component, while at low wavenumbers it equals the dark matter one. The breake occurs at $k\sim 10^{-2}\hinvMpc$, though it is not a sharp feature.}\label{fig:linear_nu_transfer}
\end{figure}

We now illustrate the one-loop $P_{\rm diff,\,dm}$, in Fig.~\ref{fig:Pdmdiff}. We see that at high wavenumber, $k\gtrsim 0.1 \hinvMpc$, $P_{\rm diff,\,dm}$ approaches $-f_\nu P_{\rm dm,\,dm}$. This is expected from the suppression at linear level of $\delta_\nu$ at high wavenumbers. However, at lower wavenumbers, the curve does not go to zero, as one would naively expect from the behavior of $\delta_{\rm diff}^{(1)}$, but rather it approaches, $-\frac{1}{2}f_\nu P_{\rm dm,\,dm}$. 
The reason of this is the following. At lower $k$'s, much lower than the break of the linear power spectrum associated to the free streaming length, the slope of the linear power spectrum is so steep  that some loops are dominated by wavenumbers beyond the scale of the free streaming length, where the linear power spectrum of $\delta_\nu$ is suppressed (and the one of $\delta_{\rm diff}$ is unsuppressed). This does not happen for all diagrams though, because in some diagrams the internal lines are represented by dark matter fields. It happens only for a fraction of order one of the diagrams, which contribute very approximately as much as half of the dark matter power spectrum.

\begin{figure}[htb!]
\centering
\includegraphics[width=16.5cm]{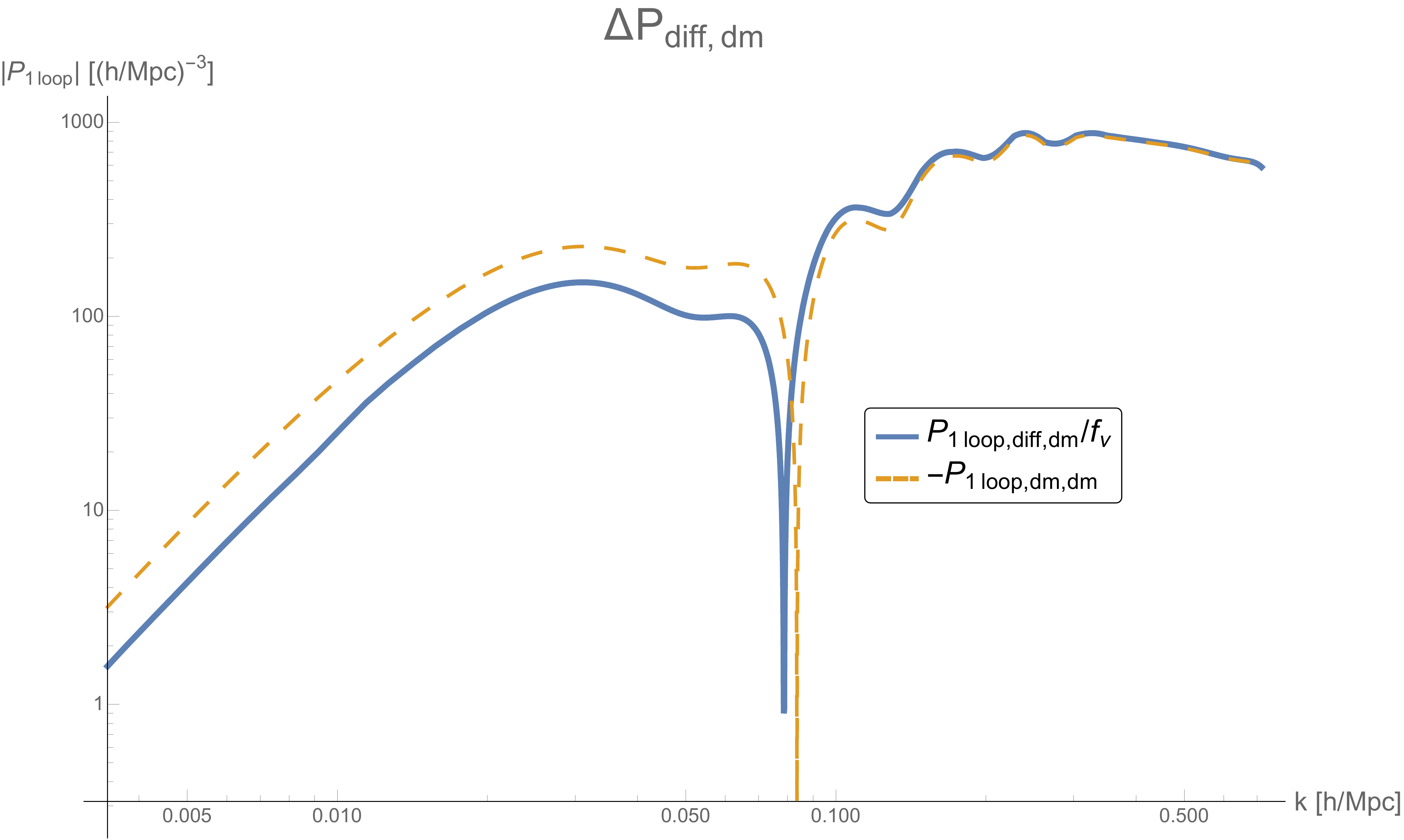}
\caption{One-loop order $P_{\rm diff,\;dm}$. For comparison, we plot the one-loop dark matter power spectrum $P_{\rm dm,\,dm}$. We see that at high wavenumbers $P_{\rm diff,\;dm}$ approaches $-f_\nu P_{\rm dm,\,dm}$. At lower wavenumbers, below the free streaming length scale, $P_{\rm diff,\;dm}$ gets smaller than $f_\nu P_{\rm dm,\,dm}$, but it is only order one suppressed with respect to this one: it does not become parametrically smaller. }\label{fig:Pdmdiff}
\end{figure}

Next, in Fig.~\ref{fig:Pdmdiff_counter}, we present the contribution of the counterterms to $P_{\rm diff,\;dm}$. There are five different curves, according to the six counterterms that are being used, $c_s^2$, $C_{\rm vev},\; C_{\rm res,\; Iso},\; C_{\rm res,\; Ani},$ and $c_s^2\ f_{\rm \nu, \,slow}$, and, for comparison purposes, we also include the usual $c_s$ counterterm for the dark matter power spectrum. We can see that the term in $c_s$ has a different shape, but a comparable size, to the dark matter term in $c_s$. Instead, the remaining counterterm diagrams have  a  significantly smaller size. This can be explained  by the fact that these terms appear only as associated to the UV limits where the field that is evaluated at high momentum is a neutrino field. The suppression of the linear fluctuations then implies the suppression for the counterterm. This is consistent with what we see in the effective Boltzmann equation in~(\ref{eq:Boltzone_eff5}), where counterterms for fast neutrinos are suppressed by $\frac{H}{\knl^3v}\sim\frac{1}{\knl^2}\frac{\kfs}{\knl}$.
Furthermore, we can see how the various contributions in $C_{\rm vev},\; C_{\rm res,\; Iso},$ and $C_{\rm res,\; Ani}$ are quite comparable with each other. Indeed, the ratio between the size of these terms and the $c_s$ term can be used as an estimate of the size of $f_{\nu,\rm \, slow}$. In fact, we did include the slow neutrinos in the calculation of these counterterms, as we discussed that whatever we add here is degenerate with the contribution from the term in $f_{\rm \nu,\;slow}$:
 the results of  Fig.~\ref{fig:Pdmdiff_counter} suggest that for the mass of neutrinos that we consider, $f_{\rm \nu, \,slow}$ is already quite small, less than $10^{-1}$~(\footnote{This fact can be used as a prior for $f_{\rm \nu,\;slow}$ when fitting the observational data.}). By using the Poisson equation, one can see that this is consistent with taking $\vnl\sim 10^{-3}$, as expected from the size of the gravitational potential at the non-linear scale.

\begin{figure}[htb!]
\centering
\includegraphics[width=12.5cm]{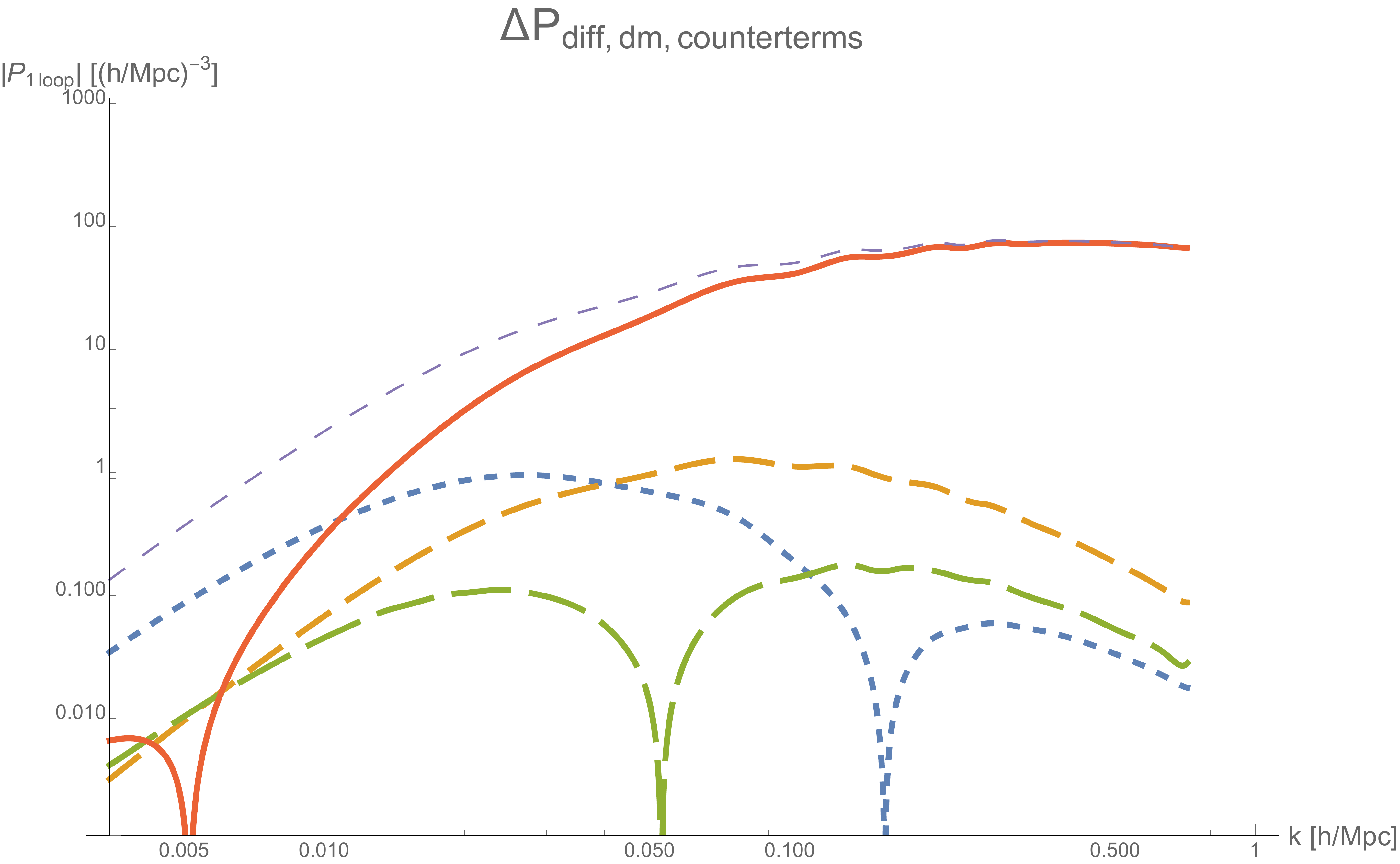}\hspace{0.01cm}
\includegraphics[width=4.cm]{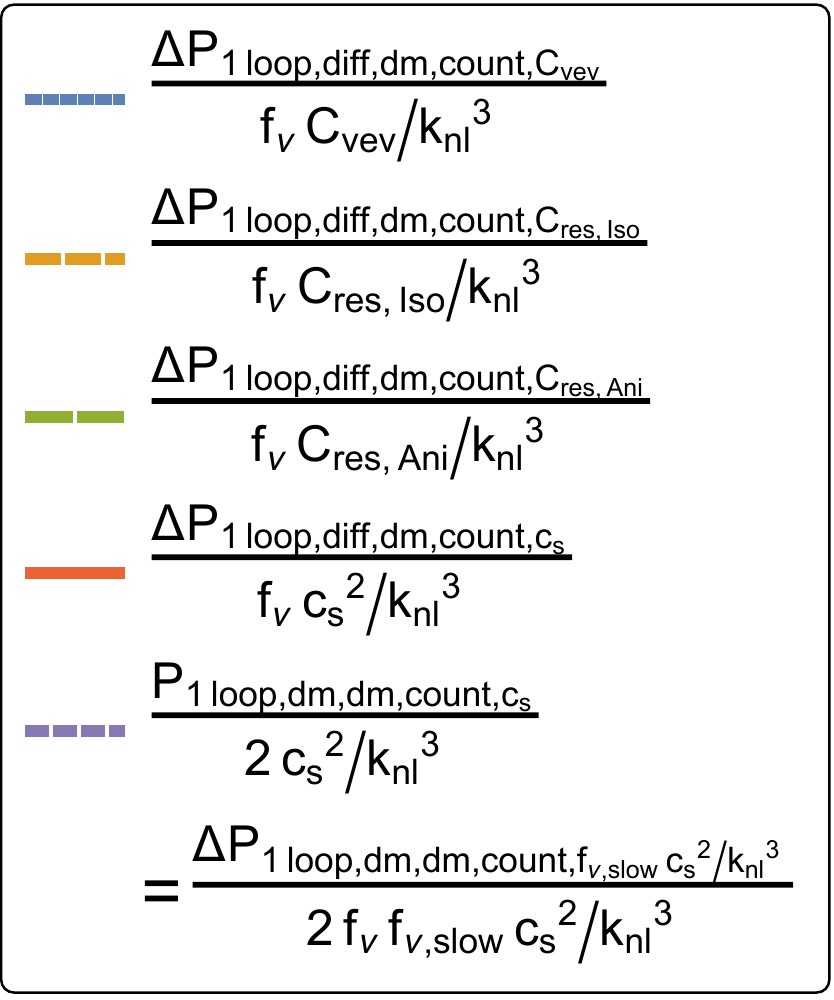}
\caption{Contribution of the counterterms diagrams for $P_{\rm diff,\;dm}$. We see that apart for the term in $c_s$ and in $f_{\rm \nu,\;slow}$, the additional counterterms are expected to give a small contribution.}\label{fig:Pdmdiff_counter}
\end{figure}

In Fig.~\ref{fig:DeltaPdmdm} we then present the contribution of neutrinos to $P_{\rm dm,\,dm}$ that we called $\Delta P_{\rm dm,\,dm}$. Again, we notice a similar structure to $P_{\rm diff,\,dm}$, as the source of the dark matter perturbation is $\delta_{\rm diff}$, as this is what appears in the Poisson equation. However, there is an important difference: the influence of the linear $\delta_{\rm diff}$ on the linear $\delta_{\rm dm}$ is logarithmically enhanced in the length of time  from matter radiation equality to the present time. As we discussed, this enhancement represents the difference in the growth factor that occurs in a cosmology with a certain $\Omega_{\rm NR}$, but where a fraction $f_\nu$ does not clusters. Strictly speaking, such a logarithmic enhancement might lead to a break of perturbation theory for the linear solution if the length of time is too long, and the effect needs to be resummed. In our case, the effect is still sufficiently small that perturbation theory  holds, however since our equations are non-relativistic they do not extend to very early times.  We therefore evaluate the logarithmically enhanced diagrams by using the linear solution for $\delta_{\rm dm }$ provided by a relativistic Boltzmann solver. Notice that the secular enhancement of the effect of a perturbation is typical whenever we expand perturbatively a term in the linear equations of motion~\footnote{Another typical example is when we solve perturbatively in the mass the propagation of a scalar field.}.

In our case, this means that at high wavenumbers, $\Delta P_{\rm dm,\,dm}$ is expected to be larger than $-f_\nu P_{\rm dm,\,dm}$ by a factor of two times $\log(a_{\rm equality}/a_0)\sim 8$. Indeed, we see that   $\Delta P_{\rm dm,\,dm}$ is about $-14 f_\nu \cdot P_{\rm dm,\,dm}$ at high wavenumbers. At lower wavenumbers the same phenomenon that we described in the case of $P_{\rm diff,\,dm}$ has the consequence that $\Delta P_{\rm dm,\,dm}$ does not vanish at lower $k$'s, but it is about a fraction of order one of $-14f_\nu P_{\rm dm,\,dm}$ (numerically, it seems to be a number close to $-8 f_\nu P_{\rm dm,\,dm}$, but the functional form is quite different).

\begin{figure}[htb!]
\centering
\includegraphics[width=16.5cm]{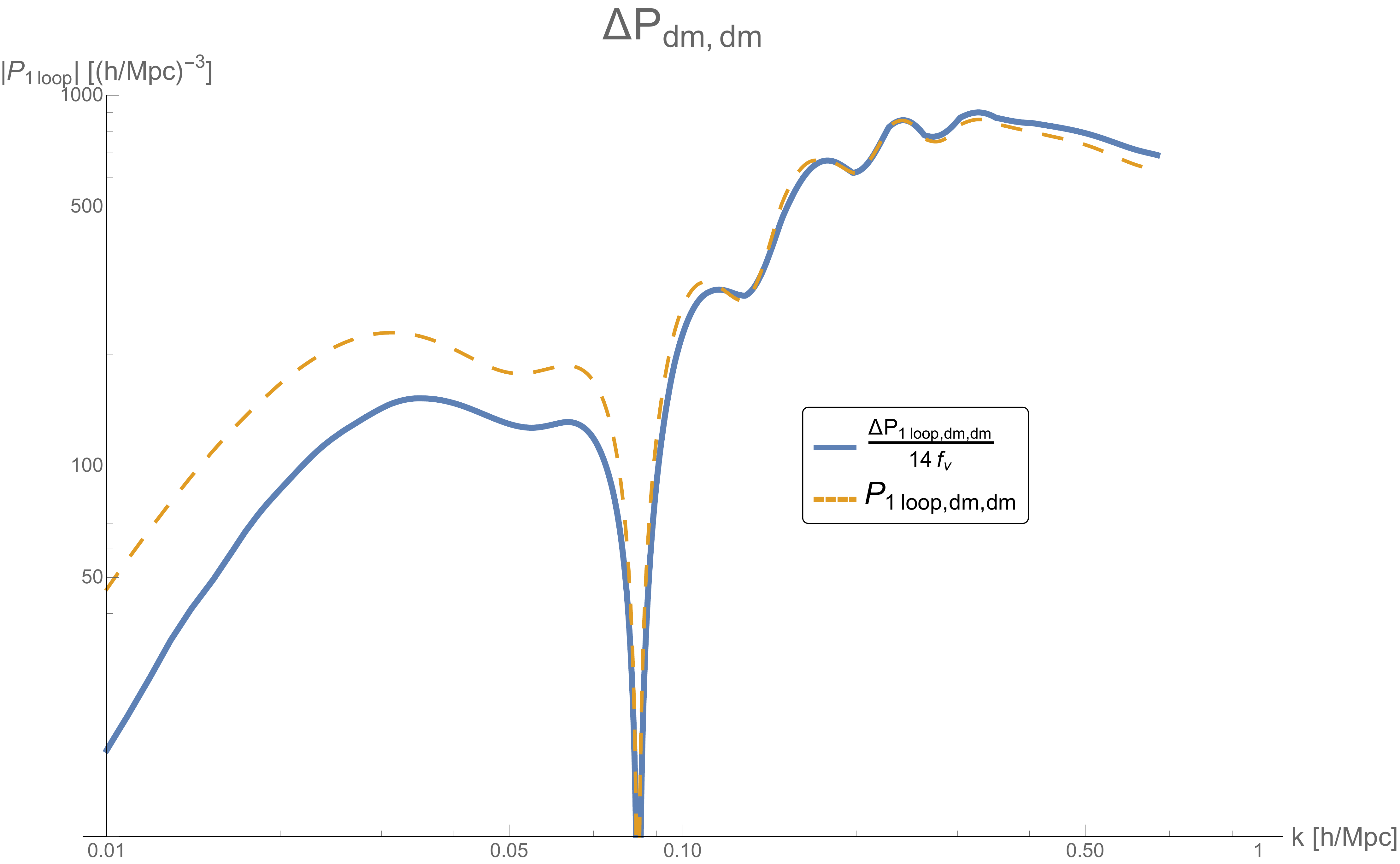}
\caption{$\Delta P_{\rm dm,\,dm}$ is plotted divided by a factor of $14f_\nu$, so that it is numerically similar to $P_{\rm dm,\,dm}$ at high wavenumbers. The factors of 14 is the result of the logarithmic enhancement of the effect of $\delta_{\rm diff}$ on $\delta_{\rm dm}$. At lower wavenumbers, only a fraction of the diagrams becomes suppressed, so that $\Delta P_{\rm dm,\,dm}$ does not become  negligible.}\label{fig:DeltaPdmdm}
\end{figure}

In Fig.~\ref{fig:DeltaPdmdm_counter}, we present the contribution of the counterterms to the calculation of  $\Delta P_{\rm dm,\,dm}$. Contrary to the case of $P_{\rm diff,\;dm}$, we see that the contribution is much larger. However, one can argue that most of the contribution comes from the dark matter contribution of (\ref{eq:dmdmcounternuresponsedm}). In fact, if we remove the contribution of~(\ref{eq:dmdmcounternuresponsedm}) we obtain the plot in Fig.~\ref{fig:DeltaPdmdm_counter_onlynu}. In this second plot, we see that the contribution from the counterterms acting on the neutrino fields is much smaller. The only term that is still quite large is the $c_s$ term, which indeed involves a counterterm acting on the dark matter field in the internal lines. This result is in agreement with that we saw for $P_{\rm diff,\;dm}$. On the other hand, as we had anticipated,  the contribution of  (\ref{eq:dmdmcounternuresponsedm}) is degenerate with the ordinary dark-matter $c_s$ term, up to higher derivative term. Notice however that the combination of dark matter UV fields that enters in these counterterms is not the same as the one that defines~$c_s$. 
Therefore, this contribution can be encoded in a correction of order $f_\nu$ to the $c_s$ coefficient of the dark matter field, $c_s^2\to c_s+f_\nu \Delta c_s^2$~(\footnote{Notice furthermore that the $c_s$ counterterm for the dark matter does not receive order $f_\nu$ corrections independently of the one we include here. It is therefore important that $\Delta c_s^2$ is expressed in terms of $C_{\rm vev}, \; C_{\rm res, \; Iso}$ and $C_{\rm res,\; Ani}$, so that now new parameter is included.}), where $\Delta c_s^2$ can be expressed in terms of $C_{\rm vev}, \; C_{\rm res, \; Iso}$ and $C_{\rm res,\; Ani}$, and it is not expected that $\Delta c_s^2=c_s^2$. From the plot we expect $\Delta c_s^2\sim {\cal O}(10) C_{\rm res,\,Iso}$, where we take $C_{\rm vev}, \; C_{\rm res, \; Iso}$ and $C_{\rm res,\; Ani}$ as comparable~\footnote{It might appear somewhat surprising the fact that the contribution of $C_{\rm res, \; Iso}$ is larger than the other ones and of the $c_s$ dark-matter counterterm, if they all take similar numerical values. This is mainly due to a somewhat unfortunate choice that we made in~(\ref{eq:approx}) when we defined the tensors multiplying $C^{(2)}_2$ and $C^{(2)}_3$, related respectively to $C_{\rm res,\,Iso}$ and $C_{\rm res,\,Ani}$, to be respectively $\delta_{ij}$ and $\d_i\d_j/\d^2$. Probably, a more fortunate choice would be to define these tensors to be $\delta_{ij}/3$ and $\d_i\d_j/\d^2-\delta_{ij}/3$, so that the tensor multiplying $C_{\rm res,\,Ani}$ would be traceless, and the one multiplying $C_{\rm res,\,Iso}$ would have unit trace. This choice amounts to redefine $C_{\rm res,\,Iso}\to 3\, C_{\rm res,\,Iso}\;,\ \ C_{\rm res,\,Ani}\to C_{\rm res,\,Ani}-3\, C_{\rm res,\,Iso}$. Upon this redefinition, the three contribution in $C_{\rm vev}, \; C_{\rm res, \; Iso}$ and $C_{\rm res,\; Ani}$ comparable, and their contribution similar is size to the one of the usual $c_s$ dark-matter counterterm for similar numerical values. We leave to use this more fortunate choice of basis to future work.}.

\begin{figure}[htb!]
\centering
\includegraphics[width=16.5cm]{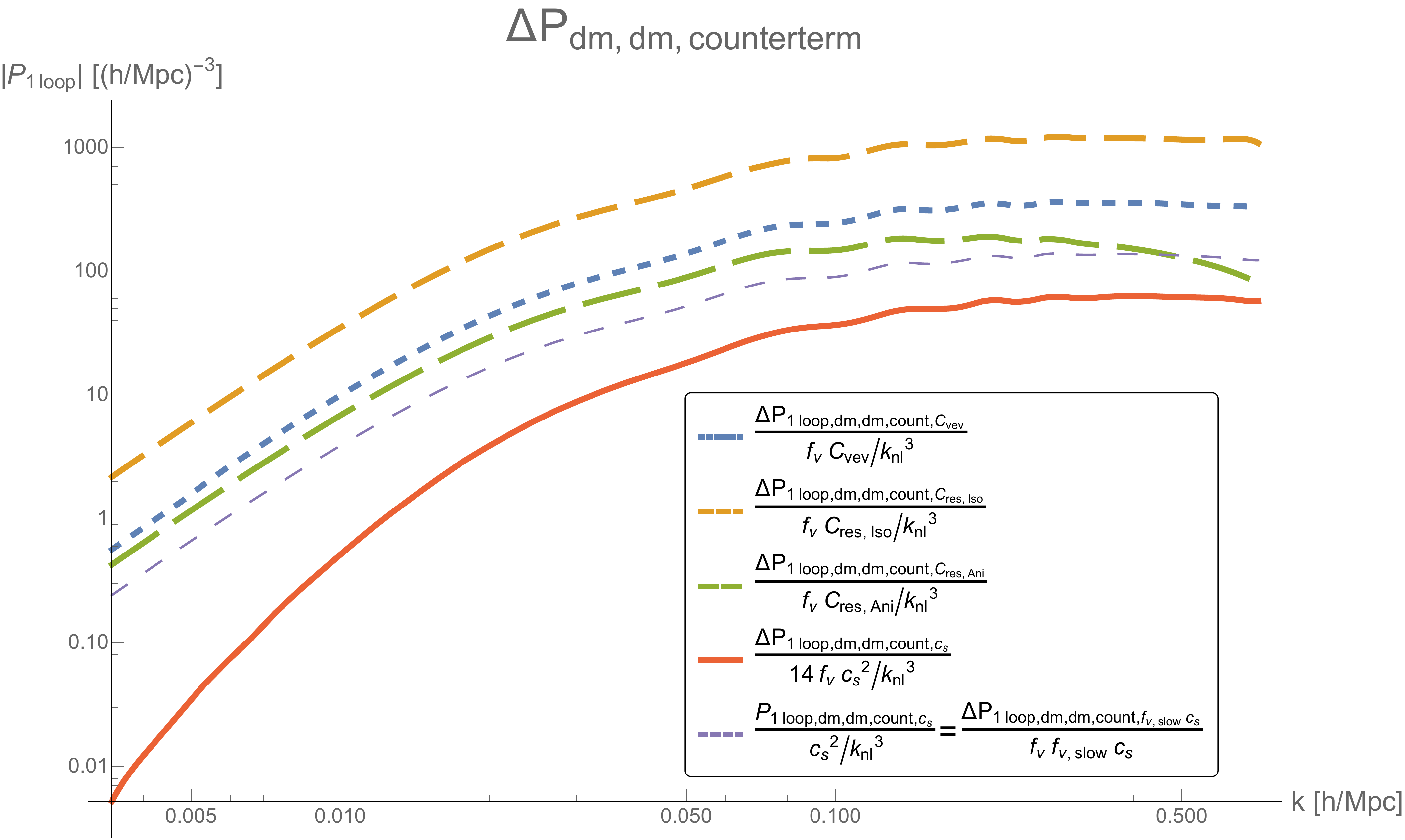}
\caption{Contribution of the counterterms to $\Delta P_{\rm dm,\,dm}$. We see that the contribution is rather larger with respect to what we saw for $P_{\rm diff,\;dm}$, but most of this contribution has the functional form of a ordinary dark-matter $c_s$ term that originates from the dark matter contribution contained inside the renormalization of $\delta_{\rm dm}\d_i \delta_{\rm dm}$.}\label{fig:DeltaPdmdm_counter}
\end{figure}

\begin{figure}[htb!]
\centering
\includegraphics[width=12.5cm]{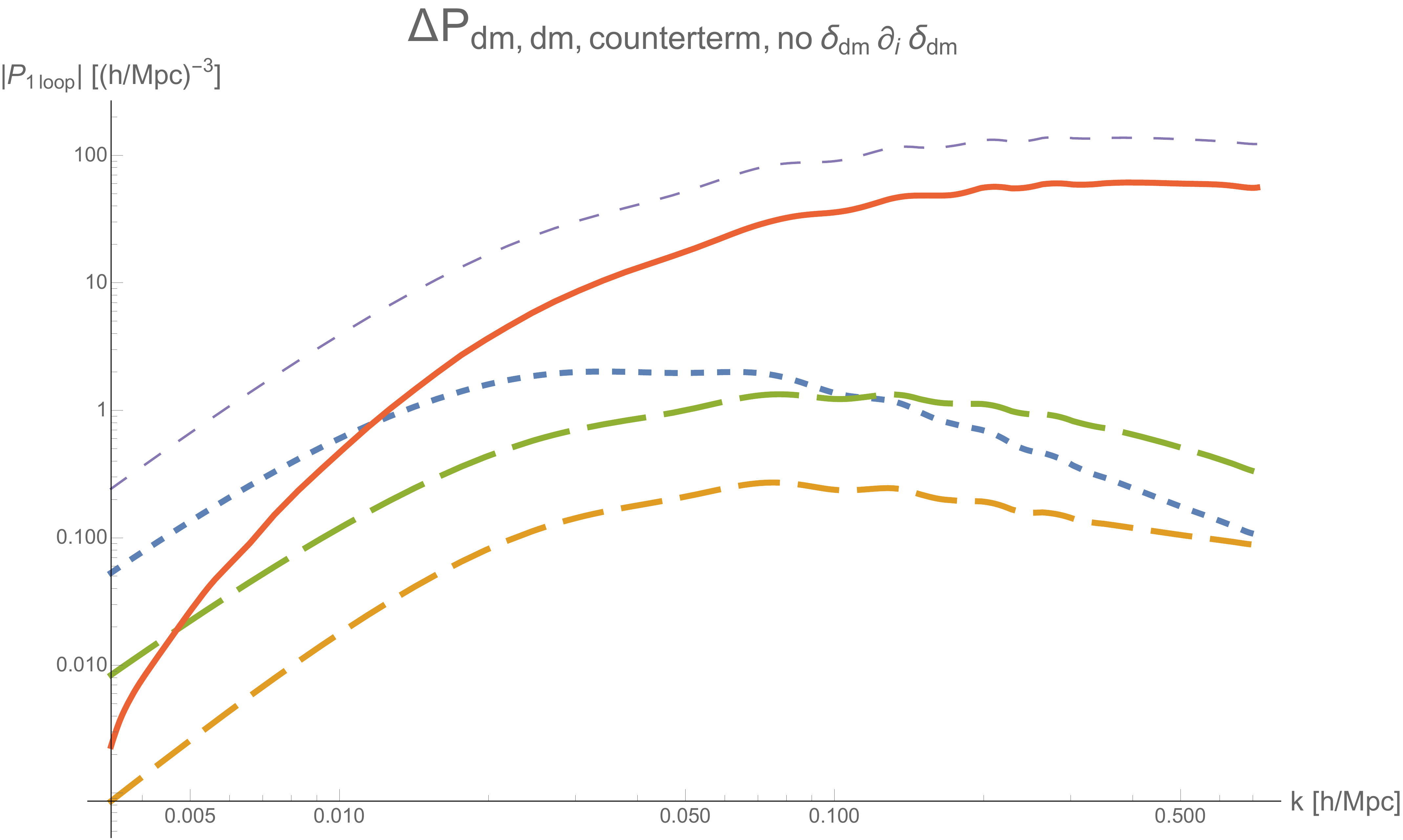}\hspace{0.01cm}
\includegraphics[width=4.cm]{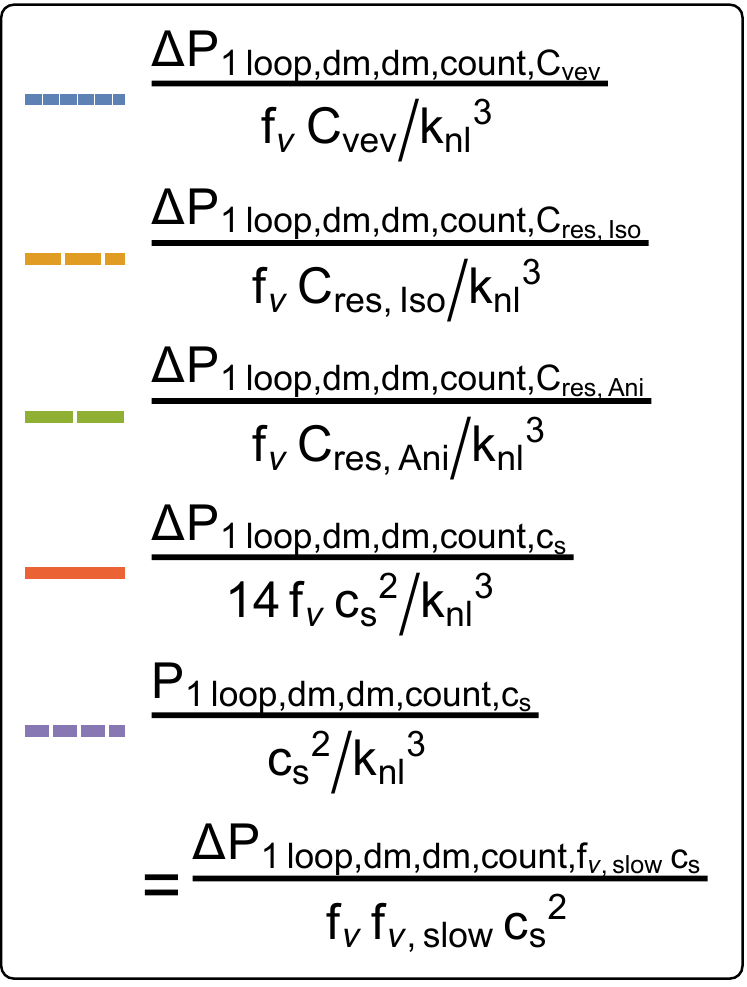}
\caption{Contribution of the counterterms $\Delta P_{\rm dm,\,dm}$ without including the contribution of the renormalization of the dark matter fields inside $\delta_{\rm dm}\d_i\delta_{\rm dm}\subset \delta_{\rm dm}\d_i\delta_{\rm diff}$, in (\ref{eq:dmdmcounternuresponsedm}). The remaining contribution, and in particular the one not known in terms of the $c_s$ parameter, is quite small. The contribution from the renormalization of the dark matter field can be taken into account by a shift $c_s^2\to c_s+f_\nu \Delta c_s^2$ of the dark matter $c_s$-counterterm, where $\Delta c_s^2$ is known in terms of $C_{\rm vev}, C_{\rm res,\, Iso}$ and $C_{\rm res,\, Ani}$.}\label{fig:DeltaPdmdm_counter_onlynu}
\end{figure}

The sum of the contributions to $P_{\rm diff,\,dm}$ and $\Delta P_{\rm dm,\,dm}$, leads us to present in Fig.~\ref{fig:Ptotal} the contribution to the total power spectrum $\Delta P_{\rm total}$. As for the case of $\Delta P_{\rm dm,\,dm}$, we plot the contribution divided by $16f_\nu$ so that the result is similar to $P_{\rm dm,\,dm}$. This enhancement is due to the logarithmic enhancement of the effect at linear level of $\delta_{\rm diff}$ on $\delta_{\rm dm}$, as we discussed earlier when describing $\Delta P_{\rm dm,\,dm}$. The gross feature resulting shape is similar to the one we described for $\Delta P_{\rm dm,\,dm}$: the shape is very similar to $-16 f_\nu P_{\rm dm,\,dm}$ at high wavenumbers, and, at lower wavenumbers, it approaches some intermediate value (very approximately $-8 f_\nu P_{\rm dm,\,dm}$). A third curve that we present in the figure is the one obtained by including in  $\Delta P_{\rm total}$ only the diagrams that are logarithmically enhanced~\footnote{This is a truncation of the calculation, not a systematic approximation that can be recovered order by order.}. We see that this approximation give results that are only about 30\% accurate. Indeed notwithstanding the enhancement of a few diagrams the  diagrams that are not enhanced are quite numerous and significant.

\begin{figure}[htb!]
\centering
\includegraphics[width=16.5cm]{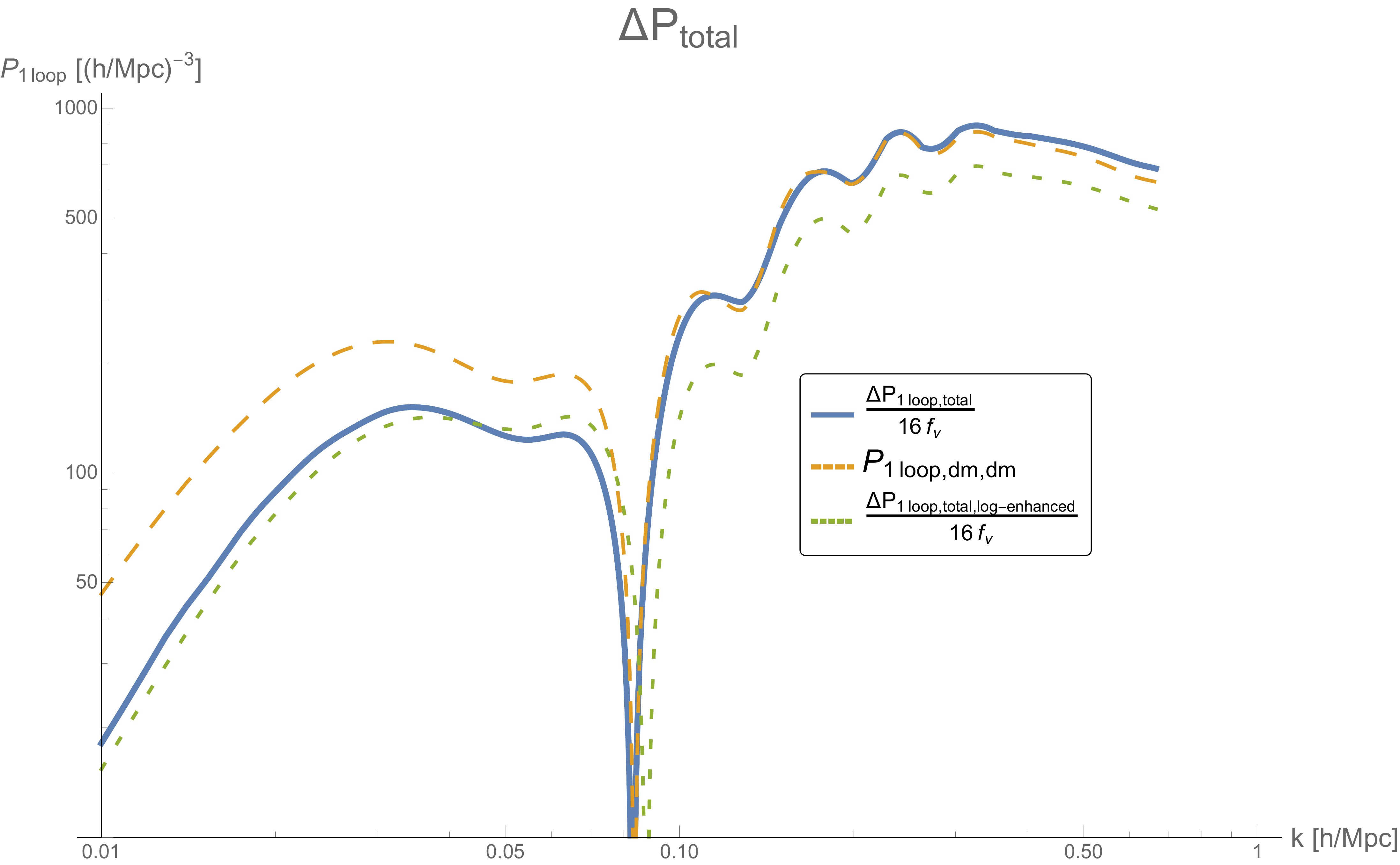}
\caption{The one-loop correction to the total power spectrum in the presence of neutrinos, $\Delta P_{\rm total}$. At high wavenumbers, the contribution is quite similar to $-16f_\nu P_{\rm dm,\,dm}$. At low wavenumbers, some diagrams are enhanced by the fact that $\delta_{\rm diff}$ does not vanish at high internal wavenumbers. We also plot the curve obtained by keeping only the log-enhanced terms, which is accurate to only about 30\%.}\label{fig:Ptotal}
\end{figure}

Finally, in Fig.~\ref{fig:Ptotal_counter}, we present the contribution of the counterterms to $\Delta P_{\rm total}$. As expected from the discussion for $\Delta P_{\rm dm,\,dm}$, the contribution of the counterterms is quite large. However, most of this comes from the contribution of the renormalization of $\delta_{\rm dm}\d_i\delta_{\rm dm}$, whose functional form is the same as the one of the $c_s$ dark-matter counterterm. Without including this contribution, represented by (\ref{eq:dmdmcounternuresponsedm}), we obtain the results of Fig.~\ref{fig:Ptotal_counter_only_nu}.  We see that in this case the contribution is rather small.  A comparable term is the one associated to the $c_s$ term. It is very simple to keep this term as well. Two of the diagrams that contribute to the $\Delta P_{\rm dm,\,dm}$, which is the leading term up to a 15\% correction, simply correspond to computing the counterterm for ordinary dark matter using, for the linear dark matter power spectrum, the one in the presence of neutrinos. There is a third diagram contributing to this term, given by~(\ref{eq:cs_non_corrspoendent}), represented on the left of Fig.~\ref{fig:perturbative_nu_diagram_counter_3}. This diagram corresponds to sourcing $\delta_{\rm diff}$ by $c_s$ dark matter term, and does not have a simple correspondent in a dark-matter only calculation. However the contribution of this third diagram is very small and can be neglected within some accuracy.   After these terms, all the remaining ones are more than an order of magnitude smaller than the one that we kept.

In summary, at the level of the counterterms, we see that the contribution of the counterterms is small if we do not include the contribution from the renormalization of $\delta_{\rm dm}\d_i\delta_{\rm dm}$ and the new counterterm whose coefficient is $c_s^2$. Therefore two approaches emerge. At leading order, one can neglect all the counterterms of $P_{\rm diff,\, dm}$ and the ones in $\Delta P_{\rm dm,\,dm}$, but just allow for a shift $c_s^2\to \;c_{s, \, f_\nu}^2= c_s^2+f_\nu \Delta c_s^2+f_\nu f_{\nu,\rm slow} c_s^2$ in the dark matter $c_s$-counterterm, where $\Delta c_s^2$ is known in terms of $C_{\rm vev}, \; C_{\rm res, \; Iso}$ and $C_{\rm res,\; Ani}$, and also include the new counterterms in $c_s^2$. Since this last one is already of order $f_\nu$,  at this order in $f_\nu$ one can use directly $c_{s, \, f_\nu}^2$ here. Therefore, this procedure amounts to adding no parameters to the dark-matter-only calculation (the only parameter being~$c_{s, \, f_\nu}^2$). If instead one wished to include the much smaller counterterms, and renormalize the full one-loop calculation, one should include all the remaining counterterms that we computed, which are shown in Fig.~\ref{fig:Ptotal_counter}. This calculation corresponds to adding three parameters to the dark-matter-only calculation ($C_{\rm vev}, \; C_{\rm res, \; Iso}$, $C_{\rm res,\; Ani}$), as $ f_{\nu,\rm slow}$ has been reabsorbed in the definition of $c_{s, \, f_\nu}^2$.

\begin{figure}[htb!]
\centering
\includegraphics[width=16.5cm]{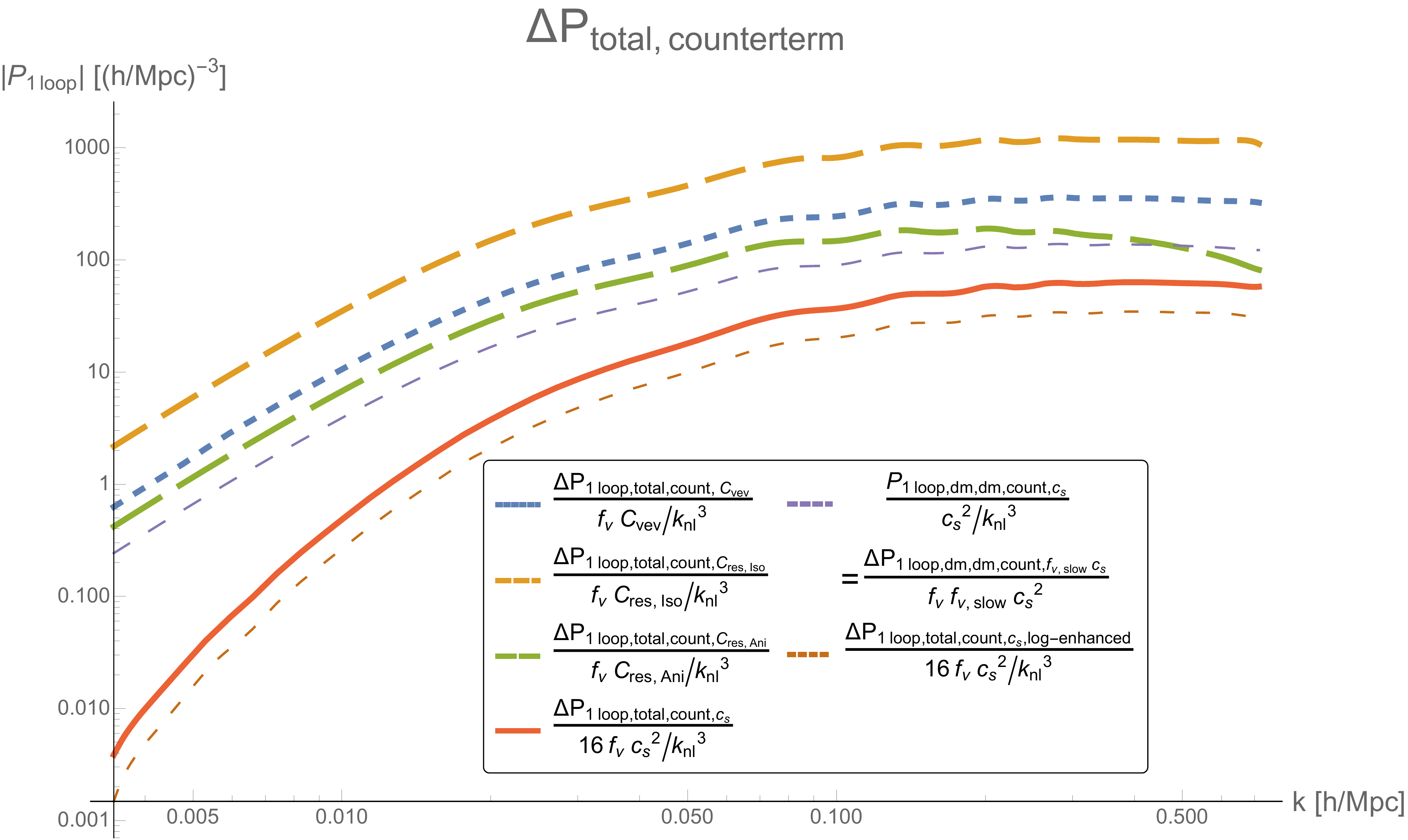}
\caption{Contribution of the counterterms to $\Delta P_{\rm total}$. The dashed orange line is obtained by keeping only the log-enhanced diagrams.}\label{fig:Ptotal_counter}
\end{figure}

\begin{figure}[htb!]
\centering
\includegraphics[width=12.5cm]{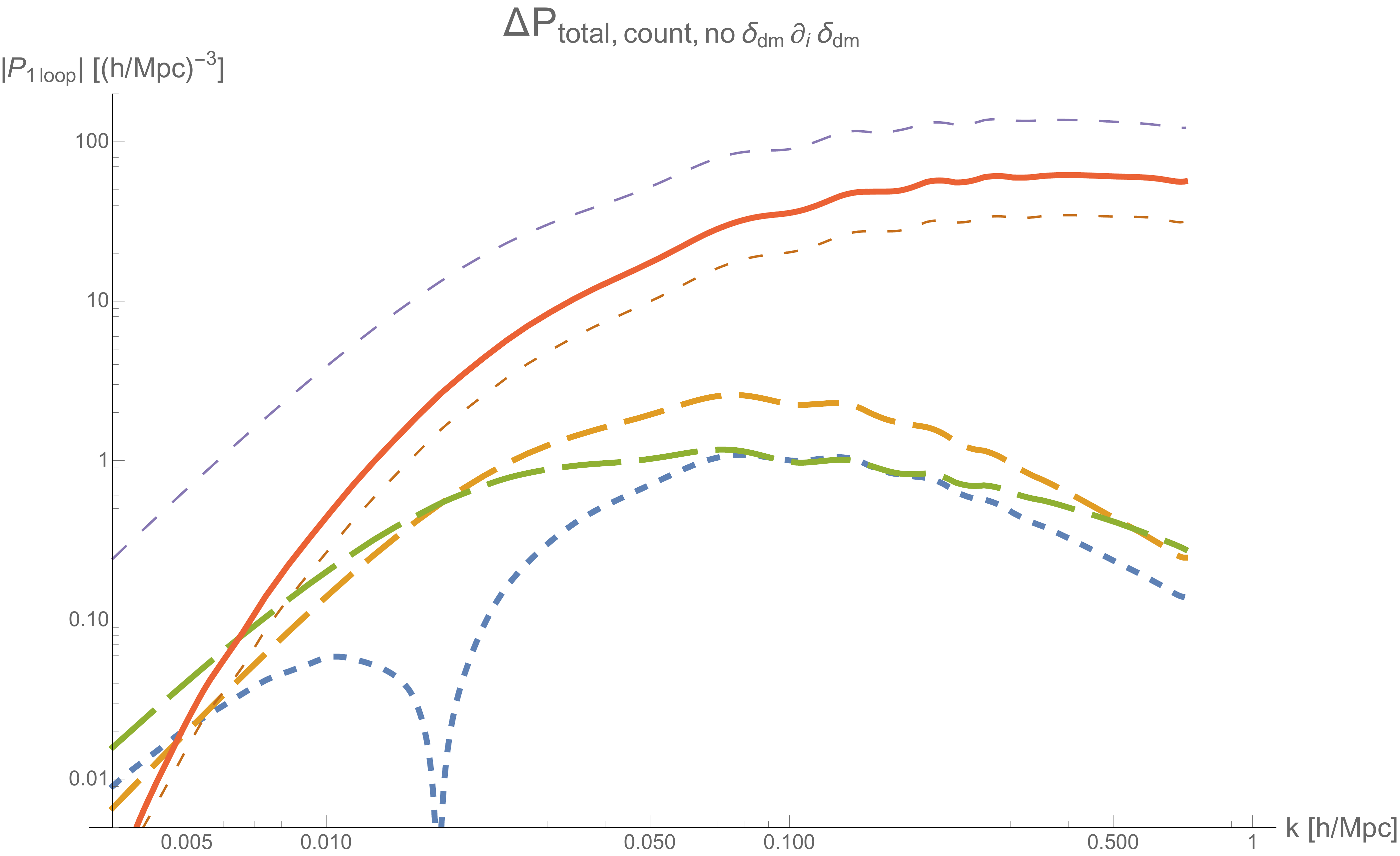}\hspace{0.01cm}
\includegraphics[width=4.cm]{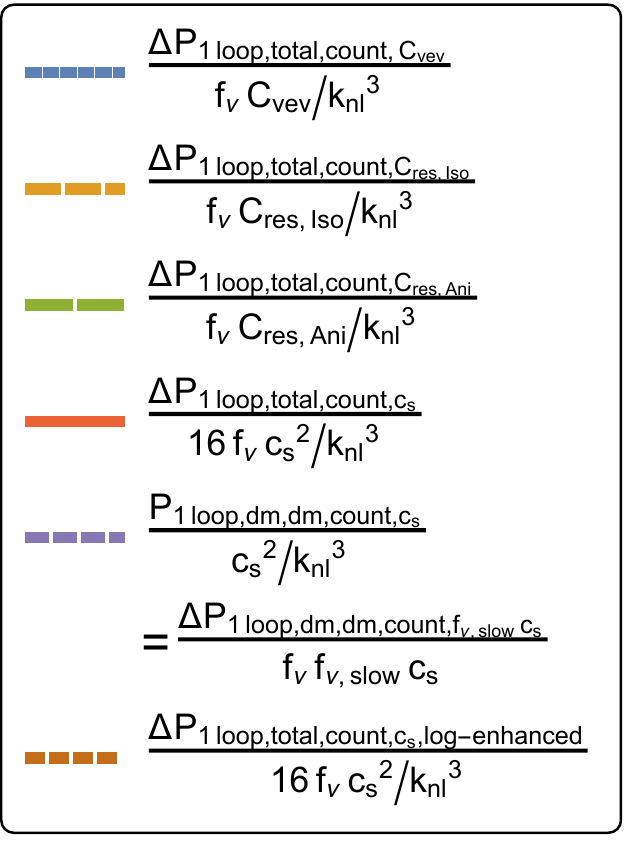}
\caption{Contribution of the counterterms to $\Delta P_{\rm total}$ excluding the contributions from the renormalization of the purely dark matter terms in (\ref{eq:dmdmcounternuresponsedm}). One sees that the remaining contribution is rather small.}\label{fig:Ptotal_counter_only_nu}
\end{figure}

\subsection*{Checks}

Given the large number of diagrams that we have computed,  the cautious reader might be skeptical about the results due to potential mistakes. Here we present a series of rather powerful checks that make us reasonably confident of the absence of sizable mistakes in our calculation.

We start with $P_{\rm diff, dm}$. In this case, if we set $\tilde f^{[0]}=1$, we should obtain the one-loop $P_{\rm dm,\,dm}$. Our calculation passes this check with percent accuracy. 

Next, we move to the more complex $\Delta P_{\rm dm,\,dm}$. The integral diagrams are very simple expressions in terms of the diagrams that lead to  $P_{\rm diff, dm}$, which have already been checked. A mistake in them is somewhat unlikely. Next, we have the two diagrams  $P_{\rm dm,\,dm,\,(3,1)}^{{\rm non-integral,\; }f^{[1]}}$ and $P_{\rm dm,\,dm,\,(2,2)}^{{\rm non-integral,\; }f^{[1]}}$, represented in Fig.~\ref{fig:perturbative_nu_dm_diagram_3}. In this case, if we substitute $\delta_{\rm diff}$ with $\delta_{\rm dm}$, we should obtain   $P_{\rm dm,\,dm}$, which we do with percent accuracy~\footnote{This is a non-trivial check because we had to modify the public code of~\cite{Lewandowski:2016yce} to account for the $k$-dependence of the growth factor.}. Next are the two diagrams of Fig.~\ref{fig:perturbative_nu_dm_diagram_5}. In this case, if we substitute $\delta^{(2)}_{\rm dm\leftarrow \nu}$ and $\Theta^{(2)}_{\rm dm\leftarrow \nu}$ with respectively $\delta^{(2)}_{\rm dm}$ and $\Theta^{(2)}_{\rm dm}$, we obtain $P_{\rm dm,\,dm}$, which we do with percent accuracy. Furthermore, it is easy to realize that if, in the same diagram, we set $\tilde f^{[0]}=1$, we {\it do not} obtain $P_{\rm dm,\,dm}$, because the input $\left.\Theta^{(2)}_{\rm dm\leftarrow\nu}\right|_{\tilde f^{[0]}=1}= \delta^{(2)}_{\rm dm}$ is not equal to $\Theta^{(2)}_{\rm dm}$, as it lacks the contribution from the quadratic terms in the continuity equation. Upon adding these quadratic terms, the diagram obtained by setting $\tilde f^{[0]}=1$ gives $P_{\rm dm,\,dm}$ with percent accuracy, as it should.

This shows that all the  ordinary diagrams not coming from counterterms have passed some rather non-trivial checks. Finally, we notice that most of the counterterms diagrams are obtained as simple manipulations of the analogous ordinary diagrams, again, making the occurrence of mistakes in this case quite unlikely as well. Still, we make our Mathematica code publicly available at the `EFTofLSS repository' website, and we hope the interested readers will point to us  any mistake they find.

\section{Conclusions}

We have developed a formalism that allows us to predict the clustering of matter in the weakly non-linear regime  in the presence of massive neutrinos. Since the free streaming length of dark matter is about the non-linear scale, it can be safely described as an effective fluid-like system for $k\lesssim \knl$. This is not the case for neutrinos.  These can be split into two families. Slow neutrinos, whose free streaming length, $1/\kfs$, is shorter than the non-linear scale, can be described as an effective fluid-like system similar to dark matter. Instead, fast neutrinos,  which have a free streaming length much longer than the non-liner scale, need to be described through their  Boltzmann equation in order to have a description that does not break at $\kfs\ll \knl$. We then solve perturbatively the coupled  equations at leading order in $f_\nu$ and perturbatively in $k/\knl\ll1$. We point out that the size of the parameter that controls the perturbative expansion is different for `fast' and `slow' neutrinos: the gravitational interactions are small for fast neutrinos at all wavenumbers, while they are small for slow neutrinos only for $k\ll\knl$.

When solving the coupled set of equations products of dark matter fields that are not renormalized by the usual dark matter counterterms appeared. For fast neutrinos, we renormalize these terms in two ways. First we directly added suitable counterterms to both the dark matter effective equations and also the perturbative solution of the neutrino Boltzmann equation. Second we developed an effective Boltzmann equation that contains  new terms that are able to correctly parametrize the effect of short distance physics on long distance observables in an expansion in $k/\knl\ll1$ and  $\kfs/\knl\ll 1$. In both procedures we also had to include an additional speed-of-sound-like counterterm to account for the difference in the nature of the perturbative expansion between slow and fast neutrinos.   In this way perturbative results can correctly describe the effect at long distances induced by short-distance uncontrolled non-linearities.

We have used the first procedure, that we find somewhat more efficient, to compute the correction to the total matter power spectrum $\Delta P_{\rm total}$ at leading order in $f_\nu$. It consists of two parts. One is the cross correlation $P_{\rm diff, \, dm}$ between $\delta_{\rm diff}=\delta_{\nu}-\delta_{\rm dm}$ and $\delta_{\rm dm}$, and the other, $\Delta P_{\rm dm,\,dm}$, is the correction to the dark matter power spectrum that is due to the different evolution of dark matter in the presence of neutrinos. The contribution in $\Delta P_{\rm dm,\,dm,}$ is numerically larger because it encodes the correction to the growth of dark matter perturbations in a cosmology with massive neutrinos: this is enhanced by the logarithm of the redshift of matter-radiation equality. Overall, the correction due to neutrinos  is about $16 f_\nu$ times the dark matter one-loop power spectrum at wavenumbers higher than the typical neutrino free streaming length. It is about half of that at lower wavenumbers.

In comparison to the pure dark-matter calculation there are several new counterterms that are needed. However, the leading ones have either the same functional form of the ordinary $c_s^2$ dark-matter counterterm (and can be accounted for by rescaling the prefactor), or have a different functional form but the prefactor is the same $c_s^2$ of dark matter (and, at the order in $f_\nu$ at which we work, one can safely use the rescaled $c_s^2$ here). The remaining counterterms are expected to be much smaller, negligible to a first approximation. 

Our formalism opens up a series of interesting ways to proceed. Of course, it will be interesting to compute higher order correlation functions and higher $N$-point functions. In addition it would be interesting to compare our predictions with numerical simulations and with analytical models for the clustering of neutrinos. We leave this to future work.

\subsubsection*{Acknowledgments}

We thank Ethan Nadler for help with CAMB. LS is partially supported by DOE Early Career Award DE-FG02-12ER41854. MZ is supported by NSF grants AST-1409709 and PHY-1521097 and by the Canadian Institute for Advanced Research (CIFAR) program on Gravity and the Extreme Universe. 

\appendix
\section*{Appendix}
\section{The Effective Boltzmann Equation\label{app:EffectiveBoltzmann}}

We are here going to see how the counterterms for products of fields that we have defined in~(\ref{eq:product_renorm}) allow us to define an Effective Boltzmann equation whose counterterms are able to correct for the UV-dependence of the diagrams that we obtain in perturbation theory. 

In order to simplify the notation, we write  $f^{[3]}(\vec k, \vec v, s)_{\rm  UV}$ as the distribution function that results from the Green's function acting on a source, $ f^{[3]}_{\cal S}(\vec k, \vec v, s)_{\rm UV}$:
\bea\label{eq:f3inapp}
&& f^{[3]}(\vec k, \vec v, s)_{\rm  UV} = \int_0^s ds_1\, a(s_1)^2  \;  e^{- i\, \vec k \cdot\vec v\, (s-s_1)}\;   f^{[3]}_{\cal S}(\vec k, \vec v, s_1)_{\rm  UV}\ ,
\eea
and focus directly on $ f^{[3]}_{\cal S}(\vec k, \vec v, s)_{\rm  UV}$. Before we proceed, it is first suggestive to write the general source $f^{[3]}_{\cal S}(\vec k, \vec v, s)_{\rm UV}$ by performing a few Fourier transforms to obtain:
\bea\label{eq:source_real}
&&f^{[3]}_{\cal S}(\vec k, \vec v, s)_{\rm UV}=\int^s ds_1\; a(s_1)^2\, \int^{s_1} d s_2\; a(s_2)^2\, \int d^3 x_1\; e^{- i\,\vec k \cdot\vec x_1} \\ \nn
&&\qquad\left.\frac{\d  \Phi(\vec x_1,s)}{\d x^{i_1}}\right|_{\vec x_1} \frac{\d}{\d v^{i_1}}\left\{\left.\frac{\d  \Phi(\vec x_2,s_1)}{\d x^{i_2}}\right|_{\vec x_2=\vec x_1-\vec v  (s-s_1)} \frac{\d}{\d v^{i_2}}\left[\left.\frac{\d  \Phi(\vec x_3,s_2)}{\d x^{i_3}}\right|_{\vec x_3=\vec x_1-\vec v  (s-s_2)} \frac{\d f_0(v)}{\d v^{i_3}}\right]\right\}\ .
\eea
Eq.~(\ref{eq:source_real}) is instructive because it explicitly shows that the perturbative solution involves products of fields evaluated at the same location, which need to be renormalized according to the expressions we provided in~(\ref{eq:product_renorm}).

In order to check how the renormalization of products of fields that we described in~(\ref{eq:product_renorm}) can correct for any UV dependence that we obtain in perturbation theory~\footnote{We stress that this is just a check, as renormalization is a well established concept whose failure would be a major revolution in Physics.}, in this appendix we will  analyze the various uncontrolled UV contributions that occur in perturbation theory and check that these have exactly the same functional form that is obtained by using one of the counterterms that we described in~(\ref{eq:product_renorm}). In turn, this will allow us to introduce additional terms in the Boltzmann equation contributing to the solution in the same functional form as the terms from~(\ref{eq:product_renorm}).

There are several contributions in perturbation theory. Let us start first by focussing on $f^{[3]}$, while we will discuss later the contributions that come from considering the non-linear gravitational field in $f^{[1]}$ and $f^{[2]}$. Let us therefore consider the contribution to $f^{[3]}$ from the phase space where at least one of the wavenumbers is beyond the non-linear scale. There are four possibilities for the wavenumbers to be above the non-linear scale: $(a)$: $\{q_1,q_2\}\gtrsim \knl$; $(b)$: $\{q_2,q_3\}\gtrsim \knl$; $(c)$: $\{q_1,q_3\}\gtrsim \knl$; and $(d)$: $\{q_1,q_2,q_3\}\gtrsim \knl$. Let us consider them one at a time and let us start with case $(a)$: $\{q_1,q_2\}\gtrsim \knl$.

We can extract the UV-dependence of the perturbative diagrams by taking the expectation value over the short modes. From (\ref{eq:third_order_sol}), for case $(a)$,  we obtain
\bea\label{eq:source_a_initial}
&& f^{[3]}_{\cal S}(\vec k, \vec v, s)_{\rm (a), UV} =\int_{q_1 \gtrsim \knl} \frac{d^3 q_1}{(2\pi)^3} \;   i\, q_1^{i_1}\frac{\d}{\d v^{i_1}}\left[ \int_0^{s_1} ds_2\, a(s_2)^2  \right.   \\ \nn
&&\quad  \;  e^{- i\, (-\vec q_1+\vec k)\cdot\vec v\, (s_1-s_2)}   i\,  (-q_1^{i_2})\,\left(\frac{3}{2}\right)^2 a(s_1)^2 a(s_2)^2 H(s_1)^2  \Omega_{\rm dm}(s_1) H(s_2)^2  \Omega_{\rm dm}(s_2)\frac{P_{\delta\delta}(q_1,s_1,s_2)_0}{q_1^4}  \\ \nn
&&\quad \left. \times\ \frac{\d}{\d v^{i_2}} f^{[1]}(\vec k,\vec v,s_2)\right]\ , 
\eea
where $P_{\delta\delta}(q_1,s_1,s_2)_0$ is the linear expression for $\langle\delta(\vec q_1,s_1)\delta(\vec q_2,s_2)\rangle=(2\pi)^3\delta_D^{(3)}(\vec q_1+\vec q_2)\; P_{\delta\delta}(q_1,s_1,s_2)$, and we used the Poisson equation
at leading order in $f_\nu$.

\subsubsection*{`Fast' and `Slow' neutrinos}

The structure of the perturbative solution will be radically different if either some of the Green's functions oscillate as we perform the time integrations, or they do not.  As we did in the main text, also here it is therefore useful to split the neutrinos in `fast' and `slow'. We remind that `fast' neutrinos are those whose free-streaming length, $1/\kfs\sim v/H$, is longer than the non-linear scale, and `slow' neutrinos are the remaining ones. 
For fast neutrinos, as long as a mode is shorter than the non-linear scale, the associated Green's function oscillates in an Hubble time: $k v /(a H)\gtrsim 1$.

\subsection{Fast Neutrinos}

Let us start by considering fast neutrinos, $v\gtrsim a \vnl$. In this case, for modes shorter than the non-linear scale, the time-integral receives most of its contribution from times close to the final time, as otherwise the Green's function rapidly oscillates. This means that we can Taylor expand the remaining part of the integrand, which has time scales of order Hubble, around the final time, and bring it out of the integral.  In formulas, defining $ s_2=s_1-\Delta s_2$, we can write
\bea\label{eq:boltz_a_1}\nn
&& f^{[3,111]}_{\cal S}(\vec k, \vec v, s_1)_{\rm (a),\, UV, \, fast_\nu} =\\ \nn
&&\qquad= \left(\frac{3}{2}\right)^2 \int_{q_1 \gtrsim \knl} \frac{d^3 q_1}{(2\pi)^3}      \frac{q_1^{i_1} q_1^{i_2}}{q_1^4}\, a(s_1)^2   \frac{\d}{\d v^{i_1}} \int_{-\infty}^{0} d\Delta s_2\, a(s_1-\Delta s_2)^4 e^{- i\, (-\vec q_1+\vec k)\cdot\vec v\, \Delta s_2} \\ \nn
&&  \qquad  H(s_1)^2  \Omega_{\rm dm}(s_1) H(s_1-\Delta s_2)^2  \Omega_{\rm dm}(s_1-\Delta s_2)  P_{\delta\delta}(q_1,s_1,s_1-\Delta s_2)_0  \  \frac{\d}{\d v^{i_2}} f^{[1]}(\vec k,\vec v, s_1-\Delta s_2)\\ \nn
&&\qquad\simeq \left(\frac{3}{2}\right)^2 a(s_1)^6  \Omega_{\rm dm}(s_1)^2H(s_1)^4   \frac{\d}{\d v^{i_1}}  \int_{q_1 \gtrsim \knl} \frac{d^3 q_1}{(2\pi)^3} \frac{q_1^{i_1} q_1^{i_2}}{q_1^4}      \\ 
&& \qquad    \int_{-\infty}^{0} d\Delta s_2   \;\left[   e^{- i\, (-\vec q_1+\vec k)\cdot\vec v\, \Delta s_2} +{\cal O}\left(\frac{\d_s}{\knl v}\right) \right]  \; P_{\delta\delta}(q_1,s_1,s_1)_0 \, \frac{\d}{\d v^{i_2}} f^{[1]}( \vec k,\vec v,s_1) \ ,
\eea
where in the second step we have Taylor expanded around $s_1$ the functions that depend on time on a Hubble scale, and we have taken the leading term. The subleading terms are included in the terms that scale as $\frac{\d_s}{\knl v}$, where the scale of the denominator will be explained shortly.

We can now perform the integral on the angle of $\vec q_1$, $\hat q_1$~\footnote{It is useful to use the usual trick of writing $\int d^2\hat q\; \hat q^i \hat q^j e^{ i\, q\, \hat q \cdot \vec v }= \alpha\, \delta^{ij}+ \beta\, \hat v^i \hat v^j $, and solving for $\alpha$ and $\beta$ by contracting with $\delta^{ij}$ or $\hat v^i \hat v^j$. One can alternatively write $\int d^2\hat q\; \hat q^i \hat q^j e^{ i\, q\, \hat q \cdot \vec v }=-\frac{1}{q^2}\frac{\d^2}{\d v^i \d v^j} \int d^2\hat q\; e^{ i\, q\, \hat q \cdot \vec v }$.}, obtaining
\bea\nn\label{eq:boltz_a_2}
&& f^{[3,111]}_{\cal S}(\vec k, \vec v, s)_{\rm (a),\, UV, \, fast_\nu} \simeq\\ \nn
&&\qquad \left(\frac{3}{2}\right)^2 \Omega_{\rm dm}(s_1)^2 a(s_1)^6 H(s_1)^4\,   \frac{\d}{\d v^{i_1}}   \int_{-\infty}^{0} d\Delta s_2      \int_{q_1 \gtrsim \knl} \frac{d^3 q_1}{(2\pi)^3}  \frac{P_{\delta\delta}(q_1,s,s)_0}{4\pi q_1^2}\; e^{- i\, \vec k\cdot\vec v\, \Delta s_2}  \\ \nn
&&\qquad  \left[ \left(\left(-\frac{2 q_1 v \Delta s_2  \cos(q_1 v \Delta s_2 ) - 2 \sin(q_1 v \Delta s_2 )}{(q_1 v \Delta s_2 )^3}\right) \delta^{i_1 i_2} +\right.\right.\\ \nn
&&\qquad\left.\left.\left(\frac{6 q_1 v  \Delta s_2 \cos(q_1 v \Delta s_2 ) - 6 \sin(q_1  v \Delta s_2) +2 
   (q_1 v  \Delta s_2)^2 \sin(q_1  v\Delta s_2)}{(q_1  v\Delta s_2)^3}\right) \hat v^{i_1} \hat v^{i_2}\right)  \right.\\ 
   &&\qquad
   \left.
   +{\cal O}\left(\frac{\d_s}{\knl v}\right) \right]  \frac{\d}{\d v^{i_2}} f^{[1]}( \vec k,\vec v,s)   \ .
\eea
The time integral is peaked for $\Delta s_2\sim \frac{1}{q v}\lesssim \frac{1}{\knl v}$ due to the rapid oscillations. This explains the suppression factor of the time derivatives $\frac{\d_s}{\knl v}$ that we anticipated earlier, and justifies our Taylor expansion of the integrand in~(\ref{eq:boltz_a_1}). Furthermore, this realization allows us to Taylor expand the phase $e^{- i\, \vec k\cdot\vec v\, \Delta s_2}\simeq 1-i \vec k\cdot\vec v\, \Delta s_2+\ldots $, each term contributing as hierarchical powers in $k/\knl\ll 1$. We are therefore led to~\footnote{We have used that
\bea\label{eq:boltz_a_2time}
&&  \int_{-\infty}^{0} d\Delta s_2   \left(-\frac{2 q_1 v \Delta s_2  \cos(q_1 v \Delta s_2 ) - 2 \sin(q_1 v \Delta s_2 )}{(q_1 v \Delta s_2 )^3}\right)\\ && \nn\qquad\qquad\qquad\qquad\qquad\qquad\qquad\ =\frac{1}{q_1v}\int_{-\infty}^{0} d x   \left(-\frac{2x  \cos(x ) - 2 \sin (x) )}{x^3}\right) =\frac{\pi}{2 q_1 v}\ ,\\ \nn
&&  \int_{-\infty}^{0} d\Delta s_2  \frac{6 q_1 v  \Delta s_2 \cos(q_1 v \Delta s_2 ) - 6 \sin(q_1  v \Delta s_2) +2 
   (q_1 v  \Delta s_2)^2 \sin(q_1  v\Delta s_2)}{(q_1  v\Delta s_2)^3} =-\frac{\pi}{2 q_1 v}\ .
\eea}
\bea\nn\label{eq:boltz_a_3}
&& f^{[3,111]}_{\cal S}(\vec k, \vec v, s)_{\rm (a),\, UV, \, fast_\nu} \simeq  \left(\frac{3}{2}\right)^2 \Omega_{\rm dm}(s_1)^2 a(s_1)^6 H(s_1)^4\,          \left[\int_{q_1 \gtrsim \knl} \frac{d^3 q_1}{(2\pi)^3}  \frac{P_{\delta\delta}(q_1,s,s)_0}{4\pi q_1^3}\right]    \\ 
&&  \qquad\qquad\qquad  \frac{\d}{\d v^{i_1}}\left[ \frac{\pi}{2 v} \left( \delta^{i_1 i_2} -  \hat v^{i_1} \hat v^{i_2}\right)  \frac{\d}{\d v^{i_2}} f^{[1]}( \vec k,\vec v,s) +{\cal O}\left(\frac{\d_s}{\knl v},\; \frac{k}{\knl}\right) \right]    \ .
\eea

Expression (\ref{eq:boltz_a_3}) represents an actual contribution to the answer we obtain in perturbation theory. However, as we said, this part of the calculation involves the gravitational potential evaluated at wavenumbers higher than $\knl$, which is clearly not under perturbative control. For this reason, this contribution is incorrect and we need to add the counterterms in~(\ref{eq:approx}). Indeed, as expected on theoretical grounds, the contribution from the UV terms of $(2\pi)^3\delta^{(3)}_D(\vec q+\vec  q')P_{\delta\delta}(q,s,s)_0=\langle[\delta(\vec q,s)]_\R[\delta(\vec q',s)]_\R\rangle$ is degenerate with the one that we obtain from using the $C^{(2)}_1$ counterterm: $(2\pi)^3\delta^{(3)}_D(\vec q+\vec q') \int dt' C^{(2)}_1(s,s,t')\subset \langle[\delta(\vec q,s)\delta(\vec q',s)]_\C\rangle$. These two terms have the same dependence in time and $\vec q$. Therefore the counterterm in $C_1^{(2)}$ is capable of correcting the UV dependence in perturbation theory that comes from $P_{\delta\delta}(q,s,s)_0$. 

An equivalent formulation to explicitly inserting the counterterm in $C_1^{(2)}$ in the correlation function is provided by realizing that its contribution to the solution is equivalent to adding to the Boltzmann equation in~(\ref{eq:Boltzone}) a suitable term. The result in (\ref{eq:boltz_a_3}) corresponds exactly to the contribution that we would obtain in perturbation theory if we modified the Boltzmann equation~(\ref{eq:Boltzone}) by adding a suitable counterterm:
\bea\label{eq:Boltzone_eff1}
&&\frac{\d f(\vec x,\vec v,\tau)}{\d\tau}+ \frac{v^i}{a} \frac{\d f(\vec x,\vec v,\tau)}{\d x^i}-   a \frac{\d\Phi}{\d x^i}\, \frac{\d f(\vec x,\vec v,\tau)}{\d v^i}+\\ \nn
&&\qquad\alpha_0\cdot a^6 \Omega_{\rm dm}(a)^2 \frac{H^4}{\knl^3}   \frac{\d}{\d v^{i_1}} \left[ \frac{\Pi_2(\hat v)^{i_1 i_2}}{v}  \frac{\d}{\d v^{i_2}} +{\cal O}\left(\frac{\d_\tau}{\knl v},\; \frac{\d_x}{\knl}\right) \right] f( \vec x,\vec v,\tau)=0\ ,
\eea
where we defined the projection tensor
\be
\Pi_2(\hat v)^{i_1 i_2}=\delta^{i_1 i_2}-\hat v^{i_1}\hat v^{i_2}\ .
\ee
In~(\ref{eq:Boltzone_eff1}), $\alpha$, related to $\int^\tau d\tau' C^{(2)}_1(\tau,\tau,\tau')$, is an order one number which is supposed to cancel the mistake that we obtain in perturbation theory  from wavenumbers shorter than the non-linear scale and to give the correct result. As $C^{(2)}_1$, it cannot be computed analytically, and needs to be fitted to observational or numerical data~\footnote{One could split $\alpha$ as the sum of a counterterm~$\alpha_c$ and a renormalized term~$\alpha_r$, where 
\be
\frac{\alpha_c}{\knl^3}\sim \int_{q\gtrsim \knl} d^3q \frac{P(q)}{q^3}\ .
\ee
}.

To verify our argument, let us try to solve~(\ref{eq:Boltzone_eff1}) perturbatively in the term in counterterm in~$\alpha$. The leading terms corresponds to plugging $f^{[0]}(v)$ in the counterterm. This contributes only to the $\vec k=0$ mode, which is uninteresting~\footnote{It would not contribute to the number density of the zero mode as, upon integration over $\vec v$, this term is a total derivative.}. Explicitly, we obtain
\bea\label{eq:counter_a_zero}
&& f_{\rm \alpha,\; (0)}( \vec v, s) = \int_0^s ds_1\, a(s_1)^2  \;     f_{\alpha,\, {\cal S},\,(0)}(  \vec v, s_1)\ , \\ \nn
&&f_{\alpha,\, {\cal S},\,(0)}( \vec v, s_1)= \alpha(s_1)\cdot a(s_1)^6  \Omega_{\rm dm}(s_1)^2 \frac{H^4}{\knl^3}   \frac{\d}{\d v^{i_1}} \left[ \frac{\Pi_2^{i_1i_2}(\hat v)}{v}  \frac{\d}{\d v^{i_2}} \right] f^{[0]}( v)\ ,
\eea
where we used the retarded Green's function at $\vec k=0$.

At the next order in $\vec \d \Phi$, there are two ways to use the term in $\alpha$. The first is to plug $f^{[1]}$ in the term in $\alpha$, and apply to it the retarded Green's function. This corresponds to the perturbative solution where we first apply the perturbative vertex of the standard Boltzmann equation $\sim \frac{\d\Phi}{\d x^i}\frac{\d f}{\d v^i}$, where we substitute $f\to f^{[0]}$. This generates a perturbed solution to which we subsequently apply the counterterm in $\alpha$, generating what we call a source $ f^{[3]}_{\alpha,\, {\cal S},\,(a)}$:
\bea\label{eq:UV_perturbed_sol_a}
&& f^{[3]}_{\rm \alpha,\; (a)}(\vec k, \vec v, s) = \int_0^s ds_1\, a(s_1)^2  \;  e^{- i\, \vec k \cdot\vec v\, (s-s_1)}\;   f^{[3]}_{\alpha,\, {\cal S},\,(a)}(\vec k, \vec v, s_1)\ ,\\ \nn
&& f^{[3]}_{\alpha,\, {\cal S},\,(a)}(\vec k, \vec v, s_1)= i k^i \Phi(k, s_1) \frac{\d}{\d v^i} f_{\alpha,\, {\cal S},\,(0)}(\vec k, \vec v, s_1)\ .
\eea
 This source is exactly of the form (\ref{eq:boltz_a_3}), so it can correct the UV dependence of this contribution. 

However, it is clear that once we include the counterterm in $\alpha$ in the Boltzmann equation, we can have another contribution: one where we first insert the counterterm in $\alpha$ with $f\to f^{[0]}$, obtaining $f_{\rm \alpha,\,(0)}( \vec v, s)$ of (\ref{eq:counter_a_zero}). We can then substitute $f_{\rm \alpha,\,(0)}(  \vec v, s)$ into the standard vertex $\sim \frac{\d\Phi}{\d x^i}\frac{\d f_\alpha}{\d v^i}$, obtaining a solution of the form
\bea\label{eq:UV_perturbed_sol_b}
&& f^{[3]}_{\rm \alpha,\; (b)}(\vec k, \vec v, s) = \int_0^s ds_1\, a(s_1)^2  \;  e^{- i\, \vec k \cdot\vec v\, (s-s_1)}\;   f^{[3]}_{\alpha,\, {\cal S},\,(b)}( k, \vec v, s_1)\ ,
\eea
where 
\bea\label{eq:UV_perturbed_sol_source_b}
&& f^{[3]}_{\alpha,\, {\cal S},\,(b)}(\vec k, \vec v, s_1)=  i\, k^{i_1}\Phi(\vec k,s_1)\; \frac{\d}{\d v^{i_1}}  f_{\rm \alpha,\; (0)}(  \vec v, s_1)\\ \nn
&&= i\, k^{i_1}\Phi(\vec k,s_1)\; \frac{\d}{\d v^{i_1}}  \int_0^{s_1} ds_2\, a(s_2)^2  \;   \alpha(s_2)\cdot a(s_2)^6  \Omega_{\rm dm}(s_2)^2\frac{H(s_2)^4}{\knl^3}   \frac{\d}{\d v^{i_2}} \left[ \frac{\Pi_2(\hat v)^{i_2i_3}}{v}  \frac{\d}{\d v^{i_3}} f^{[0]}( v)  \right]\ .
\eea
In order for our argument to be correct, it better be that there is a perturbative contribution with the functional form as (\ref{eq:UV_perturbed_sol_source_b}). As the labelling suggests, this is the contribution in perturbation theory associated to case $(b)$ of phase space: $\{q_2,q_3\}\gtrsim \knl$. 
Let us indeed check this contribution. After taking the expectation value over the short modes in (\ref{eq:third_order_sol}) and changing variables to $s_3=s_2-\Delta s_3$, we obtain:
\bea
&& f^{[3,111]}_{{\cal S}}(\vec k, \vec v, s_1)_{(b),\,{\rm UV},\, {\rm fast}_\nu}=  i\, k^{i_1}\Phi(\vec k_1,s_1)  \int_{q_2\gtrsim \knl} \frac{d^3 q_2}{(2\pi)^3}\; q_2^{i_2} q_2^{i_3}    \\ \nn
&&\qquad \frac{\d}{\d v^{i_1}} \int_0^{s_1} ds_2\, a(s_2)^2  \;       \, a(s_2)^6 \Omega_{\rm dm}(s_2)^2H(s_2)^4\frac{P_{\delta\delta}(q_2,s_2,s_2)_0}{ q_2^2}  \frac{\d}{\d v^{i_2}} \int_{-\infty}^{0} d\Delta s_3\,   \;  e^{i\, \vec q_2\cdot\vec v\, \Delta s_3}       \frac{\d f^{[0]}(v)}{\d v^{i_3}}\ .
\eea
The structure of  the integral over $\hat q_2$ and then over $\Delta s_3$ is identical to what we encounter in~(\ref{eq:boltz_a_2}) (and done in (\ref{eq:boltz_a_2time})), so we easily obtain
 \bea\nn
&& f^{[3,111]}_{{\cal S}}(\vec k, \vec v, s_1)_{(b),\,{\rm UV},\, {\rm fast}_\nu}=  i\, k^{i_1}\Phi(\vec k_1,s_1)  \frac{\d}{\d v^{i_1}} \int_0^{s_1} ds_2\, a(s_2)^8  \;         \left[\int_{q_2\gtrsim \knl} \frac{d^3 q_2}{(2\pi)^3}\; \frac{P_{\delta\delta}(q_2,s_2,s_2)_0}{4\pi q_2^3}\right]   \\ 
&&\qquad\qquad\qquad  \Omega_{\rm dm}(s_2)^2 H(s_2)^4 \frac{\d}{\d v^{i_2}}\left[\frac{\pi}{2 v} \Pi_2(\hat v)^{i_2i_3}        \frac{\d f^{[0]}(v)}{\d v^{i_3}}\right] \ .
\eea
This has exactly the same functional form of (\ref{eq:UV_perturbed_sol_source_b}) which means that the same value of $\alpha$ that was chosen to fix the contribution for case $(a)$,  will automatically fix the UV dependence of the phase space that we identified as case $(b)$. Notice in particular how the time dependence for $\alpha$ that is chosen in~(\ref{eq:UV_perturbed_sol_a}) to fix case $(a)$, will automatically fix case $(b)$ as well. This is of course consistent with the fact that $\alpha$ originates from the same counterterm $C_1^{(2)}$.

Having taken care of cases $(a)$ and $(b)$, we are now ready to consider the contribution from perturbation theory from the phase space labelled by $(c)$: $\{q_1,q_3\}\gtrsim \knl$. Again, starting from~(\ref{eq:third_order_sol}), and taking the expectation value over the short modes, we obtain the following expression for the resulting source: 
\bea\label{eq:sol_source_c}
&& f^{[3,111]}_{{\cal S}}(\vec k, \vec v, s_1)_{(c),\,{\rm UV}, \, {\rm fast}_\nu}=\\ \nn 
&& =\int_0^{s_1} ds_2\, a(s_2)^2   \int_0^{s_2} ds_3\, a(s_3)^2\;   \frac{\d}{\d v^{i_1}} \int \frac{d^3 q_1}{(2\pi)^3}\;q_1^{i_1} q_1^{i_3}    \;  e^{- i\, (\vec k-\vec q_1)\cdot\vec v\, (s_1-s_2)}   i\,  k_2^{i_2}\, \Phi( \vec k_2,s_2) \\ \nn
&&\qquad\qquad \quad \frac{\d}{\d v^{i_2}}  \;  e^{i\, \vec q_1\cdot\vec v\, (s_2-s_3)}    \frac{\d f^{[0]}(v)}{\d v^{i_3}}\;  a(s_1)^2 a(s_3)^2 \left(\frac{3}{2}\right)^2 H(s_1)^2  \Omega_{\rm dm}(s_1) H(s_3)^2  \Omega_{\rm dm}(s_3)\frac{P_{\delta\delta}(q_1,s_1,s_3)_0}{ q_1^4}\\ \nn
&&\simeq  \int_{-\infty}^{0} d\Delta s_2\,   \int_{-\infty}^{0} d\Delta s_3\, \;   \frac{\d}{\d v^{i_1}} \int \frac{d^3 q_1}{(2\pi)^3}\;q_1^{i_1} q_1^{i_3}    \;  e^{ i\, \vec q_1\cdot\vec v\, \Delta s_2} \left[\frac{\d}{\d v^{i_2}}  \;  e^{i\, \vec q_1\cdot\vec v\, \Delta s_3}  \right.   \\ \nn
&&\qquad\qquad \quad \left.  \; i\,  k^{i_2}\, \Phi( \vec k,s_1 )\, a(s_1)^8 \left(\frac{3}{2}\right)^2 H(s_1)^4  \Omega_{\rm dm}(s_1)^2 \frac{P_{\delta\delta}(q_1,s_1,s_1)_0}{ q_1^4}\,  \frac{\d f^{[0]}(v)}{\d v^{i_3}}\right]\; \\ \nn
&&=  \int_{-\infty}^{0} d\Delta s_2\,   \int_{-\infty}^{0} d\Delta s_3\, \;   \frac{\d}{\d v^{i_1}} \int \frac{d^3 q_1}{(2\pi)^3}\;q_1^{i_1} q_1^{i_3}    \;  e^{ i\, \vec q_1\cdot\vec v\, \left(\Delta s_2+\Delta s_3\right)}     \left( \frac{\d}{\d v^{i_2}}- i\, q_1^{i_2} \Delta s_2\right)    \\ \nn
&&\qquad\qquad \quad   \; i\,  k^{i_2}\, \Phi( \vec k,s_1 )\, a(s_1)^8 \left(\frac{3}{2}\right)^2 H(s_1)^4  \Omega_{\rm dm}(s_1)^2\frac{P_{\delta\delta}(q_1,s_1,s_1)_0}{ q_1^4}\,  \frac{\d f^{[0]}(v)}{\d v^{i_3}}\; \ .
\eea
Analogously to what we did earlier, we have approximated the time of evaluation of the non-oscillating integrand to $s_1$, and have Taylor expanded the oscillating phase in~$k/q_1\lesssim k/\knl\ll 1$.
After changing variables $x_{\{2,3\}}= q_1 v \Delta s_{\{2,3\}}$, we can perform the integration over $\hat q_1$, and  subsequently the integration over $x_{\{2,3\}}$, to obtain:
\bea\label{eq:sol_source_c2}
&& f^{[3,111]}_{{\cal S}}(\vec k, \vec v, s_1)_{(c),\,{\rm UV}, \, {\rm fast}_\nu}=\\ \nn 
&&=  \int_{-\infty}^{0} d x_2\,   \int_{-\infty}^{0} d x_3\, \;   \frac{\d}{\d v^{i_1} } \int \frac{d^3 q_1}{(2\pi)^3}\;q_1^{i_1} q_1^{i_3} \frac{1}{(q_1 v)^2}   \;  e^{ i\, \hat q_1\cdot\hat v\, \left(x_2+x_3\right)}     \left(\frac{\d}{\d v^{i_2}}- i\, \frac{\hat q_1^{i_2}}{v} x_2\right)    \\ \nn
&&\qquad\qquad \quad   \; i\,  k^{i_2}\, \Phi( \vec k,s_1 )\, a(s_1)^8 \left(\frac{3}{2}\right)^2 H(s_1)^4  \Omega_{\rm dm}(s_1)^2\frac{P_{\delta\delta}(q_1,s_1,s_1)_0}{ q_1^4}\,  \frac{\d f^{[0]}(v)}{\d v^{i_3}}\; \\ \nn
&&= \Omega_{\rm dm}(s_1)^2 \left[ \int \frac{d^3 q_1}{(2\pi)^3}\;    \frac{P_{\delta\delta}(q_1,s_1,s_1)_0}{ 4\pi \,q_1^4}\right] \frac{\d}{\d v^{i_1} } \\ \nn
&&\frac{2}{ v^2} \left( \left[c_1 \delta^{i_1 i_3}+c_2 \hat v^{i_1}\hat v^{i_3}\right]   \frac{\d}{\d v^{i_2}}+ \frac{1}{v}\left[c_3 \left(\delta^{i_1 i_3} \hat v^{i_2}+\delta^{i_2 i_1} \hat v^{i_3}+\delta^{i_3 i_2} \hat v^{i_1}\right) +c_4 \hat v^{i_1}\hat v^{i_2}\hat v^{i_3}\right]\right)    \\ \nn
&&\qquad\qquad \quad   \; i\,  k^{i_2}\, \Phi( \vec k,s_1 )\, a(s_1)^8 \left(\frac{3}{2}\right)^2 H(s_1)^4   \frac{\d f^{[0]}(v)}{\d v^{i_3}}\ ,
\eea
where $c_1=1,\quad c_2\simeq -2,\quad c_3\simeq 1,\quad c_4\simeq -5$.

The uncontrolled UV contribution is again degenerate with the contribution of the counterterm~$C^{(2)}_1$.  In turns, including this contribution amounts to modify the Boltzmann equation as follows:
\bea\label{eq:Boltzone_eff2}
&&\frac{\d f(\vec x,\vec v,\tau)}{\d\tau}+ \frac{v^i}{a} \frac{\d f(\vec x,\vec v,\tau)}{\d x^i}-   a \frac{\d\Phi}{\d x^i}\, \frac{\d f(\vec x,\vec v,\tau)}{\d v^i}+\\ \nn
&&\quad a^6 \Omega_{\rm dm}(a)^2 \frac{H^4}{\knl^3}   \frac{\d}{\d v^{i_1}} \left[\alpha_0\cdot \frac{\left( \delta^{i_1 i_2} -  \hat v^{i_1} \hat v^{i_2}\right)}{v}  \frac{\d}{\d v^{i_2}}\right.+\\ \nn
&&\left.\quad\quad\frac{4}{\pi}\alpha_0\cdot a^2 \frac{H}{ \knl\, v}   \left( \frac{\left[c_1 \delta^{i_1 i_3}+c_2 \hat v^{i_1}\hat v^{i_3}\right]}{v}   \frac{\d}{\d v^{i_2}}+ \frac{\left[c_3 \left(\delta^{i_1 i_3} \hat v^{i_2}+\delta^{i_2 i_1} \hat v^{i_3}+\delta^{i_3 i_2} \hat v^{i_1}\right) +c_4 \hat v^{i_1}\hat v^{i_2}\hat v^{i_3}\right]}{v^2}\right) \right.   \\ \nn
&&\left.\qquad\qquad\qquad \quad    \frac{\d \Phi( \vec x,s )}{\d x^{i_2}}\,      \frac{\d }{\d v^{i_3}}+{\cal O}\left(\frac{\d_s}{\knl v},\; \frac{\d_{x}}{\knl}\right) \right]f( \vec x,\vec v,s)= 0\ ,
\eea
with indeed a coefficient that is known in terms of~$\alpha$. To correct the UV contribution of~(\ref{eq:sol_source_c2}), we need to use the new term once in perturbation theory.   We notice immediately that the new term in $\alpha$ is suppressed with respect to the former term in $\alpha$ by a factor of $\frac{H}{ \knl\, v}\lesssim \frac{\d_s}{\knl\,v }\ll 1$, which makes this contribution comparable to the ones suppressed by ${\cal O}\left(\frac{\d_s}{\knl v}\right)$ that originated by the contributions of kind $(a)$ and $(b)$ and that we did not explicitly compute (but we could have, if we wished to go to higher orders). Though our aim is to develop a treatment that is, at least in principle, valid to arbitrary high order we will limit ourselves to explicitly write only the leading counterterms. We will therefore drop the higher-order term in $\alpha$ for the rest of the paper. 

Before moving on, notice interestingly that some of the terms in higher order in $\d_s/(\knl v)$ are associated to taking different moments of the time integrals of the counterterms of the correlation functions (such as $C^{(2)}_1$). This is the mirroring, at the level of the Boltzmann effective equation, of how we are sensitive to the time dependence and the time-non-locality of the the counterterms.

We are finally led to consider the contribution from the region $(d)$: $\{q_1,q_2,q_3\}\gtrsim \knl$. Taking the expectation value over the short modes, we obtain: 
\bea\label{eq:f3stoch}
&& f^{[3,111]}_{{\cal S}}(\vec k, \vec v, s_1)_{(d),\,{\rm UV},\, {\rm fast}_\nu}=\\ \nn
&&\quad\int_{q_1\gtrsim \knl} \frac{d^3 q_1}{(2\pi)^3}\int_{q_2\gtrsim \knl} \frac{d^3 q_2}{(2\pi)^3}\int_{q_3\gtrsim \knl} \frac{d^3 q_3}{(2\pi)^3}\; (2\pi)^3\delta^{(3)}_D(\vec k-\vec q_1-\vec q_2-\vec q_3)\;   \\ \nn
&&\qquad\qquad \frac{\d}{\d v^{i_1}} \int_0^{s_1/(\knl v)} dx_2  \;  e^{- i\, \frac{(\vec q_2+\vec q_3)}{\knl}\cdot\hat v\, (x_1-x_2)} \frac{1}{(\knl\, v)^2} \frac{\d}{\d v^{i_2}} \int_0^{x_2} dx_3  \;  e^{- i\, \frac{\vec q_3}{\knl}\cdot\hat v\, (x_2-x_3)}  \\ \nn
&&\qquad  a(s_1)^4 i\, \frac{q_1^{i_1}}{q_1^2} \frac{q_2^{i_2}}{q_2^2} \frac{ q_3^{i_3}}{q_3^2} \Omega_{\rm dm}(s_1)^3H(s_1)^6\langle \delta(\vec q_1,s_1) \delta( \vec q_2,s_1)   \delta(\vec q_3,s_1)\rangle_0 \frac{\d f^{[0]}(v)}{\d v^{i_3}}=0 \ .
\eea
where $x_{\{2,3\}}= s_{\{2,3\}} \knl\, v$. The result vanishes because in the free limit the short wavelength fields are Gaussian. Again, this contribution should be degenerate with the contribution from counterterms, which will not vanish (indeed, short wavelength fields are not Gaussian at all). If we replace  $\langle \delta(\vec q_1,s_1) \delta( \vec q_2,s_1)   \delta(\vec q_3,s_1)\rangle_0$ with $\langle [\delta(\vec q_1,s_1) \delta( \vec q_2,s_1)   \delta(\vec q_3,s_1)]_\C\rangle$, we see that we have several different contributions. There is a non vanishing expectation value, proportional to $C^{(3)}_1$. However, this contribution affects only the zero mode of the density fluctuations, which is of scarce interest (and in particular it does not contribute to the overall density perturbation at zero momentum). The terms in $C^{(3)}_{\{2,3\}}$ correspond to the correction of the deformation of this expectation value in the presence of long-wavelength fields. Upon substituting these terms in (\ref{eq:f3stoch}) one could perform the integrals over $\hat q$ and $\Delta s$ as we did before, deriving the effective modification to the Boltzmann equation. We will discuss these kind of response terms later for the two-point function, while we do not discuss the ones associated to the three-point function more explicitly here. In fact, we immediately see that the structure of the two time integrals, each one oscillating very fast, makes this term suppressed by a double power of $(\knl\, v)$, and so it is subleading with respect to the terms we consider here. Another interesting contribution comes from the stochastic term, which can be similarly treated and that we will discuss later on.

At this point, we have analyzed how the various contributions from the third order perturbative solution $f^{[3]}$ in~(\ref{eq:third_order_sol}) are degenerate with the corresponding contribution of counterterms for the correlation functions of dark matter fields, and how their effect can be incoded by modifying the Boltzmann equation to include additional terms of the form given in~(\ref{eq:Boltzone_eff2}). So far, we have not included in the perturbative solution the contribution from the non-linearities of the gravitational field. To be explicit, if we wish to go to third order in the fluctuations, we could insert a third-order $\Phi^{(3)}$ in the linear solution $f^{[1]}$ in~(\ref{eq:first_order_sol}), or a second order $\Phi^{(2)}$ in the quadratic solution $f^{[2]}$ in~(\ref{eq:second_order_sol}). These perturbative solutions will contain convolution integrals involving high wavenumber modes, and, similarly to the earlier case, they also will not be under perturbative control. Counterterms, re-absorbable in some new terms in the Boltzmann equation, need to be able to correct for these mistakes. Let us analyze these contributions one at the time.

When we plug $\Phi^{(3)}$ in the linear solution $f^{[1]}$ in~(\ref{eq:first_order_sol}), and we focus on long-wavelength $f^{[1]}$, then we evaluate $\Phi^{(3)}$ at the same long-wavelength. In this case, the counterterms present in the equations that describe the dark matter dynamics are able to give the correct $\Phi$ at long wavelength. There is no additional correction needed.

When instead we plug $\Phi^{(2)}$ in the quadratic solution $f^{[2]}$, and we focus on long wavelength  $f^{[2]}$, we obtain several contributions. Let us explore them in more detail. We can write
\bea\label{eq:second_order_sol_three}\nn
&& f^{[2,21+12]}(\vec k, \vec v, s) =-\int \frac{d^3 q_1}{(2\pi)^3}\int \frac{d^3 q_2}{(2\pi)^3}\; (2\pi)^3\delta^{(3)}_D(\vec k-\vec q_1-\vec q_2)\int_0^s ds_1\, a(s_1)^2  \int_0^{s_1} ds_2\, a(s_2)^2\\ \nn
&&\qquad\qquad\qquad q_1^{i_1} q_2^{i_2} \;  e^{- i\, (\vec q_1+\vec q_2)\cdot\vec v\, (s-s_1)}    
\frac{\d}{\d v^{i_1}}  \left( e^{- i\, \vec q_2\cdot\vec v\, (s_1-s_2)}  \frac{\d f^{[0]}(v)}{\d v^{i_2}} \right) \\ 
&&\qquad\qquad\qquad \left[\Phi^{(2)}(\vec q_1,s_1)     \, \Phi^{(1)}( \vec q_2,s_2)+\Phi^{(1)}(\vec q_1,s_1)     \, \Phi^{(2)}( \vec q_2,s_2)\right]\\ \nn
&&\quad=-\int \frac{d^3 q_1}{(2\pi)^3}\int \frac{d^3 q_2}{(2\pi)^3}\; (2\pi)^3\delta^{(3)}_D(\vec k-\vec q_1-\vec q_2)\int_0^s ds_1\, a(s_1)^2  \int_0^{s_1} ds_2\, a(s_2)^2\\ \nn
&&\quad\qquad q_1^{i_1} q_2^{i_2} \;  e^{- i\, (\vec q_1+\vec q_2)\cdot\vec v\, (s-s_1)}    
\frac{\d}{\d v^{i_1}}  \left( e^{- i\, \vec q_2\cdot\vec v\, (s_1-s_2)}  \frac{\d f^{[0]}(v)}{\d v^{i_2}} \right) \left(\frac{3}{2}\right)^2 \frac{H(s_1)^2  \Omega_{\rm dm}(s_1) H(s_2)^2  \Omega_{\rm dm}(s_2)}{q_1^2 q_2^2}\\ \nn
&&  \qquad\qquad\int \frac{d^3 q_3}{(2\pi)^3}\left[D(s_1)^2 F_2( \vec q_3,\vec q_1-\vec q_3) \delta^{(1)}(\vec q_3) \delta^{(1)}(\vec q_1- \vec q_3)   \;  \, \delta^{(1)}( \vec q_2,s_2)\right.\\ \nn 
&&\quad\qquad\qquad\qquad\left.+\delta ^{(1)}(\vec q_1,s_1)   \;  \, D(s_2)^2 F_2( \vec q_3,\vec q_2-\vec q_3) \delta^{(1)}(\vec q_3) \delta^{(1)}(\vec q_2- \vec q_3)\right]\ ,
\eea
where we used the Poisson's equation, and as usual wrote
\be
\delta^{(2)}(\vec k,t)=D(t)^2\int d^3 q\; F_2(\vec q, \vec k-\vec q)\; \delta^{(1)}(\vec q)\,\delta^{(1)}(\vec k-\vec q)\ .
\ee
where $D$ is the growth factor.

Similar to what we did earlier, we can now focus on the associated source term, and on the contribution from UV modes. We fist consider the case where one mode is long-wavelength. We take the expectation value over the short modes, and pull out of the time-integral the slow varying terms. After  algebra passages that are similar to the ones that we did earlier, for the source we obtain
\bea\label{eq:second_order_sol_three_source}
&& f^{[2,21+12]}_{\cal S}(\vec k, \vec v, s_1)_{{\rm UV, fast}_{\nu}} \simeq-\int_{q_1\gtrsim \knl} \frac{d^3 q_1}{(2\pi)^3}  \int_0^{s_1} ds_2\, a(s_1)^2\\ \nn
&&\quad\quad q_1^{i_1} (\vec k-\vec q_1)^{i_2} \;  
\frac{\d}{\d v^{i_1}}  \left( e^{- i\, (\vec k-\vec q_1)\cdot\vec v\, (s_1-s_2)}  \frac{\d f^{[0]}(v)}{\d v^{i_2}} \right) \left(\frac{3}{2}\right)^2 \frac{H(s_1)^4}{q_1^2 (\vec k-\vec q_1)^2} \Omega_{\rm dm}(s_1)^2 \\ \nn
&&  \quad\quad 2D(s_1)^3 \left[ F_2( \vec q_1-\vec k,\vec k) P_{\delta\delta}(|\vec k-\vec q_1|,s_1,s_1)_0\right.
\left.+P_{\delta\delta}(q_1,s_1,s_1)_0  \;   F_2(- \vec q_1,\vec k) \right] \delta^{(1)}(\vec k) +{\cal O}\left( \frac{\d_s}{\knl\, v}\right)\\ \nn
&&\quad\simeq \int_{q_1\gtrsim \knl} \frac{d^3 q_1}{(2\pi)^3} a(s_1)^2  \int_0^{s_1} ds_2\,  q_1^{i_1}  q_1^{i_2} \;  
\frac{\d}{\d v^{i_1}}  \left( e^{i\,\vec q_1\cdot\vec v\, (s_1-s_2)}  \frac{\d f^{[0]}(v)}{\d v^{i_2}} \right) \left(\frac{3}{2}\right)^2 \frac{H(s_1)^4}{q_1^4} \Omega_{\rm dm}(s_1)^2\\ \nn
&&\qquad  \quad2D(s_1)^3\left[F_2( \vec q_1,\vec k)+F_2( -\vec q_1,\vec k)\right] \,
P_{\delta\delta}(q_1,s_1,s_1)_0\, \delta^{(1)}(\vec k) +{\cal O}\left(\frac{k}{\knl}, \frac{\d_s}{\knl\, v}\right)
\eea
where in the second passage we have Taylor expanded in $k/q_1\lesssim k/\knl\ll 1$. In the same limit, we have that
\be\label{eq:F2limit}
\lim_{k/q\to 0}F_2(\vec q, \vec k)=\frac{1}{2}\frac{\vec q\cdot\vec k}{k^2}+\left(\frac{5}{7}+\frac{2}{7}\frac{(\vec q\cdot\vec k)^2}{q^2 k^2}\right) \ .
\ee
By plugging (\ref{eq:F2limit}) into  (\ref{eq:second_order_sol_three_source}), we can now perform the integral over $\hat q_1$ and over $s_2$. We obtain
\bea\label{eq:second_order_sol_three_source_2}\nn
&& f^{[2,21+12]}_{\cal S}(\vec k, \vec v, s_1)_{{\rm UV, fast}_{\nu}} \simeq \frac{9}{2}a(s_1)^2 D(s_1)^3 \Omega_{\rm dm}(s_1)^2 H(s_1)^4 \delta^{(1)}(\vec k)  \int_{q_1\gtrsim \knl} \frac{d^3 q_1}{(2\pi)^3} \frac{P_{\delta\delta}(q_1,s_1,s_1)_0}{4\pi q_1^3}\, \\ \nn
&& \frac{\pi}{7}\frac{\d}{\d v^{i_1}} \left[ 5 \left( \frac{(\delta^{i_1i_2}-\hat v^{i_1}\hat v^{i_2})}{v} \right)\right.  \\ \nn
&&  \left. \frac{1}{2 v}  \left(\left(\delta^{i_1 i_2}+2\hat k^{i_1}\hat k^{i_2}\right)-\left(\delta^{i_1 i_2} (\hat k\cdot\hat v)^2+\hat v^{i_1}\hat v^{i_2}+2 \hat v\cdot \hat k \left(\hat v^{i_1}\hat k^{i_2}+\hat v^{i_2}\hat k^{i_1}\right)\right)+3\left(\hat v^{i_1}\hat v^{i_2} (\hat v\cdot\hat k)^2\right) \right)\right.\\ 
&&\left.+{\cal O}\left(\frac{k}{\knl}, \frac{\d_s}{\knl\, v}\right)\,\right]\frac{\d f^{[0]}(v)}{\d v^{i_2}}\ .
\eea
We can see that there is a contribution proportional to $\delta(\vec k)$, without overall factors of $q/k$ in front of it.  The origin of this contribution is related to the fact that when we consider the expectation value of two short wavelength fields $\delta(\vec q_1)\delta(\vec q_2)$, at some higher order in perturbation theory, we are actually to consider this expectation value in the background of a long-wavelength field.

As expected, there must be a counterterm able to correct for the mistake coming from using perturbation theory in this regime. This is indeed what the counterterms in $C_{\{2,3\}}^{(2)}$ in the third line of (\ref{eq:approx}) account for. If used in perturbation theory when evaluating $\langle\delta(\vec q,s)\delta(\vec q',s)\rangle\supset \langle[\delta(\vec q,s)\delta(\vec q',s)]_\C\rangle$, these terms would indeed give the same functional contribution as we find here~\footnote{The perturbative solution used to derive (\ref{eq:second_order_sol_three_source_2}) obtained a specific functional dependence on the tidal tensor $\d_i\d_j\Phi$, relating the two possible tensorial contributions and associating to them a specific time dependence. This functional form goes beyond what imposed purely by the symmetries of the problem and by the part of the solution for which we can rely on perturbation theory, and cannot be trusted. The counterterms $C_{\{2,3\}}^{(2)}$ are indeed able to fix this completely.}.

As much as the term parametrized by $\alpha$ in~(\ref{eq:Boltzone_eff2}) was able to account for the perturbation-theory mistake when approximating the expectation value of two short-wavelength gravitational fields with their value in the absence of long modes, quite logically, the contribution in the second line of (\ref{eq:second_order_sol_three_source_2}) has exactly the same functional form that we obtain if promote the parameter  $\alpha$ in (\ref{eq:Boltzone_eff2}) to be a functional of the long wavelength dark matter fields
\bea\label{eq:alpha_fun}
&&\alpha(\tau)\qquad \to\qquad \alpha\left(\tau,\d_i\d_j\Phi(\vec x,\tau)_{\rm past\;light \;cone},\ldots\right)\\ \nn
&&\ =\alpha_0(\tau)+\int^\tau d\tau' \left[\frac{\alpha_1(\tau,\tau')}{H}\, \d^2\Phi(\xfl(\vec x,\tau,\tau'),\tau')+\Delta_{\rm stoch}(\xfl(\vec x,\tau,\tau'),\tau')+ {\cal O}\left((\d_i\d_j\Phi)^2,\frac{\d_i}{\knl}\right)\right],
\eea
where $\Delta_{\rm \Phi, stoch}$ represents the stochastic term (related to the time integral of $\epsilon_{\rm stoch.}$), and we use perturbatively this term using for $f$ the unperturbed distribution. In general the right hand side of~(\ref{eq:alpha_fun}) represents the most general, local in space but non-local in time, dependence on the dark matter field that is allowed by diff. invariance. 

The contribution to the third line of~(\ref{eq:second_order_sol_three_source_2}) instead corresponds to a term with a different tensorial structure, that we can account for in the Boltzmann equation by introducing a new term. We are therefore led to upgrade~(\ref{eq:Boltzone_eff2}) to 
\bea\label{eq:Boltzone_eff3}
&&\frac{\d f(\vec x,\vec v,\tau)}{\d\tau}+ \frac{v^i}{a} \frac{\d f(\vec x,\vec v,\tau)}{\d x^i}-   a \frac{\d\Phi}{\d x^i}\, \frac{\d f(\vec x,\vec v,\tau)}{\d v^i}+\\ \nn
&& a^6 \Omega_{\rm dm}(a)^2 \frac{H^4}{\knl^3}\frac{\d}{\d v^{i_1}}\\ \nn
&&\qquad\left\{\left(\alpha_0(\tau)+\int^\tau d\tau'\left[ \frac{\alpha_1(\tau,\tau')}{H}\, \d^2\Phi(\xfl,\tau')+ \Delta_{\rm stoch,1}(\xfl,\tau')+{{\cal O}}\left((\d^2\Phi)^2, \Delta_{\rm stoch}\d^2\Phi,\; \frac{\d_x}{\knl}\right) \right]\right)\right.  \\ \nn
&&\quad\qquad\qquad\qquad\qquad\times\ \left.  \left[ \frac{\Pi_2(\hat v)^{i_1 i_2}}{v}  +{\cal O}\left(\frac{\d_\tau}{\knl v},\; \frac{\d_x}{\knl}\right) \right] \right. \\ \nn
&&\qquad\quad\qquad\qquad\quad\left.+\left(\int^\tau d\tau' \frac{\beta_1(\tau,\tau')}{H} \d_{i_3}\d_{i_4}\Phi(\xfl,\tau')+3+{{\cal O}}\left((\d^2\Phi)^2, \Delta_{\rm stoch}\d^2\Phi\; \frac{\d_x}{\knl}\right)\right)\right. \\ \nn
&&\quad\qquad\qquad\qquad\qquad \times \ \left. \left[\frac{\Pi_4(\hat v)^{i_1 i_2 i_3 i_4}}{v}+{\cal O}\left(\frac{\d_\tau}{\knl v},\; \frac{\d_x}{\knl}\right)\right]\right\} \frac{\d}{\d v^{i_2}}f( \vec x,\vec v,\tau)=0\ .
\eea
where we defined the projection tensor
\bea
&&\Pi_4(\hat v)^{i_1 i_2 i_3 i_4}=\left(\delta^{i_1 i_2}\delta^{i_3 i_4} +2 \delta^{i_1 i_3}\delta^{i_2 i_4}\right)\\ \nn
&&\ \quad\qquad\qquad-\left(\delta^{i_1 i_2} \hat v^{i_3}\hat v^{i_4}+\hat v^{i_1}\hat v^{i_2}\delta^{i_3 i_4}+2 \hat v^{i_4} \left(\hat v^{i_1}\delta^{i_2 i_3}+\hat v^{i_2}\delta^{i_1 i_3}\right)\right)+3\left(\hat v^{i_1}\hat v^{i_2} \hat v^{i_3}\hat v^{i_4}\right) \ .
\eea
Let us add an important observation. Contrary to the counterterms that we introduced in~(\ref{eq:Boltzone_eff2}), the new counterterms introduced in~(\ref{eq:Boltzone_eff3}) have been found by using perturbation theory for the short-wavelength gravitational field, which is not trustable. Because of this, we indeed allowed a generic coefficient $\beta$, but one might wonder how we can trust the particular tensorial structures in~(\ref{eq:Boltzone_eff3}), given by $\Pi_2(\hat v)$ and $\Pi_4(\hat v)$. However, we can notice that the structure of $\Pi_{\{2,4\}}$ is dictated by simply using the Green's function of the neutrinos, which we can trust, and the fact that the response of the short wavelength fields can be proportional to $\d_i\d_j\Phi$. The derivation therefore can be trusted.  Indeed, we would obtain the same result by using the terms in $C_{\{2,3\}}^{(2)}$ in the third line of~(\ref{eq:approx}) when computing $\langle\delta(\vec q,s)\delta(\vec q',s)\rangle$.  This observation makes it clear that the effective Boltzmann equation that we obtain is not the most general one that we would obtain if we assumed complete ignorance about the UV behavior, apart for some symmetries to be realized. Indeed, it is strongly affected by the fact that we assume we can trust perturbation theory for fast neutrinos in an external gravitational fields, at all wavenumbers.

As implicitly already affirmed in~(\ref{eq:Boltzone_eff3}), one finally notices that the same tensorial structure arises if one uses the stochastic counterterms present in~(\ref{eq:approx}), as well as those present in $[\delta]_\C$. At leading order, the stochastic terms present in the two-point function in~(\ref{eq:approx}) correct for some of the UV dependence that we obtain when we consider both modes in $f^{[2]}$ to be in the UV, while those in~(\ref{eq:counter3}) correct for the UV dependence in $f^{[3]}$ when all the three modes are taken beyond the linear regime. The stochastic terms are characterized by a Poisson distribution (the stochastic term of $[\delta]_\C$ must be proportional to two additional derivatives because $\delta$ is a conserved quantity). This Poisson statistics, plus translation invariance, ensures that, upon taking expectation values in correlation functions of the stochastic terms in $f^{[2]}$ or in $f^{[3]}$ with other stochastic terms, the angular dependence on the UV momenta is the same as the one that was present in the non-stochastic counterterms, and therefore integration over the angles of the UV momenta and over time will lead to the same tensorial structure $\Pi_{\{2,3,\ldots\}}(\hat v)$ as for the non-stochastic counterterms, as already assumed in~(\ref{eq:Boltzone_eff3}).

Coming back to~(\ref{eq:second_order_sol_three_source}), we should consider one specific contribution that naively comes at next-to-leading order in expanding the slow-varying integrand around the latest time, but that is actually boosted by a factor of $q/k\gtrsim \knl/k$, and so can become important for $k\lesssim \kfs$. Such a term is 
\bea\label{eq:second_order_sol_three_source_next}\nn
&& f^{[2,21+12]}_{\cal S}(\vec k, \vec v, s_1)_{{\rm UV, fast}_{\nu},{\rm subl.}}\quad \supset \quad \frac{9}{2} a(s_1)^2 H(s_1)^4 \Omega_{\rm dm}(s_1)^2\int_{q_1\gtrsim \knl} \frac{d^3 q_1}{(2\pi)^3} \frac{q_1^{i_1}  q_1^{i_2}}{q_1^4} P_{\delta\delta}(q_1,s_1,s_1)_0\, \\ 
&&\quad\quad  \;  \int^0_{-\infty} d\Delta s_2 
\frac{\d}{\d v^{i_1}}  \left( e^{- i\, (\vec k-\vec q_1)\cdot\vec v\, \Delta s_2}  \frac{\d f^{[0]}(v)}{\d v^{i_2}} \right)  \\ \nn
&&  \quad\quad D(s_1)^2 \frac{\d D(s_1)}{\d s_1}\Delta s_2  \frac{1}{2}\frac{\vec q_1\cdot\vec k}{k^2}  \delta^{(1)}(\vec k) \ ,
\eea
where we have focused on the leading term in $k/q\lesssim k/\knl\to 0$ that does not cancel anymore from the combination $\left[F_2( \vec q_1,\vec k)+F_2( -\vec q_1,\vec k)\right] $.  As usual, we can perform the integral over $\hat q$ and $\Delta s_2$, obtaining
\bea\label{eq:second_order_sol_three_source_next2}\nn
&& f^{[2,21+12]}_{\cal S}(\vec k, \vec v, s_1)_{{\rm UV, fast}_{\nu},{\rm subl.}}\quad \supset \quad \frac{9\pi}{4}\, a(s_1)^2 H(s_1)^4 \Omega_{\rm dm}(s_1)^2\int_{q_1\gtrsim \knl} \frac{d^3 q_1}{(2\pi)^3} \frac{ P_{\delta\delta}(q_1)_0}{4\pi q_1^3}\,   \\ 
&&\quad\quad  \frac{\d}{\d v^{i_1}}\left(  \frac{\Pi_3(\hat v)^{i_1 i_2 i_3}}{v^3}      \frac{\d f^{[0]}(v)}{\d v^{i_2}} \right)   D(s_1)^2 \frac{\d D(s_1)}{\d s_1}  \frac{1}{2}i \frac{k^{i_3}}{k^2}  \delta^{(1)}(\vec k) \ ,
\eea
where
\be
\Pi_3(\hat v)^{i_1 i_2 i_3}= \left(\delta^{i_1 i_2} \hat v^{i_3}+\delta^{i_3 i_1} \hat v^{i_2}+\delta^{i_2 i_3} \hat v^{i_1}\right) -3\, \hat v^{i_1}\hat v^{i_2}\hat v^{i_3}\ .
\ee
Now, we remind that $\vdm$ has the following expression at linear order:
\be
\vdm^{(1)}(k,s)= a(s_1)\frac{\d D(s_1)}{\d s_1}i \frac{\vec k}{k^2} \delta^{(1)}(\vec k)\ .
\ee 
Therefore, we can write (\ref{eq:second_order_sol_three_source_next2}) as
\bea\label{eq:second_order_sol_three_source_next3}\nn
&& f^{[2,21+12]}_{\cal S}(\vec k, \vec v, s)_{{\rm UV, fast}_{\nu},{\rm subl.}}\quad \supset \quad  \frac{9\pi}{8}\, a(s_1) H(s_1)^4 \Omega_{\rm dm}(s_1)^2\int_{q_1\gtrsim \knl} \frac{d^3 q_1}{(2\pi)^3} \frac{ P_{\delta\delta}(q_1,s_1,s_1)_0}{4\pi q_1^3}\,   \\ 
&&\quad\quad  \frac{\d}{\d v^{i_1}}\left(  \frac{\Pi_3(\hat v)^{i_1 i_2 i_3}}{v^3}   \cdot   \frac{\d f^{[0]}(v)}{\d v^{i_2}} \right)   v_{\rm dm}^{i_3}(\vec k,s_1)  \ ,
\eea
We notice that this term depends on the same UV-uncontrolled parameter that was present in~(\ref{eq:boltz_a_3}). Therefore, the coefficient of the counterterm that renormalizes the term in~(\ref{eq:second_order_sol_three_source_next3}) must be known in terms of the one that renormalized~(\ref{eq:boltz_a_3}), which we called either $C_{1}^{(2)}$ at the level of the renormalization of the dark matter correlation function, or  $\alpha$ at the level directly of the Boltzmann equation. The tensorial structure $\frac{\Pi_3(\hat v)^{i_1 i_2 i_3}}{v^2}$ in (\ref{eq:second_order_sol_three_source_next2}) is also very suggestively the one that emerges from $\frac{\Pi_2^{i_1i_2}(\hat v)}{v}$ if we replace the velocity with the relative velocity: $\vec v\to \vec v\r=\vec v-\vdm$ and take the leading term in $v_{\rm dm}/v\lesssim \vnl/v\ll 1$ (which is the order we are working on in fluctuating fields).  The numerical coefficient is also the correct one. We therefore deduce that in order to renormalize this term, we simply need to upgrade our counterterms in the Boltzmann equation to be a function of the relative velocity. Indeed the need of such an upgrade could have been understood directly in terms of symmetries. The following set of transformations keep the Boltzmann equation invariant:
\bea\label{eq:symmetry}
&&\tilde t=t\\ \nn
&& \tilde x^i=x^i+C^i(t)\\ \nn
&& \tilde v^i =v^i+\dot C^i(t)\\\nn
&& \tilde f(\tilde x,\tilde v,\tilde t)= f(x,v,t)\\ \nn
&& \d_i\Phi =\d_i\Phi+ \ddot C^i(t)\ .
\eea
 Under these transformation the velocity is not invariant, but relative velocity is, and therefore the new counterterms in the Boltzmann equation should depend on $v$ through $v\r$.
 
 As we have discussed, each counterterm in the Boltzmann equation must correspond to a counterterm in the dark matter correlation function of~(\ref{eq:deltafluid}). Indeed, the promotion of the term in $\alpha$ of $\vec v$ to $\vec v\r$ is simply identified by using the second line of~(\ref{eq:deltafluid}) multiplied with the term in $C_{2}^{(2)}$ in~(\ref{eq:approx}). One clearly sees that the two terms are related with no relative freedom. 
 
We are therefore led to write
\bea\label{eq:Boltzone_eff4}
&&\frac{\d f(\vec x,\vec v,\tau)}{\d\tau}+ \frac{v^i}{a} \frac{\d f(\vec x,\vec v,\tau)}{\d x^i}-   a \frac{\d\Phi}{\d x^i}\, \frac{\d f(\vec x,\vec v,\tau)}{\d v^i}+a^6 \frac{H^4}{\knl^3} \Omega_{\rm dm}(a)^2\;\frac{\d}{\d v^{i_1}} \\\ \nn
&&\quad \left\{\left(\alpha_0(\tau)+\int^\tau d\tau' \left[\frac{\alpha_1(\tau,\tau')}{H}\, \d^2\Phi(\xfl,\tau')+ \Delta_{\rm stoch,1}(\xfl,\tau')+{{\cal O}}\left((\d^2\Phi)^2, \Delta_{\rm stoch}\d^2\Phi,\; \frac{\d_x}{\knl}\right)\right] \right)\right.  \\ \nn
&&\qquad\qquad\qquad\qquad\times\ \left.  \left[ \frac{\Pi_2(\hat v\r)^{i_1 i_2}}{v\r}  +{\cal O}\left(\frac{\d_\tau}{\knl v},\; \frac{\d_x}{\knl}\right) \right] \right. \\ \nn
&&\qquad\qquad\quad\left.+\left(\int^\tau d\tau' \frac{\beta_1(\tau,\tau')}{H} \d_{i_3}\d_{i_4}\Phi(\xfl,\tau')+\frac{\d_{i_3 i_4}}{\d^2} \Delta_{\rm stoch,2}(\xfl,\tau')+{{\cal O}}\left((\d^2\Phi)^2, \Delta_{\rm stoch}\d^2\Phi\; \frac{\d_x}{\knl}\right)\right)\right. \\ \nn
&&\qquad\qquad\qquad\qquad \times \ \left. \left[\frac{\Pi_4(\hat v\r)^{i_1 i_2 i_3 i_4}}{v\r}+{\cal O}\left(\frac{\d_\tau}{\knl v},\; \frac{\d_x}{\knl}\right)\right]+\ldots\right\} \frac{\d}{\d v^{i_2}}f( \vec x,\vec v,\tau)=0\ ,
\eea
where $\dots$ includes terms that are subleading in  $\knl / (H\, v)\simeq \vnl/v\ll1$, whose form can be computed in a very analogous way.

\subsubsection*{Mini-Summary for Fast Neutrinos}

Eq.~(\ref{eq:Boltzone_eff4}) represents our effective Boltzmann equation to describe the dynamics of neutrinos characterized by a velocity $v$ is faster than the velocity whose associated free streaming length is of order the non-linear scale, $\vnl$. It was derived by noticing that, at leading order in $f_\nu$, the perturbative solution of the Boltzmann equation for neutrinos involves products of dark matter fields evaluated at the same location. These products are sensitive to uncontrolled contributions from modes inside the non-linear regime, and can be corrected by the addition of suitable counterterms that renormalize them.  In the limit $\vnl/v\ll1$, the contribution of these counterterms can be encoded in suitable modifications of the Boltzmann equation that amounts to the inclusion of additional operators that need to be added to the solution in a perturbative manner. 

The perturbative expansion is organized in powers of the fluctuating fields $\delta$,  $k/\knl\ll1$ and $\vnl/v\ll1$. This leads to the radical simplification of the counterterms that we see in~(\ref{eq:Boltzone_eff4}). Several ingredients indeed went into~(\ref{eq:Boltzone_eff4}) taking that form. In particular we highlight a couple. First, it is assumed that the perturbative expansion is $\vnl/v$ is convergent for the velocity of interest, independently of the wavenumber. Given that $\vnl/v\lesssim 1$ for all the velocities of interest, this seems to be a safe assumption. Second, this expansion in $\vnl/v\ll1$ allows us to be quite insensitive to the unknown time and momentum dependence of the UV contribution of the diagram and the counterterms, so that we can perform several integrations that are present in the perturbative expansion, obtaining the very constrained and explicit structure of the counterterms that is characteristic of~(\ref{eq:Boltzone_eff4}).

The recipe to construct the effective Boltmann equation~(\ref{eq:Boltzone_eff4}) at the desired order is very simple. One solves the perturbative series for the ordinary Boltzmann equation; one then substitute the UV-sensitive terms with the counterterms provided around~(\ref{eq:approx}); and finally performs the integrals over the angles and time by Taylor expanding the slow-varying fields around the latest time. By trusting perturbation theory for the fast neutrinos at all $k$'s, this gives a form of the effective Boltzmann equation that is much more constrained than what would be allowed purely in terms of symmetries.

\subsection{Slow Neutrinos\label{app:slowneutrinos1}}

The main ingredients of the construction of the effective Boltzmann equation for fast neutrinos are that, first, we could trust perturbation theory for all wavenumbers, and second, we could perform some of the convolutions over the momenta and time explicitly. In the case of slow neutrinos, we could attempt to do the same process. In fact, we can start from the UV contribution for $f^{[3,111]}$ coming from the phase space region $(a)$: $\{q_1,q_2\}\gtrsim \knl$,  that we have in~(\ref{eq:source_a_initial}). We could then notice that for slow neutrinos, $v\Delta s\lesssim 1/\knl$, which implies that, for $k\lesssim \knl$, the phase of the Green's function is not oscillating, and can be Taylor expanded in $k/\knl\ll 1$. From~(\ref{eq:source_a_initial}), we obtain
\bea\label{eq:source_a_initial_slow}\nn
&& f^{[3,111]}_{\cal S}(\vec k, \vec v, s_1)_{\rm (a), UV,\,\rm slow_\nu} \simeq \int_{q_1 \gtrsim \knl} \frac{d^3 q_1}{(2\pi)^3} \;   i\, q_1^{i_1}\frac{\d}{\d v^{i_1}}\left[ \int_0^{s_1} ds_2\, a(s_2)^2 \;\left[1-i\, \vec k \cdot\vec v(s_1-s_2)+\ldots\right] \right.   \\ \nn
&&\quad  \;  e^{i\,\vec q_1\cdot\vec v\, (s_1-s_2)}   i\,  (-q_1^{i_2})\,\left(\frac{3}{2}\right)^2 a(s_1)^2 a(s_2)^2 H(s_1)^2  \Omega_{\rm dm}(s_1) H(s_2)^2  \Omega_{\rm dm}(s_2)\frac{P_{\delta\delta}(q_1,s_1,s_2)_0}{q_1^4}  \\ 
&&\quad \left. \times\ \frac{\d}{\d v^{i_2}} f^{[1]}(\vec k,\vec v,s_2)\right]\ .
\eea
For the case of fast neutrinos, we were able to integrate this expression in $\hat q$ and $s_2$. This was useful because it allowed us to determine the functional dependence on $\vec v$ in terms of just one, or a few, coefficients. We could clearly integrate~(\ref{eq:source_a_initial_slow}) in $\hat q_1$, obtaining expressions similar to what we found in~(\ref{eq:boltz_a_2}). For the leading term, we would obtain one term proportional to $\delta^{i_1i_2}$ and another to $\hat v^{i_1}\hat v^{i_2}$, with oscillating-in-time prefactors that need to be integrated in time. These prefactors will have the crucial difference with respect to those of~(\ref{eq:boltz_a_2}) that now the correlation function $P_{\delta\delta}(q_1,s_1,s_2)_0$, as well as other time dependent factors, are also part of the integrand: the velocity can be so small that the phase of the Green's function does not oscillate, so that we can not bring these factors out of the integral.  This creates a problem because while we know the time dependence of $P_{\delta\delta}(q_1,s_1,s_2)_0$ for $q_1\gtrsim \knl$, this time dependence is incorrect and is indeed corrected by the counterterm $C_1^{(2)}$ from~(\ref{eq:approx}), of which we do not know the time dependence. This would live us with a source of the form
\bea\label{eq:source_a_initial_slow2}\nn
&& f^{[3,111]}_{\cal S}(\vec k, \vec v, s)_{\rm (a), UV,\,\rm slow_\nu} \simeq \frac{\d}{\d v^{i_1}}\left(g_1(v,s) \delta^{i_1 i_2}+g_2(v,s) \hat v^{i_1}\hat v^{i_2}+{\cal O}\left(\frac{\d_i}{\knl}\right)\right) \frac{\d}{\d v^{i_2}} f^{[1]}(\vec k,\vec v,s)\ ,\\
\eea
where in integrating $f^{[1]}$ in time, we have used the fact that slow neutrinos do not move more than a distance equivalent to the non-linear scale during an Hubble time.
The unknown functions $g_1(v,s)$ and $g_2(v,s)$ are quite harmful because once we wish to compute the overall density, we need to integrate over $\vec v$. But (\ref{eq:source_a_initial_slow2}) has an even worse problem. Similarly to what we did for fast neutrinos, it is obtained by trusting perturbation theory for the slow neutrinos. As we discussed at the end of Sec.~\ref{sec:perturbativebegin}, perturbation theory for the neutrinos amounts for $k\gtrsim \kfs$ to expanding in insertions of the gravitational force, which scales as $(\vnl/v)^2$. For slow neutrinos this parameter is order (and potentially even much larger than) one. Trusting the perturbative expansion amounts to assuming that the radius of convergence of the perturbative series is larger the value of the parameter that is of interest. While we assumed that this was the case of fast neutrinos (where $\vnl/v\ll 1$), we do not feel confident that we should assume this also for slow neutrinos~\footnote{Notice that if we were able to assume the validity of perturbation theory for slow neutrinos, it would follow that we should assume it to be valid also for dark matter particles, as, after the first few gravitational interactions, dark matter particles acquire a velocity of order $\vnl$ and behave as slow neutrinos.}. By repeated use of perturbation theory, we would obtain an expression that, extremely schematically, takes the form
\bea\label{eq:source_a_initial_slow3}\nn
&& f^{[3,\ldots]}_{\cal S}(\vec k, \vec v, s)_{\rm (a), UV,\,\rm slow_\nu,\rm schem.} \\ \nn
&&\qquad \sim  \frac{\d}{\d v^{i_1}} \sum_n \int^s d s_1\,\ldots \int^{s_{n-1}} ds_n\; C^{i_1\ldots i_n}(\hat v) (\Delta v)^n \frac{\d^n}{\d v^{i_2}\ldots v^{i_n}} f^{[1]}(\vec k,\vec v,s_n)\ . 
\eea
where $\Delta v$ is the change of velocity due to one insertion of the gravitational potential $\Delta v^{i}\sim \d_i\Phi \cdot H^{-1}$. It is pretty clear that perturbation theory is trying to reconstruct the initial velocity of the particle that at time $s$ has velocity $\vec v$ and position $\vec x$.
Now, if we imagine to include all the counterterms, we can write the general resulting source associated to the UV modes as
\bea\label{eq:source_a_initial_slow4}
&& f^{[3,\ldots]}_{\cal S}(\vec k, \vec v, s)_{\rm UV,\,\rm slow_\nu} \sim\frac{H^3}{\knl^2} \frac{\d}{\d v^{i_1}} \\ \nn
&& \  \  g^{i_1 i_2}\left.\left(\vec v\r,\d_i\d_j\Phi(\vec x,s,\frac{\d^i\d^j}{\d^2}\Delta(\vec x,s)_{\rm stoch},\frac{\d}{\d v^i},\frac{1}{\knl}\frac{\d}{\d x^i},H(s),s\right) \frac{\d}{\d v^{i_2}} f^{[1]}(\vec k,\vec v,s)\right|_{\rm past\;light\;cone}\ ,
\eea
where $g^{i_1,i_2}$ is a dimensionless two-tensor with times scales of order $H$. Notice  that, in the limit $\knl\to\infty$, the source starts proportional to $1/\knl^2$, while in the case of fast neutrinos it started as $1/\knl^3$. This is due to the fact that, while for fast neutrinos the time integral leads to a factor of order $1/(\knl v)$, here it gives a factor of $H^{-1}$.

Notice that since we are inserting an infinite set of correlation functions, it appears to be not quite useful to simply consider the renormalization of the product of dark matter fields as in~(\ref{eq:approx}). Instead, it is rather simple to modify the Boltzmann equation by adding a counterterm that can correct for the UV dependence encoded  by the function~$g^{i_1i_2}$. This can be done by upgrading the Boltzmann equation to the following form
\bea\label{eq:Boltzone_eff5}
&&\frac{\d f(\vec x,\vec v,\tau)}{\d\tau}+ \frac{v^i}{a} \frac{\d f(\vec x,\vec v,\tau)}{\d x^i}-   a \frac{\d\Phi}{\d x^i}\, \frac{\d f(\vec x,\vec v,\tau)}{\d v^i}+a^6 \Omega_{\rm dm}(a)^2\cdot\frac{H^3}{\knl^2}\cdot\frac{\d}{\d v^{i_1}}\left\{ \frac{H}{\knl} \right.\\ \nn
&&\left.\left\{\left(\alpha_0(\tau)+\int^\tau d\tau'\left[ \frac{\alpha_1(\tau,\tau')}{H}\, \d^2\Phi(\xfl,\tau')+ \Delta_{\rm stoch,1}(\xfl,\tau')+{{\cal O}}\left((\d^2\Phi)^2, \Delta_{\rm stoch}\d^2\Phi,\; \frac{\d_x}{\knl}\right) \right]\right)\right.\right.  \\ \nn
&&\quad\qquad\qquad\qquad\times\ \left.  \left[ \frac{\Pi_2(\hat v\r)^{i_1 i_2}}{v\r}  +{\cal O}\left(\frac{\d_\tau}{\knl v},\; \frac{\d_x}{\knl}\right) \right] \right. \\ \nn
&&\quad\quad\left.+\left(\int^\tau d\tau' \left[\frac{\beta_1(\tau,\tau')}{H} \d_{i_3}\d_{i_4}\Phi(\xfl,\tau')+\frac{\d^{i_3 i_4}}{\d^2} \Delta_{\rm stoch,2}(\xfl,\tau')+{{\cal O}}\left((\d^2\Phi)^2, \Delta_{\rm stoch}\d^2\Phi\; \frac{\d_x}{\knl}\right)\right]\right)\right. \\ \nn
&&\quad\qquad\qquad\qquad \times \ \left. \left[\frac{\Pi_4(\hat v\r)^{i_1 i_2 i_3 i_4}}{v\r}+{\cal O}\left(\frac{\d_\tau}{\knl v},\; \frac{\d_x}{\knl}\right)\right]+\ldots\right\}\cdot\frac{\d}{\d v^{i_2}}\left(\Theta(v_\r-a \vnl)f( \vec x,\vec v,\tau)\right)\\ \nn
&&+\;  \left. g^{i_1 i_2}\left(\vec v\r,\d_i\d_j\left.\Phi(\vec x,s)\right|_{\rm past\;light\;cone},\frac{\d^i\d^j}{\d^2}\Delta(\vec x,s)_{\rm stoch},\frac{\d}{\d v^i},\frac{1}{\knl}\frac{\d}{\d x^i},H(s),s\right)  \right\}\\ \nn
&&\quad\qquad\qquad\qquad\qquad\qquad\qquad\qquad\qquad\qquad\qquad\cdot\left.\frac{\d}{\d v^{i_2}}\left(\Theta(a \vnl-v_\r)f( \vec x,\vec v,\tau)\right)\right|_{\rm past\;light\;cone}=0,
\eea
with the constraint that $\left.g^{ij}\right|_{\vec v_\r\to \hat v_r \vnl}$ is equal to minus the same analogous factor that multiplies the fast neutrinos, evaluated at the same non-linear velocity, so that we ensure conservation of the overall number of neutrinos.

Eq.~(\ref{eq:Boltzone_eff5}) represents our effective Boltzmann equation, capable of correcting the mistakes that uncontrolled short-distance dark-matter dynamics induces on the dynamics of neutrinos. Let us comment of two aspects. First, for slow neutrinos, clearly this expression appears quite useless. This is the statement that it is indeed very hard to compute the velocity distribution for slow neutrinos. However we were  able to efficiently make use of the term in $g^{ij}$ when calculating the overall density distribution of neutrinos at $k\ll \knl$ by mapping it into a fluid counterterm. We will do something similar also later on. Second, we notice that the counterterms for slow neutrinos have been separated from those of the fast neutrinos by a suitable $\Theta$-function. Clearly, the expressions for fast neutrinos are valid only for $v\gg\vnl$, and so one might wonder how large is the mistake from pushing the counterterms all the way to $v\sim \vnl$. The answer is that, since we are always interested in $k\lesssim \knl$ and since neutrinos with $v\sim \vnl$ have a free streaming length of order $1/\knl$, not including these neutrinos in the slow ones amounts just to a slight redefinition of the counterterms for the slow neutrinos, that are unknown anyway. So the division of the counterterms does not seem to be an issue, as well as, similarly, any worry of double counting the neutrinos with~$v\simeq\vnl$.

\subsection{Evaluation of the counterterm diagrams using the Effective Boltzmann equation}

\subsubsection{Fast neutrinos \label{sec:counter_fast_boltz}}

For illustrative purposes, we will only compute some of the counterterm diagrams for $P_{\rm diff,\, dm}$, using the effective Boltzmann equation, as we already computed the full answer in the main text using a different formalism. This will allow us to illustrate some non-trivial manipulations in the evaluation of these counterterms. We will evaluate them at linear order in the long wavelength fields, to cancel the UV dependence of the $P_{13}$ diagrams, which is the leading one. Let us start from the counterterms for the fast neutrinos, and with the algebriacally easier associated diagram. Let us therefore consider the contribution of the counterterm in $\alpha_1$. Since the $\alpha_1$ counterterm is already linear in the long wavelength fields, we can replace $f\to f^{[0]}$. We obtain 
\bea\nn
&&P_{\rm diff,\,dm,\,\,\alpha_1}^{f^{[0]}}(k,a_0)=\int d^3 v \int^{a_0}\frac{da_1}{\calH(a_1)} \; e^{-i \, \vec k\cdot\vec v (s-s(a_1))}\,\frac{a_1^2\calH(a_1)^4\Omega_{\rm dm}(a_1)^2}{\knl^3} \alpha_1(a_1,a_2)\,\frac{\d}{\d v^{i_1}} \\\nn
&&\ \  \left[\int^{a_1}\frac{d a_2}{\calH(a_2)}\left(\frac{3}{2}\right) \calH(a_2)^2 \Omega_{\rm dm}(a_2) D(a_2) D(a_0) \,P_{\rm dm,11}(k,a_0)\, \frac{\Pi_2^{i_1i_2}\,(\hat v)}{v} \frac{\d\left(\Theta(v-a_1\vnl) f^{[0]}(v)\right)}{\d v^{i_2}}\right]\ . \\ 
\eea
We can integrate the two derivatives with respect to the velocity by parts, and simplify the resulting expression using that 
\be
\frac{\d}{\d v^{i_2}}\frac{\Pi_2^{i_1i_2}\,(\hat v)}{v} = -4\frac{\hat v^i_1}{v^2} \ .
\ee
We obtain
\bea\nn
&&P_{\rm diff,\,dm,\,\,\alpha_1}^{f^{[0]}}(k,a_0)=\\ \nn
&&\quad=\int d^3 v \int^{a_0}\frac{da_1}{\calH(a_1)} \; e^{-i \, \vec k\cdot\vec v (s-s(a_1))}\,\frac{a_1^2\calH(a_1)^4\Omega_{\rm dm}(a_1)^2}{\knl^3}\; \int^{a_1}\frac{d a_2}{\calH(a_2)}\;\; f^{[0]}(v) \Theta(v-a_1\vnl) \\\nn
&&\quad  \alpha_1(a_1,a_2) \left(-i k^{i_1}\right)\left(\frac{3}{2}\right) \calH(a_2)^2 \Omega_{\rm dm}(a_2) D(a_2) D(a_0) \,P_{\rm dm,11}(k,a_0)\, \left[ -4\frac{\hat v^{i_1}}{v^2}-i k^{i_2}(s-s_1)  \frac{\Pi_2^{i_1i_2}\,(\hat v)}{v} \right]\ .
\eea
We can now perform the integral over $\hat v$~\footnote{The following expressions are useful for these calculations:
\bea
&& \int \frac{d^2\hat v}{4\pi} \; e^{-i \vec k\cdot \vec v \Delta s}= J_0(k v \Delta s)\ , \\ \nn
&& \int \frac{d^2\hat v}{4\pi} \;  \hat v^i  e^{-i \vec k\cdot \vec v \Delta s}= -i\, \hat k^i\, J_1(k v \Delta s)\ , \\ \nn
&&\int \frac{d^2\hat v}{4\pi} \;  \hat v^i \hat v^j  e^{-i \vec k\cdot \vec v \Delta s}= \frac{\delta^{ij}}{3} \left(J_0(k v \Delta s)+J_2(k v \Delta s)\right)-\hat k^i \hat k^j J_2(k v \Delta s)\ ,\\ \nn
&& \int \frac{d^2\hat v}{4\pi} \;  \hat v^i  \hat v^j \hat v^k e^{-i \vec k\cdot \vec v \Delta s}= i\,\left[-\frac{1}{5} \left(\hat k^i \delta^{jk}+\hat k^j \delta^{ki}+\hat k^k \delta^{ij}\right)\, J_3(k v \Delta s)\,-\hat k^i\hat k^j\hat k^k \; \left(J_1(k v \Delta s)-J_3(k v \Delta s)\right)\right]\ ,
\eea
where $J_i(x)$ is the spherical Bessel function of the first kind.
}, to finally obtain
\bea
&&P_{\rm diff,\,dm,\,\alpha_1}^{f^{[0]}}(k,a_0)=P_{\rm dm,11}(k,a_0)\int d^3 v\; f^{[0]}(v)  \int^{a_0}da_1 \; \int^{a_1}d a_2\;\\ \nn
&&\quad \,  \,\frac{a_1^2\calH(a_1)^4\Omega_{\rm dm}(a_1)^2}{\knl^3}\; \left(\frac{3}{2}\right) \calH(a_2)^2 \Omega_{\rm dm}(a_2) D(a_2) D(a_0) \,   \alpha_1(a_1,a_2) \,\Theta(v-a_1\vnl) \\\nn
&&\quad   \left[ \frac{4k}{v^2} J_1(k v(s-s(a_1)))-\frac{2}{3}\frac{k^2 (s-s(a_1))}{v^2}\left(J_0(k v(s-s(a_1)))+J_2(k v(s-s(a_1)))\right)\right]\ .
\eea
The calculation associated to $\beta_1$ proceed is a very similar way, and we do not show it.

We now evaluate the contribution proportional to $\alpha_0$ and that is linear in the long wavelength fields. We have  three kinds of contributions, either if we use the~$\alpha_0$ vertex first and then use the standard gravitational vertex $\sim \d_i\Phi \d f/\d v^i$, which we denote as a diagram labelled by~$f^{[0]}$, as the~$\alpha_0$ counterterm is evaluated on $f^{[0]}$; or a second diagram, where the $\alpha_0$ vertex acts on $f^{[1]}$, and that therefore we denote with $f^{[1]}$; and a third diagram that we obtain when we expand to linear order the relative velocity and we therefore act on $f^{[0]}$ (for a reason we explain below, it is useful to group this diagram together with the one in $f^{[1]}$.).  Let us start from the first one. Since the vertex in $\alpha_0$ will be acted upon by an interaction vertex, which is linear in the fields, we can evaluate it to zeroth order. We obtain
\bea
&&P_{\rm diff,\,dm,\alpha_0}^{f^{[0]}}(k,a_0)=\\ \nn
&&\qquad=\int d^3 v \int^{a_0}\frac{da_1}{\calH(a_1)} \; e^{-i \, \vec k\cdot\vec v (s-s(a_1))} \left(\frac{3}{2}\right) \calH(a_1)^2\Omega_{\rm dm}(a_1) D(a_1) D(a_0) \,P_{\rm dm,11}(k,a_0) \frac{i k^{i_1}}{(-k^2)}\\\nn
&&\qquad \frac{\d}{\d v^{i_1}} \left[\Theta(v-a_1\vnl) \int^{a_1}\frac{d a_2}{\calH(a_2)} \frac{a_2^2\calH(a_2)^4 \Omega_{\rm dm}(a_2)^2}{\knl^3}\alpha_0(a_2)\frac{\d}{\d v^{i_2}}\left[\frac{\Pi_2(\hat v)^{i_2 i_3}}{v}\cdot \frac{\d f^{[0]}(v)}{\d v^{i_3}} \right]\right]\ .
\eea
Now, we can integrate by parts the derivative over the velocities, perform the integral in $\hat v$, and obtain
\bea
&&P_{\rm diff,\,dm,\alpha_0}^{f^{[0]}}(k,a_0)=P_{\rm dm,11}(k,a_0)\,\int d^3 v\; f^{[0]}(v)\; \int^{a_0}da_1\;\int^{a_1}da_2 \;  \\ \nn
&&\quad \left(\frac{3}{2}\right) \calH(a_1)^2\Omega_{\rm dm}(a_1) D(a_1) D(a_0)\,(s-s_1)^2\Theta(v-a_2\vnl)\frac{a_2^2\calH(a_2)^3 \Omega_{\rm dm}(a_2)^2}{\knl^3}\alpha_0(a_2)\\\nn
&&\quad \left[ 4\frac{k^2}{v^2} J_1(k v(s-s_1))-\frac{2}{3}\frac{k^2(s-s_1)}{v}\left(J_0(k v(s-s(a_1)))+J_2(k v(s-s(a_1))\right)\right]\ .
\eea

The contribution in  $P_{\rm diff,\,dm,\alpha_0}^{f^{[1]}}$ will naively give a contribution proportional to $P_{\rm dm,11}(k)$ as $k\to 0$. However, this contribution is cancelled by the contribution from expanding the relative velocity to linear order, as indeed it is this term that makes the counterterm diff. invariant. We therefore group these two contributions together. The calculation follows similar steps as before, and we obtain the final expression
\bea\label{eq:counterfastgroup}
&&P_{\rm diff,\,dm,\alpha_0}^{f^{[1]}}(k,a_0)= -P_{\rm dm,11}(k,a_0)\ \int d^3 v\; f^{[0]}(v)\;  \int^{a_0}da_1\;\int^{a_1}da_2\\ \nn
&&\quad\, \;\alpha_0(a_1)\,\frac{a_1^2\calH(a_1)^3 \Omega_{\rm dm}(a_1)^2}{\knl^3} \,\Theta(v-a_1\vnl)\, \left(\frac{3}{2}\right) \calH(a_2) \Omega_{\rm dm}(a_2) D(a_2) D(a_0)\\\nn
&&\quad \left\{-\frac{2}{v^3}\left[J_0(k v(s-s(a_2)))-J_0(k v(s-s(a_1)))\right]+ \frac{4}{v^3}\left[J_2(k v(s-s(a_2)))-J_2(k v(s-s(a_1)))\right]\right.\\ \nn
&&\quad 6\frac{k (s-s(a_1))}{v^2}  \left[J_1(k v(s-s(a_2)))-J_1(k v(s-s(a_1)))\right] \\ \nn
&&\quad  - \frac{24}{15}\frac{k (s-s(a_1))}{v^2}  \left[J_3(k v(s-s(a_2)))-J_3(k v(s-s(a_1)))\right]+4\frac{k (s-s(a_2))}{v^2} J_1(k v(s-s(a_2))) \\ \nn
&&\quad \left.-\frac{2}{3}\frac{k^2 (s-s(a_1))(s-s(a_2))}{v}\left[J_0(k v(s-s(a_2)))+J_2(k v(s-s(a_2)))\right]\right\}\ .
\eea
The only new ingredient we need in order to derive this expression is to realize that the time dependence of the dark matter velocity can be written as 
\be
\vdm(a)\propto \int^a da_1 \; \frac{3}{2}\calH(a_1) D(a_1)\ .
\ee
This allows us to write a counterterm that, as $k\to 0$, goes manifestly as $k^2P(k)$. Notice furthermore that in deriving (\ref{eq:counterfastgroup}), we have neglected to consider the factors of $\d\Phi$ that originate from Taylor expanding the $\Theta$-function of $v_\r-a\vnl$. They indeed would contribute to a term that goes as $k^0 P(k)$. However, if we are interested in properties of the whole set of neutrinos as we are in our case, any contribution that comes from these terms is cancelled by an identical and opposite one that we obtain when we Taylor expand the  $\Theta-$function for slow neutrinos, that we discuss below. 

\subsubsection{Slow neutrinos\label{app:slowneutrinos2}}

We now compute the analogous contribution for slow neutrinos. We have only one counterterm, which gives
\bea
&&P_{\rm diff,\,dm,{\rm slow}}(k,a_0)=\int d^3 v \int^{a_0}\frac{da_1}{\calH(a_1)} \; e^{-i \, \vec k\cdot\vec v (s-s(a_1))}\;  \frac{\d}{\d{v^{i_1}}} \\\nn
&& \left\langle \left[ g^{i_1 i_2}\left(\vec v\r,\d_i\d_j\left.\Phi(\vec x,s)\right|_{\rm past\;light\;cone},\frac{\d^i\d^j}{\d^2}\Delta(\vec x,s)_{\rm stoch},\frac{\d}{\d v^i},\frac{1}{\knl}\frac{\d}{\d x^i},H^2(s),s\right) \frac{\d \left(\Theta(a \vnl-v_\r) f\right)}{\d v^{i_2}} \right]_{\vec k}\right. \\ \nn
&&\left. \qquad \cdot\  \delta(\vec k',a_0) \right\rangle'\\ \nn
&&=\int d^3 v \int^{a_0}\frac{da_1}{\calH(a_1)} \; e^{-i \, \vec k\cdot\vec v (s-s(a_1))}\; i\, k^{i_1} (s-s(a_1))\;  \\\nn
&& \left\langle \left[ g^{i_1 i_2}\left(\vec v\r,\d_i\d_j\left.\Phi(\vec x,s)\right|_{\rm past\;light\;cone},\frac{\d^i\d^j}{\d^2}\Delta(\vec x,s)_{\rm stoch},\frac{\d}{\d v^i},\frac{1}{\knl}\frac{\d}{\d x^i},H^2(s),s\right) \frac{\d \left(\Theta(a \vnl-v_\r) f\right)}{\d v^{i_2}} \right]_{\vec k}\right. \\ \nn
&&\left. \qquad \cdot\  \delta(\vec k',a_0) \right\rangle'
\eea
where $\langle\ldots\rangle'$ means that we dropped the momentum-conservation $\delta$-function. In the second step, we integrated the derivative in velocity by parts.

Since we are interested in $k\ll \knl$ and $v\lesssim \vnl(a)$, we can Taylor expand the Green's function, to obtain 
\bea
&&P_{\rm diff,\,dm,{\rm slow}}(k,a_0)=\int d^3 v \int^{a_0}\frac{da_1}{\calH(a_1)} \; i\, k^{i_1}(s-s(a_1))\;  \\\nn
&& \left\langle \left[g^{i_1 i_2}\left(\vec v\r,\d_i\d_j\left.\Phi(\vec x,s)\right|_{\rm past\;light\;cone},\frac{\d^i\d^j}{\d^2}\Delta(\vec x,s)_{\rm stoch},\frac{\d}{\d v^i},\frac{1}{\knl}\frac{\d}{\d x^i},H(s),s\right) \frac{\d \left(\Theta(a \vnl-v_\r) f\right)}{\d v^{i_2}} \right]_{\vec k}\right.\\ \nonumber
&&\quad\left.\cdot\ \delta(\vec k',a_0) \right\rangle+{\cal O}(\left(k/\knl\right)^2) \ .
\eea
We can formally evaluate the integral in time, and obtain, at leading order in $k/\knl$ and $\delta$,
\bea\label{eq:finalcounterslowBoltz}
&&P_{\rm diff,\,dm,{\rm slow}}(k,a_0)=c_{1,\rm slow}(a_0)\, f_{\nu,\rm slow}(a_0)\; \frac{k^2}{\knl^2} P_{\rm dm,11}(k,a_0) .
\eea
where we remind that $f_{\nu,\rm slow}(a)\sim\int d^3 v\; f_\nu^{[0]}(v)\; \Theta(a \vnl-v)$,
and $c_{1,\rm slow}$ is an order one number. Notice that (\ref{eq:finalcounterslowBoltz}) is manifestly of order $k^2 P(k)$, for $k\ll\knl$~\footnote{It might be unclear that the terms in $v_\r$, that contain $\d\Phi$, will not lead to terms in $k^0P(k)$. Here is a schematic proof, at leading order in $k\ll \knl$. The idea is the same that lead to the cancellation of the analogous term in (\ref{eq:counterfastgroup}): for any diagram that comes from considering a~$\d\Phi$ inside a $v_\r$ in $g^{ij}$ and $f^{[0]}$, there is an analogous one coming from using $v_\r=v$ and $f^{[1]}$. Schematically
\bea
&&\int d^3 v \; g^{ij}(v_\r,\ldots) \frac{\d f}{\d v^j}\ \supset \ \int d^3 v  \;\left[g^{ij}(v,\ldots) \frac{\d f^{[1]}}{\d v^j}+\frac{\d g^{ij}(v,\ldots)}{\d v^l} v_{\rm dm}^l  \frac{\d f^{[0]}}{\d v^j}\right]\\ \nn
&&\sim \int d^3 v  \;\left[g^{ij}(v,\ldots) \frac{\d}{\d v^j} \left[\int da' e^{-i\vec k\cdot\vec v (s(a)-s(a'))} \frac{3}{2}\calH(a')^2 \d^l\Phi(a')\frac{\d f^{[0]}}{\d v^l}\right]\right.\\ \nn
&&\qquad\qquad\left.+\frac{\d g^{ij}(v,\ldots)}{\d v^l}  \left(\int da' \frac{3}{2}\calH(a')^2 \d^l\Phi(a')\right)  \frac{\d f^{[0]}}{\d v^j}\right]\\\nn
 &&\sim \int d^3 v  \;\left[g^{ij}(v,\ldots)  \int da' \frac{3}{2}\calH(a')^2 \d^l\Phi(a')\frac{\d^2 f^{[0]}}{\d v^j\d v^l}\right.\left.- g^{ij}(v,\ldots)  \int da' \frac{3}{2}\calH(a')^2 \d^l\Phi(a')  \frac{\d^2 f^{[0]}}{\d v^j\d v^l}\right]+{\cal O}(k)\\ \nn
&&\sim {\cal O}(k)\ ,
\eea
where in the last step, in the first term we took the leading term in $k$, and in the second term we integrated by parts the $\d/\d v$.}. 
}

\bibliography{references}

 \end{document}